\documentclass[%
 reprint,
 superscriptaddress,
 amsmath,amssymb,
 aps,
 floatfix,
]{revtex4-1}
\usepackage{graphicx}
\usepackage{dcolumn}
\usepackage{bm}
\usepackage{physics}
\usepackage{mathtools}
\usepackage{amsfonts}
\usepackage{amssymb}
\usepackage{amsmath}
\usepackage[caption=false,labelformat=empty]{subfig}
\usepackage{diagbox}
\usepackage{hyperref}
\usepackage{dsfont}
\hypersetup{linktocpage,colorlinks,citecolor={blue},pdfdisplaydoctitle=true,pdfpagemode=UseOutlines,bookmarksnumbered=true}

\begin{document}

\title{A variational Monte Carlo algorithm for lattice gauge theories with continuous gauge groups: a study of (2+1)-dimensional compact QED with dynamical fermions at finite density}

\author{Julian Bender}
\affiliation{Max-Planck Institute of Quantum Optics, Hans-Kopfermann-Str. 1, 85748 Garching, Germany}
\affiliation{Munich Center for Quantum Science and Technology (MCQST), Schellingstr. 4, D-80799 München}
\author{Patrick Emonts}
\affiliation{Max-Planck Institute of Quantum Optics, Hans-Kopfermann-Str. 1, 85748 Garching, Germany}
\affiliation{Munich Center for Quantum Science and Technology (MCQST), Schellingstr. 4, D-80799 München}
\affiliation{Instituut-Lorentz, Universiteit Leiden, P.O. Box 9506, 2300 RA Leiden, The Netherlands}
\author{J. Ignacio Cirac}
\affiliation{Max-Planck Institute of Quantum Optics, Hans-Kopfermann-Str. 1, 85748 Garching, Germany}
\affiliation{Munich Center for Quantum Science and Technology (MCQST), Schellingstr. 4, D-80799 München}

\date{\today}

\begin{abstract}
Lattice gauge theories coupled to fermionic matter account for many interesting phenomena in both high energy physics and condensed matter physics. 
Certain regimes, e.g. at finite fermion density, are difficult to simulate with traditional Monte Carlo algorithms due to the so-called sign-problem.
We present a variational, sign-problem-free Monte Carlo method for lattice gauge theories with continuous gauge groups and apply it to (2+1)-dimensional compact QED with dynamical fermions at finite density. 
The variational ansatz is formulated in the full gauge field basis, i.e. without having to resort to truncation schemes for the $U(1)$ gauge field Hilbert space.
The ansatz consists of two parts: first, a pure gauge part based on Jastrow-type ansatz states (which can be connected to certain neural-network ansatz states) and secondly, on a fermionic part based on gauge-field dependent fermionic Gaussian states.
These are designed in such a way that the gauge field integral over all fermionic Gaussian states is gauge-invariant and at the same time still efficiently tractable. 
To ensure the validity of the method we benchmark the pure gauge part of the ansatz against another variational method and the full ansatz against an existing Monte Carlo simulation where the sign-problem is absent. 
Moreover, in limiting cases where the exact ground state is known we show that our ansatz is able to capture this behavior. 
Finally, we study a sign-problem affected regime by probing density-induced phase transitions. 
\end{abstract}

\maketitle

\section{Introduction\label{sec:intro}}
Gauge theories play a prominent role in different areas of physics. 
In high-energy physics, the standard model of particle physics, a gauge theory, describes three of the four fundamental forces in nature. 
At high energy scales its interactions can be treated perturbatively, however, at lower energies this approach fails and non-perturbative techniques are required~\cite{gross_asymptotically_1973,peskin_introduction_1995}. 
This naturally gives rise to lattice gauge theories as they are non-perturbative, gauge-invariant regularizations of quantum field theories~\cite{wilson_confinement_1974,kogut_hamiltonian_1975}. 
In condensed matter, lattice gauge theories emerge as low-energy effective theories of strongly correlated electron systems, e.g. quantum spin liquids or high-temperature superconductors~\cite{affleck_large-_1988,rantner_electron_2001}. 

Much progress has been made in studying lattice gauge theories, both from the high-energy physics as well as from the condensed matter side, in particular using Euclidean Monte Carlo simulations~\cite{flag_working_group_review_2014}. 
Nevertheless, certain regimes are difficult to access within this framework as fermionic theories at finite density or with an odd number of fermion flavors may suffer from the sign problem~\cite{troyer_computational_2005} and real-time dynamics are difficult to compute as Monte Carlo algorithms are usually formulated in Euclidean spacetime. 

In recent years, several approaches to this problem have received attention: a prominent example is quantum simulation where it was shown that lattice gauge theory Hamiltonians can be realized in quantum devices (e.g. ultracold atoms, trapped ions or superconducting qubits)~\cite{zohar_quantum_2016,banuls_simulating_2020}. 
The implementation of quantum simulators has been demonstrated in one dimension using trapped ions and ultracold atoms~\cite{martinez_real-time_2016, gorg_realization_2019,schweizer_floquet_2019,yang_observation_2020,mil_scalable_2020}. 
In two and more spatial dimensions the situation becomes more challenging, in particular due to appearance of magnetic interactions, leading to four-body plaquette terms on the lattice. 
There have been proposals on how to overcome this problem in quantum simulators (either by employing a digital~\cite{tagliacozzo_simulation_2013,zohar_digital_2017,bender_digital_2018,gustafson_prospects_2021,bhattacharya_qubit_2021,ciavarella_trailhead_2021} or an analog simulation scheme~\cite{zohar_simulating_2013,ott_scalable_2021}) but so far they have not been realized in experiments.

Another significant approach is based on variational ansatz states which can capture the relevant physics of the theory but at the same time can be evaluated efficiently. 
For lattice gauge theories these states either have to respect the local gauge symmetries or one has to find a reformulation of the theory in terms of gauge-invariant variables (at the cost of more complicated interactions)~\cite{raychowdhury_loop_2020,bender_gauge_2020,haase_resource_2021} such that there is a larger freedom in choosing variational states. 
One class of ansatz states are tensor networks whose one-dimensional version, matrix product states (MPS), have been successfully applied to (1+1)-dimensional Abelian and non-Abelian lattice gauge theories~\cite{banuls_mass_2013,buyens_matrix_2014,rico_tensor_2014,kuhn_non-abelian_2015,pichler_real-time_2016,banuls_efficient_2017,banuls_density_2017,silvi_finite-density_2017,buyens_real-time_2017,silvi_tensor_2019,bruckmann_o3_2019}, enabling the study of finite chemical potential scenarios and out-of-equilibrium dynamics which are not accessible in Monte Carlo simulations of Euclidean lattice gauge theory. 
In higher dimensions, tensor network methods have been applied to lattice gauge theories with a finite-dimensional gauge field Hilbert space (either by working with quantum link formulations~\cite{chandrasekharan_quantum_1997,tschirsich_phase_2019,magnifico_lattice_2021,montangero2022loop} or using a discrete gauge group~\cite{tagliacozzo_tensor_2014,emonts_variational_2020,robaina_simulating_2021}). 
Other types of ansatz states can be formulated in the infinite-dimensional Hilbert space of continuous gauge groups. 
These include periodic Gaussian states~\cite{bender_real-time_2020} (generalizations of Gaussian states that take into account the compactness of the gauge group) or neural network-based ansatz states~\cite{luo_gauge_2022,chen_simulating_2022}.

A particularly interesting model in higher dimensions is $(2+1)$-dimensional compact QED ($\text{cQED}_3$) with dynamical charges, both in the context of high-energy physics and condensed matter physics. 
From the high-energy perspective it is interesting since it is the simplest theory to discuss confinement and chiral symmetry breaking, also quintessential to our understanding of quantum chromodyanmics (QCD). 
Already without dynamical fermions, the theory has non-trivial interactions due to the appearance of four-body magnetic terms. 
It is known to confine for all couplings~\cite{polyakov_quark_1977}. 
However, upon the inclusion of dynamical fermions, the situation is less clear since the dynamical fermionic matter might lead to deconfinement. 
These phenomena are also of high relevance in condensed matter since many low-energy effective theories of two-dimensional strongly coupled electron systems can be described by massless dynamical fermions coupled to a compact $U(1)$ gauge field. 
Compared to $\mathbb{Z}_2$ lattice gauge theories where it was shown that sign-problem-free Monte Carlo simulations could be performed for an even number of fermion flavors even at non-zero density~\cite{gazit_emergent_2017}, for the $U(1)$-theory the sign-problem is only absent for an even number of fermion flavors at half-filling~\cite{xu_monte_2019}. 
Based on the above considerations, in this work, we introduce a variational method that can access the sign-problem affected regimes of $\text{cQED}_3$ with dynamical charges without truncating the $U(1)$ gauge field Hilbert space. 
The ansatz is based on a combination of a pure gauge part containing the self-interactions of the gauge field and a fermionic part which describes the dynamics of the matter degrees of freedom with the gauge field. 
The pure gauge part is a Jastrow wave function constructed out of gauge-invariant plaquette variables (its form can be connected to certain neural network quantum states~\cite{glasser_neural-network_2018}).
The choice of ansatz is motivated by an earlier proposal~\cite{bender_real-time_2020} which could approximate ground states and real-time dynamics in $\text{cQED}_3$ with static charges. 
The fermionic part is an infinite superposition of gauge-field dependent fermionic Gaussian states which are parametrized in such a way that the resulting state is gauge-invariant. 
Note that the parametrization is done in a way that the number of parameters only scales polynomially with system size. 
In a similar fashion to neural-network quantum states, expectation values are obtained using Monte Carlo sampling. 
The optimization of variational states is done via stochastic reconfiguration~\cite{sorella_wave_2005}.  

In order to verify the capabilities of the variational method we first demonstrate also numerically that gauge invariance is indeed preserved. 
In a second step it is shown that the ansatz is exact in all limiting cases. 
This includes the weak-coupling limit ($g^2 \to 0$) where the ground state is known to be a $\pi$-flux state~\cite{PhysRevLett.73.2158} and the strong-coupling limit ($g^2 \to \infty$) where the electric energy dominates and one obtains an effective fermionic theory.
Moreover, we benchmark our ansatz against other methods: first, we compare for the pure gauge theory, i.e. compact QED without fermions, the ground state energy with another variational method~\cite{bender_real-time_2020} and see agreement over the whole coupling region.
These results were recently confirmed in another variational study based on neural-network states~\cite{luo_gauge_2022}.
For compact QED with dynamical fermions, we compare with a Euclidean Monte Carlo study at zero chemical potential and two fermion flavors where the sign-problem is absent~\cite{xu_monte_2019}. 
We compute the flux energy per plaquette which agrees with ref.~\cite{xu_monte_2019} over the whole coupling region. 
We also compare fermionic correlations quantifying the degree of antiferromagnetic order in the ground state. 
By extrapolating this quantity to the thermodynamic limit it is shown that antiferromagnetic order only persists down to a coupling of $g^2_{c,\infty}=0.15(2)$ which is in qualitative agreement with ref.~\cite{xu_monte_2019} although our extrapolated value for the transition is lower. 
To demonstrate our method in a sign-problem affected regime we study density-induced phase transition for two fermion flavors at non-zero chemical potential, similar to a tensor network study in one dimension~\cite{banuls_density_2017}. 
We consider both the case of massless and massive staggered fermions and see qualitatively similar phenomena as in ref.~\cite{banuls_density_2017}.

The manuscript is structured as follows.
In section \ref{sec:model}, we introduce the model, $\text{cQED}_3$ with dynamical fermions. 
In section \ref{sec:ansatz}, we describe the variational Monte Carlo method including the gauge-invariant construction of the variational state, the numerical evaluation with Monte Carlo techniques and the adaptation of variational parameters.
In section \ref{sec:validity}, our ansatz is benchmarked against limiting cases of the model where the ground state is known and other numerical methods.
In section \ref{sec:signproblem}, we study a sign-problem affected regime by investigating density induced phase transition for two fermion flavors at non-zero chemical potential.
In section \ref{sec:conclusion}, we summarize and conclude. 

\section{The model: (\texorpdfstring{$2+1$}{2+1})-dimensional compact QED with dynamical fermions\label{sec:model}} 
\begin{figure}[t]
    \centering
    \includegraphics[width=0.8\columnwidth]{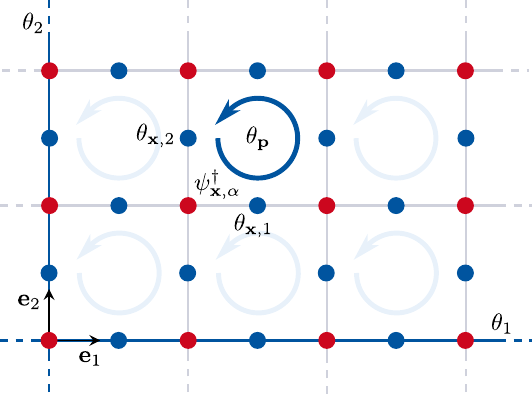}
    \caption{Naming conventions on the periodic lattice: 
    the gauge field degrees of freedom $\theta_{\mathbf{x},i}, E_{\mathbf{x},i}$ (blue) reside on the links $\mathbf{x},i$ while the fermionic degrees of freedom $\psi_{\mathbf{x},\alpha}^{\dagger}$ (red), which can come in several species $\alpha$, are located on the sites $\mathbf{x}$.
    The circular arrows on the plaquettes denote the plaquette variables $\theta_{\vb{p}}$.
    The global loops $\theta_i$ wind around the axis given by $\vb{e}_i$ and are illustrated by blue lines.}
    \label{fig:lattice}
\end{figure}

We study ($2+1$)-dimensional compact quantum electrodynamics ($\text{cQED}_3$) coupled to dynamical fermions.
The model is defined on an $L \times L$ square lattice with periodic boundary conditions. 
We work with staggered fermions~\cite{PhysRevD.11.395} which are suitable for studying chiral symmetry breaking.
The fermions can appear in several species $\alpha$ which can be subject to different chemical potentials $\mu_{\alpha}$ (in some scenarios they are also given a mass $m$). 
The Hamiltonian reads 
\begin{equation} \label{eq:full_hamiltonian}
\begin{aligned}
    H=& \frac{g^2}{2} \sum_{\mathbf{x},i} \hat{E}_{\mathbf{x},i}^2 + g_{\text{mag}} \sum_{\mathbf{p}} \left( 1 - \cos(\hat{\theta}_{\mathbf{p}}) \right) \\
    &- t \sum_{\mathbf{x},i,\alpha} \psi_{\mathbf{x},\alpha}^{\dagger} e^{i \hat{\theta}_{\mathbf{x},i}} \psi_{\mathbf{x}+\mathbf{e}_i,\alpha}  + h.c. \\
    &+ \sum_{\mathbf{x},\alpha} \left( m (-1)^{\mathbf{x}} + \mu_{\alpha} \right) \psi_{\mathbf{x},\alpha}^{\dagger} \psi_{\mathbf{x},\alpha} \\
    &\equiv H_E + H_B + H_{GM} + H_M
\end{aligned}    
\end{equation} 
where $\psi_{\mathbf{x},\alpha}$ denotes the fermionic annihilation operator for site $\mathbf{x}$ and species $\alpha$.
The gauge field operator $\hat{\theta}_{\mathbf{x},i}$ and the electric field operator $\hat{E}_{\mathbf{x},i}$ fulfill the canonical commutation relations, $[\hat{\theta}_{\mathbf{x},i},\hat{E}_{\mathbf{y},j}]=i \delta_{ij} \delta_{\mathbf{x},\mathbf{y}}$. 
Accordingly, the gauge field on a link can be represented either by an integer-valued electric field variable, $\hat{E}_{\mathbf{x},i}  \ket{E_{\mathbf{x},i}} = E_{\mathbf{x},i} \ket{E_{\mathbf{x},i}}$ ($E_{\mathbf{x},i} \in \mathbb{Z}$),  or by an element of the $U(1)$ gauge group, $\hat{\theta}_{\mathbf{x},i}  \ket{\theta_{\mathbf{x},i}} = \theta_{\mathbf{x},i} \ket{\theta_{\mathbf{x},i}}$ ($\theta_{\mathbf{x},i} \in [0,2\pi)$).
We will mostly use the group element representation throughout the manuscript.
In this representation, the electric field operator has the form $\hat{E}_{\mathbf{x},i}=-i \partial / \partial \theta_{\mathbf{x},i}$.
The plaquette operator $\hat{\theta}_{\mathbf{p}}=\hat{\theta}_{\mathbf{x},1} + \hat{\theta}_{\mathbf{x}+\mathbf{e}_1,2}-\hat{\theta}_{\mathbf{x}+\mathbf{e}_2,1} - \hat{\theta}_{\mathbf{x},2}$ is the clockwise summation of link operators around plaquette $\mathbf{p}$ where $\mathbf{x}$ is the site at the bottom left corner. 
The labelling conventions are illustrated in Fig.~\ref{fig:lattice}.
The magnetic coupling $g_{\text{mag}}$ is usually chosen to be $\frac{1}{g^2}$ but we keep it general for the moment.

The local symmetry of the model is generated by the Gauss law operators
\begin{align} \label{eq:gauss_operator}
    \hat{G}_{\mathbf{x}}= \sum_{i=1}^2 \left(\hat{E}_{\mathbf{x},i} - \hat{E}_{\mathbf{x}-\mathbf{e}_i,i} \right) - \hat{Q}_{\text{stag}}
\end{align}
where the staggered charge operator $\hat{Q}_{\text{stag},\mathbf{x}}$ is defined as
\begin{align}
     \hat{Q}_{\text{stag},\mathbf{x}} = \sum_{\alpha=1}^{N_f} \left( \psi^{\dagger}_{\mathbf{x},\alpha} \psi_{\mathbf{x},\alpha} - \frac{1}{2} ( 1 + (-1)^{\mathbf{x}} ) \right).
\end{align}
Physical states $\ket{\text{phys}}$ must be eigenstates of all Gauss law operators 
\begin{align} \label{eq:gauss_phys}
    \hat{G}_{\mathbf{x}} \ket{\text{phys}} = q_{\mathbf{x}} \ket{\text{phys}} \quad \forall \mathbf{x}
\end{align}
where eigenvalues $q_{\mathbf{x}}$ correspond to different static charge configurations.

\section{The variational method\label{sec:ansatz}} 
\begin{figure} [t]
    \centering
    \includegraphics[width=\columnwidth]{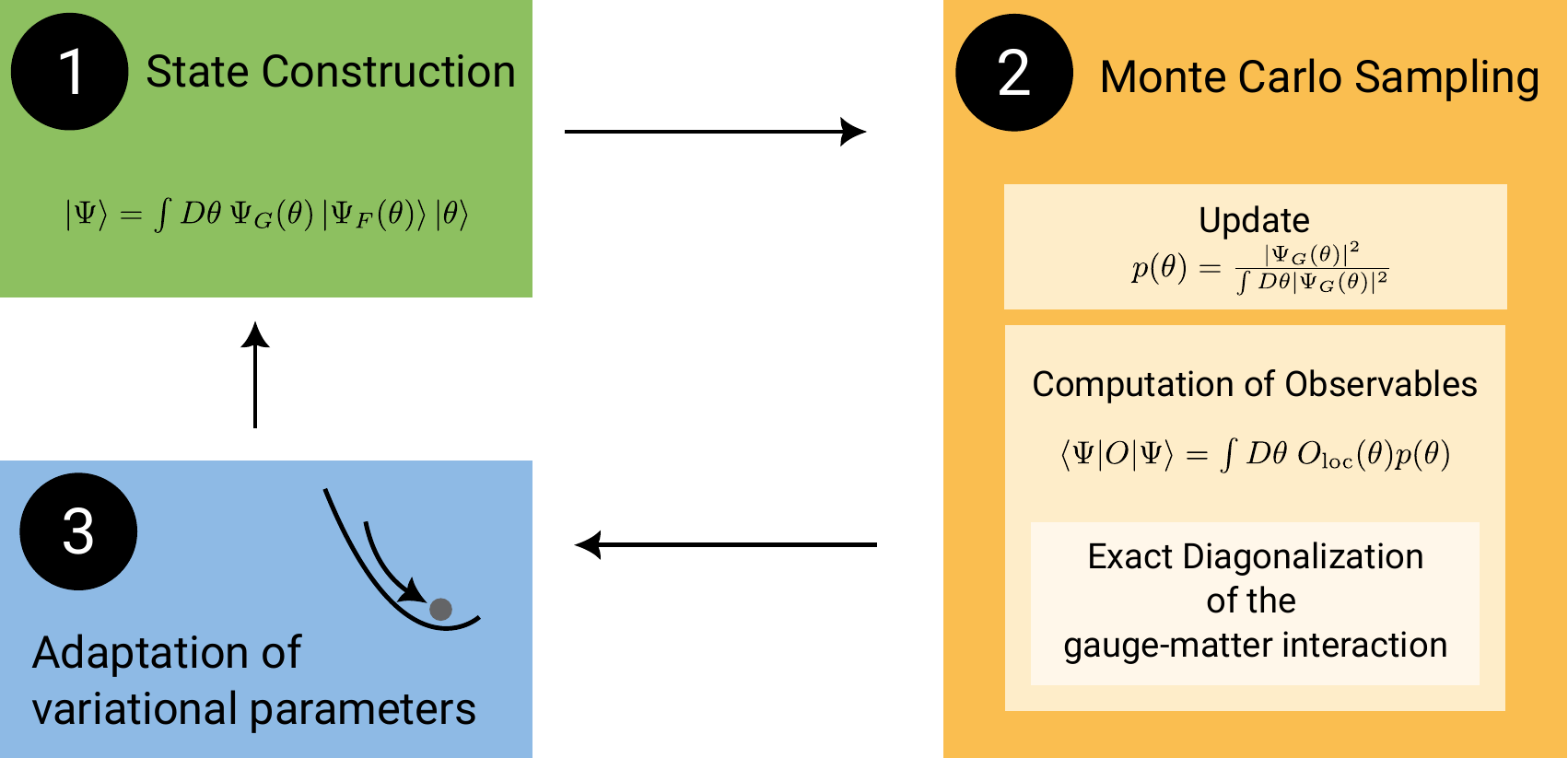}
    \caption{Scheme of variational Monte Carlo procedure: the ansatz is formulated in the full gauge field basis denoted by $\ket{\theta}$, in our case the $U(1)$ gauge group, consisting of a pure gauge part $\Psi_G(\theta)$ and gauge-field dependent fermionic Gaussian states $\ket{\Psi_F(\theta)}$. 
    Expectation values of observables $O$ can be carried out analytically w.r.t. the fermionic part (which involves the eigendecomposition of the gauge-matter interactions for fixed $\theta$).
    The resulting expressions $O_{\text{loc}}(\theta)$ are diagonal in $\theta$ and sampled with Monte Carlo techniques according to a probability distribution $p(\theta)$ in which only $\Psi_G(\theta)$ appears since the gauge-field dependent fermionic Gaussian states are normalized.
    The variational parameters are adapted according to stochastic reconfiguration.}
    \label{fig:flowchart}
\end{figure}

Since our method is based on variational Monte Carlo, we will explain it in several steps (sketched in Fig.~\ref{fig:flowchart}): we first discuss the state construction and motivate the choice of our ansatz. 
In the second step, we explain the evaluation procedure of our ansatz based on Monte Carlo sampling.
In the third and final step, we discuss the adaptation of variational parameters based on stochastic reconfiguration.

\subsection{State construction} \label{sec:state_construction}
We construct our ansatz state in the gauge field basis where states are characterized by all $U(1)$ gauge link variables $\theta_{\mathbf{x},i}$, $ \ket{\{ \theta_{\mathbf{x},i} \} } \equiv \otimes_{\mathbf{x},i} \ket{\theta_{\mathbf{x},i}}$. 
A general $\textit{gauge field}$ state can then be defined by 
\begin{align}
    \ket{\Psi_G} = \prod_{\mathbf{x},i} \int_0^{2\pi} d\theta_{\mathbf{x},i} \hspace{2pt} \Psi_G(\{ \theta_{\mathbf{x},i} \} ) \ket{\{ \theta_{\mathbf{x},i} \} }
\end{align}
where  $\Psi_G(\{ \theta_{\mathbf{x},i} \} ) $ is a function over all gauge link variables $\theta_{\mathbf{x},i}$. 

To extend the above to an arbitrary state $\ket{\Psi}$ of the $\text{cQED}_3$ model introduced in section~\ref{sec:model} (i.e. including fermions) we need to specify a fermionic Fock state $\ket{\Psi_F(\{\theta_{\mathbf{x},i}\})}$ for every gauge field configuration $\{\theta_{\mathbf{x},i}\}$:
\begin{equation} \label{eq:full_ansatz}
\begin{aligned}
    \ket{\Psi} &= \prod_{\mathbf{x},i} \int_0^{2\pi} d\theta_{\mathbf{x},i} \hspace{2pt} \Psi_G(\{ \theta_{\mathbf{x},i} \} ) \ket{\Psi_F(\{\theta_{\mathbf{x},i}\})} \ket{\{ \theta_{\mathbf{x},i} \} } \\
    &\equiv \int D\theta \hspace{2pt} \Psi_G(\theta) \ket{\Psi_F(\theta)} \ket{\theta} 
\end{aligned}    
\end{equation}
where we abbreviated for ease of notation the gauge field configurations $\{\theta_{\mathbf{x},i}\}$ as $\theta$ and the measure as $\int D\theta= \prod_{\mathbf{x},i} \int_0^{2\pi} d\theta_{\mathbf{x},i}$.
This notation will be used throughout the article.

One should note that an arbitrary state $\ket{\Psi}$ as given above is a priori not gauge-invariant. 
Thus, the gauge invariance condition for physical states in eq.~\eqref{eq:gauss_phys} severely restricts the possible choices for $\Psi_G(\theta)$ and $\ket{\Psi_F(\theta)}$.

The state $\ket{\Psi}$ defined above is completely general. 
From now on we will use the form of $\ket{\Psi}$ as the basis to construct our variational ansatz state which will be defined by specifying the pure gauge part $\Psi_G(\theta)$ and the fermionic ansatz $\ket{\Psi_F(\theta)}$.

Intuitively, the role of $\Psi_G(\theta)$ and $\ket{\Psi_F(\theta)}$ in our construction can be motivated as follows: $\Psi_G(\theta)$ is designed to approximate the ground state of the pure gauge model $H_{\text{KS}} \equiv H_B + H_E$ (the Kogut-Susskind Hamiltonian~\cite{kogut_hamiltonian_1975}) whereas $\ket{\Psi_F(\theta)}$ is designed to approximate the low-energy physics of the fermionic Hamiltonian $H_{\text{fer}} \equiv H_E + H_{GM} + H_M$ which neglects the self-interactions of the gauge field.

\subsubsection{The pure gauge part of the ansatz} 
In this section we motivate and describe the pure gauge part $\Psi_G(\theta)$ of our variational state. 
In earlier work~\cite{bender_real-time_2020} it was shown that 
\begin{equation}
    \phi_G(\theta) = \sum_{  N_{\mathbf{p}} = - \infty }^{+ \infty} e^{-\frac{1}{2} (\theta_{\mathbf{p}} - 2\pi N_{\mathbf{p}}) \alpha_{\mathbf{p}\mathbf{p'}} (\theta_{\mathbf{p'}} - 2\pi N_{\mathbf{p'}}) }
\end{equation}
is a good ansatz for the ground state of compact QED with static charges. 
It is a Gaussian in the plaquette variables $\theta_{\mathbf{p}}$ that is made periodic by an infinite sum over the integer-valued variables $N_{\mathbf{p}}$ ($\alpha_{\mathbf{p}\mathbf{p'}}$ are variational parameters). 
The periodicity is important to account for the compactness of the $U(1)$ gauge field.

Here, we would like to find an ansatz which has a similar expressive power as the states above but at the same time is suitable for a variational Monte Carlo simulation directly in $\theta$ (without resorting to the sums above). 
A useful hint is given by the Villain approximation~\cite{villain_theory_1975} which states
\begin{equation}
    e^{\gamma \left( 1 - \cos (\theta) \right)} \to \sum_{N} e^{-\frac{1}{2} \gamma (\theta - 2\pi N)^2}
\end{equation}  
for $\gamma \to \infty$. 
Therefore, a suitable ansatz state could be
\begin{equation}
    e^{- \sum_{\mathbf{p}\mathbf{p'}} \cos (\theta_{\mathbf{p}}) \alpha_{\mathbf{p},\mathbf{p'}} \cos (\theta_{\mathbf{p'}})  + \sum_{\mathbf{p}} \beta_{\mathbf{p}} \cos (\theta_{\mathbf{p}}) }
\end{equation}
with the matrix $\alpha$ and the vector $\beta$ being variational parameters. 
We will choose $\alpha$ and $\beta$ to be real since we are interested here in low-energy properties. 
For the study of real-time dynamics (which will be investigated in a future work) we would choose the variational parameters to be complex.
Apart from the cosine terms we can add sine terms, combine them in a vector 
$b(\theta)=(\cos(\theta_{\mathbf{p}_1}), ..,\cos(\theta_{\mathbf{p}_N}), \sin(\theta_{\mathbf{p}_1}), ..,\sin(\theta_{\mathbf{p}_N}))$ and generalize the above state to 
\begin{equation} \label{eq: gauge_ansatz}
    \Psi_{G}(\theta)=e^{- \frac{1}{2} b(\theta)^T \alpha b(\theta) - \beta^T b(\theta)} 
\end{equation}
which will be the variational ansatz for the pure gauge field dynamics entering the full ansatz as in eq.~\eqref{eq:full_ansatz}.  

In the case of periodic boundary conditions there are two inequivalent global non-contractible loops (inequivalent in the sense that they can not be transformed into each other by plaquette operations). 
We choose them to be $\theta_1$ (winding around the lattice along the $x_1$-axis), and $\theta_2$ (along the $x_2$-axis), respectively (see Fig.~\ref{fig:lattice}).
We incorporate them in our ansatz by expanding the vector $b(\theta)$ by the entries $\cos(\theta_1),\cos(\theta_2),\sin(\theta_1)$ and $\sin(\theta_2)$. 
This is necessary because upon the coupling of compact QED to \textit{dynamical} fermions, $\theta_1$ and $\theta_2$ become dynamical variables due to the appearance of gauge-matter interactions where the phase $e^{i \theta_{\mathbf{x},i}}$ appears. 
If expressed in terms of gauge-invariant variables, it contains contributions from both plaquette variables $\theta_{\mathbf{p}}$ and the global loops $\theta_1$ and $\theta_2$~\cite{bender_gauge_2020}. 
For static charges, the magnetic Hamiltonian is the only term depending on the gauge field variables $\theta_{\mathbf{x},i}$ which can be expressed entirely in terms of plaquette variables $\theta_{\mathbf{p}}$ such that the global loop variables only set different topological sectors (similar to the toric code).

For all our purposes it turned out that all variational parameters in $\alpha$ corresponding to the global loop variables were not relevant and that it was sufficient to only keep the global loop parameters in $\beta$ variational. 
After imposing translational invariance we thus remained with $2N+4$ variational parameters for $\alpha$ and $6$ variational parameters for $\beta$ (with $N=L^2$ the number of lattice sites).

Since $b(\theta)$ contains only closed loops the gauge field part $\Psi_G(\theta)$ as a function of $b(\theta)$ automatically preserves gauge invariance.

\subsubsection{The fermionic part of the ansatz: gauge-invariant fermionic Gaussian states}
The fermionic part of our variational ansatz $\ket{\Psi_F(\theta)}$ is a generalization of fermionic Gaussian states that can incorporate interactions between fermions and gauge fields while preserving gauge invariance. 
The overall state $\int D\theta \ket{\Psi_F(\theta)} \ket{\theta}$ is an integral over all gauge field configurations where for every gauge field configuration $\theta$ we define a fermionic Gaussian state $\ket{\Psi_F(\theta)}$.
The motivation for this construction is that the resulting state, an infinite superposition of Gaussian states, is a powerful ansatz state as it is clearly not Gaussian anymore and can capture correlations beyond the Gaussian realm. 
At the same time, we retain for every $\ket{\Psi_F(\theta)}$ the properties of fermionic Gaussian states which allows us to compute part of the expectation values analytically. 
The number of variational parameters is shown to scale only polynomially in the system size and not exponentially as the number of gauge field configurations.

Recalling that every pure fermionic Gaussian state can be represented by a unitary operator $U_{GS}$ acting on some reference state $\ket{\Psi_0}$~\cite{shi_variational_2018}, we carry out an analogous procedure for every gauge field configuration $\theta$ to construct gauge-field dependent fermionic Gaussian states as
\begin{equation} \label{eq: general fermionic ansatz}
\ket{\Psi_F \left(\theta \right)}=U_{GS}\left(\theta\right) \ket{\Psi_0} 
\end{equation}

In our method, the reference state $\ket{\Psi_0}$ will be chosen as the ground state of $H_{\text{fer}}=H_E + H_{GM} + H_M$ in the strong-coupling limit ($g^2 \to \infty$), i.e. the regime where the electric term dominates so that electric field excitations are strongly suppressed. 
In the following we will refer to strong- and weak-coupling always w.r.t. the relative strength of the electric Hamiltonian $H_E$ (quantified by its coupling constant $g^2$).
If we only consider $H_E$  in the strong-coupling limit, $\ket{\Psi_0}$ will be a Fock state where all fermions are fixed to certain sites (the exact form of the state will depend on the number of fermion flavors and the configuration of background charges). 
However, $\ket{\Psi_0}$ does not need to be Gaussian but one can also include perturbations induced by gauge-matter interactions $H_{GM}$, e.g. it is known that for two fermion flavors at half-filling the strong-coupling ground state in second-order perturbation theory is the ground state of the Heisenberg model. 
It can be shown that its properties can be incorporated in the reference state $\ket{\Psi_0}$ which can even be kept variational (for details see Appendix~\ref{sec:app_heisenberg}). 

For simplicity of the discussion, we will assume in the following a Gaussian reference state and only one flavor of staggered fermions (how the ansatz can be readily extended to multiple flavors is described in Appendix~\ref{sec:tilde_xi}). 
In the sector without background charges the reference state is chosen to be the Dirac state $\ket{D}$
\begin{equation} \label{eq:diracstate}
     \ket{\Psi_{0}} = \prod_{\mathbf{x} \in \mathcal{O}} \psi^{\dagger}_{\mathbf{x}} \ket{0} \equiv  \ket{D} 
\end{equation}
i.e. with all odd sites $\mathcal{O}$ occupied. 

The Gaussian operator $U_{GS}(\theta)$ acting on $\ket{\Psi_0}$ is defined as 
\begin{equation}
    U_{GS}\left(\theta\right)= \exp \left( \frac{i}{2} \sum_{\mathbf{x},\mathbf{y}} \psi^{\dagger}_{\mathbf{x}} \xi(\theta)_{\mathbf{x}\mathbf{y}} \psi_{\mathbf{y}}   \right)
\end{equation}
where $\xi(\theta)$ is dependent on the  gauge-field and on the variational parameters. 
The gauge-field dependence has to be chosen in a way that respects gauge invariance. 
This can be achieved by defining $\xi(\theta)$ via the eigendecomposition of the gauge-matter Hamiltonian which can be written as 
\begin{equation}
    H_{GM}= \int D\theta \hspace{2pt} \ket{\theta} \bra{\theta} \vec{\psi}^{\dagger}_{\mathbf{x}} h_{GM}(\theta)_{\mathbf{x}\mathbf{y}} \vec{\psi}_{\mathbf{y}}
\end{equation} 
with $\vec{\psi}_{\mathbf{x}} \equiv (\psi_{\mathbf{x}_1},...,\psi_{\mathbf{x}_N})^T$ a vector of all fermionic annihilation operators.
The matrix $h_{GM}(\theta)$ is hermitian and can be diagonalized for a specific gauge field configuration $\theta$ as $h_{GM}(\theta)=V(\theta) \Lambda(\theta)  V(\theta)^{\dagger} $. 
We use $V(\theta)$ to rewrite $\xi(\theta)$ as 
\begin{equation} \label{eq: xi_theta_definition}
    \xi(\theta)_{\mathbf{x}\mathbf{y}}=V(\theta)_{\mathbf{x}i} \tilde{\xi}_{ij} V(\theta)_{j\mathbf{y}}^{\dagger}
\end{equation}
with $\tilde{\xi}$ containing the variational parameters. 
Note that $\tilde{\xi}$ does not depend on the gauge field configuration and thus the number variational parameters scales quadratically with the system size (linearly for our choice of parametrization, see Appendix~\ref{sec:tilde_xi}). 
Putting everything together, the fermionic part of the ansatz for one fermion flavor takes the form
\begin{equation} \label{eq:psi_fermionic}
    \ket{\Psi_F(\theta)}= \exp \left( \frac{i}{2} \sum_{\mathbf{x},\mathbf{y}} \psi^{\dagger}_{\mathbf{x}} V(\theta)_{\mathbf{x}i} \tilde{\xi}_{ij} V(\theta)_{j\mathbf{y}}^{\dagger} \psi_{\mathbf{y}}   \right) \ket{D} 
\end{equation}
and the whole variational ansatz state $\ket{\Psi}$ is thus fully defined according to eq.~\eqref{eq:full_ansatz}. 

Gauge invariance of $\ket{\Psi}$ follows from the fact that $H_{GM}$ and its eigenstates are gauge-invariant since the construction of $\ket{\Psi_F(\theta)}$ given in eq.~\eqref{eq:psi_fermionic} is formulated in terms of these eigenstates.

Since the gauge invariance condition in eq.~\eqref{eq:gauss_phys} is local in $\theta$, every realization of the state in a Monte Carlo simulation will be gauge-invariant, i.e. even with an imperfect sampling algorithm the unphysical part of the Hilbert space is never accessed.

The motivation for the choice of ansatz above is on the one hand that it ensures gauge invariance but more importantly, by choosing the matrix $\tilde{\xi}$ appropriately, the occupation of the eigenstates of $H_{GM}$ can be tuned which allows to obtain good ground state approximations even in regimes where strong gauge field fluctuations are present. 
This has to be seen in contrast to mean-field descriptions where a certain gauge field pattern is fixed and the resulting fermionic theory is studied. 
The latter has the problem, which is particularly relevant in the study of quantum spin liquid states (where the lattice gauge theory emerges as an effective low-energy description), that it often remains unclear whether the spin liquid state is stable against gauge field fluctuations~\cite{affleck_large-_1988}.
 
The cost of working with the ansatz is that the eigendecomposition of $h_{GM}(\theta)$ needs to be carried out at every measurement step of the Monte Carlo algorithm. 
However, $h_{GM}(\theta)$ is a hermitian $N \times N$ matrix where $N$ is the number of lattice sites such that the cost is $\mathcal{O}(N^3)$ which can be done efficiently. 
Note that the number of fermion flavors does not enter as the gauge-matter interaction is the same for all flavors.
 
The fermionic ansatz state in eq.~\eqref{eq:psi_fermionic} is normalized since the Gaussian operator acting on the Dirac vacuum is unitary. 
This is beneficial for the variational Monte Carlo simulation since it will not contribute to the probability distribution that needs to be sampled. 
Thus, no sampling problems related to fermion determinants can occur in this method as opposed to action-based Monte Carlo algorithms. 

So far we have not specified the matrix $\tilde{\xi}$ in eq.~\eqref{eq:psi_fermionic} containing the fermionic variational parameters. 
For that we consider the eigendecomposition of $h_{GM}(\theta)$, denoted as $h_{GM}(\theta) \ket{w_{i}(\theta)} = \lambda_{i}(\theta) \ket{w_{i}(\theta)}$, $i \in \{1,..,N\}$. 
Assuming an $L \times L$ lattice with $L$ even, the spectrum of $h_{GM}(\theta)$ is symmetric around zero, i.e. we have $N/2$ pairs of eigenvectors $\ket{w_{k+}(\theta)}$ and $\ket{w_{k-}(\theta)}$ ($k \in \{1,..,N/2\}$) such that $\ket{w_{k+}(\theta)}$ corresponds to the eigenvalue $+\lambda_k(\theta)$ and $\ket{w_{k-}(\theta)}$ to the eigenvalue $-\lambda_k(\theta)$. 
A useful feature of these pairs is their structure in the position basis as they can be written as two vectors $\ket{w_{ke}(\theta)}$, $\ket{w_{ko}(\theta)}$ which are residing exclusively on even (respectively odd) lattice sites:
\begin{align}
    \ket{w_{k+}(\theta)}&=\frac{1}{\sqrt{2}} \left(\ket{w_{ke}(\theta)}+\ket{w_{ko}(\theta)} \right) \\
    \ket{w_{k-}(\theta)}&=\frac{1}{\sqrt{2}} \left(\ket{w_{ke}(\theta)}-\ket{w_{ko}(\theta)} \right)  \label{eq:odd_superposition}
\end{align} 
This allows us to write the strong-coupling state in eq.~\eqref{eq:diracstate}, where the fermions occupy all odd sites, as a product over all pairs $k$ where in each pair the odd superpositon is occupied, $\ket{w_{k-}(\theta)}=\frac{1}{\sqrt{2}} \left(\ket{w_{ke}(\theta)}-\ket{w_{ko}(\theta)} \right)$. 
The purpose of $\tilde{\xi}$ in eq.~\eqref{eq:psi_fermionic} is then to smoothly transform this equal superposition of $\ket{w_{k+}(\theta)}$ and $\ket{w_{k-}(\theta)}$ into a state where all $\ket{w_{k-}(\theta)}$ are occupied, corresponding to the ground state of $H_{GM}$. 
Thus, $\tilde{\xi}$ allows us to transform smoothly from the strong-coupling ground state to the weak-coupling ground state. 
For more details on $\tilde{\xi}$ and the specific choice of parametrization see Appendix~\ref{sec:tilde_xi}.

\subsection{Evaluating expectation values} \label{sec: evaluating expectation values}
In this section we describe how Monte Carlo techniques can be used to compute various expectation values for the variational ansatz presented in the previous section. 
Throughout the following discussion the variational parameters are kept fixed, their adaptation will be discussed in the next section.

For the computation of an observable $O$ with the full ansatz $\ket{\Psi}$ from eq.~\eqref{eq:full_ansatz} we obtain 
\begin{equation} \label{eq: expval_general}
\begin{aligned}
   \frac{\expval{O}{\Psi}}{\braket{\Psi}}
    &=\frac{\int D\theta \hspace{2pt} \bra{\Psi_F(\theta)} \overline{\Psi_G}(\theta) \hspace{1pt} O \hspace{1pt} \Psi_G(\theta)   \ket{\Psi_F(\theta)} }{\int D\theta \hspace{2pt} |\Psi_G(\theta)|^2  \underbrace{\braket{\Psi_F(\theta)}}_{=1}} \\ 
    &=\frac{\int D\theta \hspace{2pt} O_{\text{loc}}(\theta)  |\Psi_G(\theta)|^2 } {\int D\theta \hspace{2pt} |\Psi_G(\theta)|^2    } =\int D\theta \hspace{2pt} O_{\text{loc}}(\theta)  p(\theta) \\    
\end{aligned}    
\end{equation}
where $\ket{\Psi_F(\theta)}$ is absent in the norm since it is already normalized by construction (see  eq.~\eqref{eq:psi_fermionic}) so that the probability distribution $p(\theta)$ depends only on $\Psi_G(\theta)$.
The fermionic part of the ansatz thus only appears in the numerator for the evaluation of $O$ which is carried out analytically and only the remaining expression $O_{\text{loc}}(\theta)$ is sampled in a Monte Carlo simulation. 

We split the calculation of $O_{\text{loc}}(\theta)$ in two parts: since $O$ is a priori not diagonal in $\theta$ (e.g. all electric observables involve derivatives w.r.t. $\theta$) we first compute the action of $O$ on our ansatz $\ket{\Psi}$ which gives rise to an expression $O_{\text{fer}}(\theta)$ that is diagonal in $\theta$ but might still contain fermionic operators (e.g. due to derivatives of the fermionic ansatz $\ket{\Psi_F(\theta)}$):
\begin{equation} \label{eq: O_fer}
\begin{aligned}
O \int D\theta \hspace{1pt} \Psi_G(\theta)   \ket{\Psi_F(\theta)} \ket{\theta} 
=\int D\theta \hspace{1pt}  O_{\text{fer}}(\theta)  \Psi_G(\theta)  \ket{\Psi_F(\theta)} \ket{\theta} \\    
\end{aligned}    
\end{equation}
$O_{\text{loc}}(\theta)$ is then derived by evaluating $O_{\text{fer}}(\theta)$ w.r.t. the fermionic ansatz
\begin{equation} \label{eq: local_quantity O_loc}
\begin{aligned}
 O_{\text{loc}}(\theta) &=  \bra{\Psi_F(\theta)}  O_{\text{fer}}(\theta)   \ket{\Psi_F(\theta)} \\
 &=\frac{ \bra{\Psi_F(\theta)} \hspace{1pt} O \hspace{1pt} \Psi_G(\theta)   \ket{\Psi_F(\theta)} }{\Psi_G(\theta) } \\  
\end{aligned}    
\end{equation}
which is now a real-valued function that can be readily sampled in a Monte Carlo simulation.

The probability distribution $p(\theta)$ according to which we need to sample is only dependent on the gauge part $\Psi_G(\theta)$ defined in eq.~\eqref{eq: gauge_ansatz}:
\begin{equation} \label{eq: prob_distribution}
\begin{aligned} 
    p(\theta)&=\frac{|\Psi_G(\theta)|^2}{\int D\theta \hspace{2pt} |\Psi_G(\theta)|^2}= \frac{e^{- b^T(\theta) \alpha b(\theta) - 2\beta^T b(\theta)}}{\int D\theta \hspace{2pt} e^{- b^T(\theta) \alpha b - 2\beta^T b(\theta)}} \\
    &\equiv \frac{e^{-S(\theta)}}{\int D\theta \hspace{2pt} e^{-S(\theta)}}
\end{aligned}
\end{equation}

The method described above has to be contrasted with usual variational Monte Carlo methods~\cite{sorella_generalized_2001} where the whole trial wavefunction contributes to the probability distribution and the local quantities $O_{\text{loc}}(\theta)$ do not involve taking expectation values w.r.t. some part of the ansatz. 
 
Having discussed the general procedure, the computation of observables can be divided into three groups by level of difficulty:
the first group consists of observables that are not diagonal in $\theta$ (all electric quantities such as $H_E$) and thus first need to be brought into a diagonal form $O_{\text{fer}}(\theta)$. 
These observables are the most involved.
The second group of observables are already of that form but since $O_{\text{fer}}(\theta)$ is still a fermionic operator it needs to be evaluated w.r.t. $\ket{\Psi_F(\theta)}$ to obtain $O_{\text{loc}}(\theta)$ (e.g. $H_{GM}$ or $H_M$). 
The third group of observables is already of the form $O_{\text{loc}}(\theta)$ and can be readily sampled in a Monte Carlo simulation (e.g. $H_B$).

Two things need to be shown to demonstrate that our variational ansatz can be used efficiently: first, the efficient computation of $O_{\text{loc}}(\theta)$ and secondly, efficient sampling of the probability distribution $p(\theta)$. 
Thus, in the following, we first show exemplary for the Hamiltonian how $O_{\text{loc}}(\theta)$ is derived, i.e. we compute the local energy $H_{\text{loc}}(\theta)$. 
In a second step, we explain the Monte Carlo simulation, in particular how samples from $p(\theta)$ are generated using Metropolis algorithm.

\subsubsection{Computation of the local energy \texorpdfstring{$H_{\text{loc}}(\theta)$}{H loc} \label{sec: measurement_procedure}}
The electric Hamiltonian $H_E$ is the only term of the Hamiltonian defined in eq.~\eqref{eq:full_hamiltonian} that is not diagonal in $\theta$ (the most difficult type of observable to compute, as discussed above). 
We thus focus on $H_E$ and discuss other terms briefly at the end of this section. 

The electric Hamiltonian corresponds to second order derivatives in the gauge field variables $\theta_{\mathbf{x},i}$. 
Since our ansatz consists of a fermionic part $\ket{\Psi_F(\theta)}$ and a pure gauge part $\Psi_G(\theta)$, the electric energy has a solely fermionic contribution, a pure gauge contribution and a crossterm between the two, denoted as: 
\begin{equation} \label{eq: electric_energy}
\expval{H_E}=\expval{H_E}_{ff} + \expval{H_E}_{gg} + \expval{H_E}_{fg}
\end{equation}

We start by considering $\expval{H_E}_{gg}$, the part originating from taking twice the derivative of $\Psi_G(\theta)$ whose construction is based on the vector $b(\theta)$ (see eq.~\eqref{eq: gauge_ansatz}). 
Hence, we need to compute the derivative of $b(\theta)$ with respect to $\theta_{\mathbf{x},i }$ which gives rise to the vector
\begin{equation} \label{eq: definition b_x,i}
    \begin{aligned}
        b_{\mathbf{x},i}(\theta)= \delta_{\mathbf{p},(\mathbf{x},i)} (-&\sin(\theta_{\mathbf{p}_1}), ..,-\sin(\theta_{\mathbf{p}_N}),\\
        &\cos(\theta_{\mathbf{p}_1}), ..,\cos(\theta_{\mathbf{p}_N})) 
    \end{aligned}
\end{equation}
with 
\begin{align}
    \delta_{\mathbf{p},(\mathbf{x},i)}=\begin{cases}
    1 \quad &\text{if} \quad (\mathbf{x},i) \in \mathbf{p} \quad \text{clockwise} \\
    -1 \quad &\text{if} \quad (\mathbf{x},i) \in \mathbf{p} \quad \text{anti-clockwise} \\
    0 \quad &\text{else} \\
    \end{cases}
\end{align}
where $(\mathbf{x},i) \in \mathbf{p}$ clockwise (anti-clockwise) means that the link $(\mathbf{x},i)$ is contained in the plaquette $\mathbf{p}$ and the orientation of the link is parallel (anti-parallel) to the orientation of the plaquette. 
For periodic boundary conditions we have the additional entries $\cos(\theta_j)$ and $\sin(\theta_j)$ in $b(\theta)$ corresponding to the global loops $\theta_1$ and $\theta_2$. 
They give rise to the derivatives $-\sin(\theta_j)$ and $\cos(\theta_j)$ if $(\mathbf{x},i)$ lies on the $x_j$-axis and otherwise zero.

The electric energy of the pure gauge part and the corresponding local quantity $H_{E,gg,\text{loc}}(\theta)$ is then derived as 
\begin{equation}
    \begin{aligned}
        &\expval{H_E}_{gg} \\
         =&  \frac{\int D\theta \frac{g^2}{2}  \sum\limits_{\mathbf{x},i} \left( b^T (\theta) \alpha b_{\mathbf{x},i}(\theta) + \beta^T b_{\mathbf{x},i}(\theta) \right)^2  e^{-S(\theta)}} {\int D \theta \hspace{2pt} e^{-  S(\theta)}} \\
        \equiv& \int D\theta H_{E,gg,\text{loc}}(\theta) \hspace{1pt} p(\theta)
    \end{aligned}
\end{equation}
with the probability distribution $p(\theta)$ and 
 $S(\theta)=b^T(\theta) \alpha b(\theta) + 2\beta^T b(\theta)$, both defined in eq.~\eqref{eq: prob_distribution}. 
The part of the electric Hamiltonian acting only on $\Psi_G(\theta)$ can therefore be written in a simple diagonal form in the gauge field basis. 

It is  more difficult to compute $H_{E,ff,\text{loc}}(\theta)$, i.e. the local quantity corresponding to derivatives of the fermionic ansatz $\ket{\Psi_F(\theta)}$. 
As discussed earlier, we first derive an expression $H_{E,ff,\text{fer}}(\theta)$ that will be diagonal in $\theta$ but still contains fermionic operators (see Appendix~\ref{sec:app_el_energy} for details):
\begin{equation} \label{eq:electric_ham_in_gauge}
    \begin{aligned}
    &\expval{H_E}_{ff} \\
=& \frac{g^2}{2}\int D\theta \hspace{2pt} p(\theta) \expval{\sum_{\mathbf{x},i} - \frac{\partial^2 }{\partial \theta_{\mathbf{x},i}^2}}{\Psi_F\left(\theta\right)} \\
=&  \frac{g^2}{2} \int D\theta \hspace{2pt} p(\theta)  \sum_{\mathbf{x},i} \expval{ \vec{\psi}^{\dagger} f_{\mathbf{x},i}(\theta) \vec{\psi} \vec{\psi}^{\dagger} f_{\mathbf{x},i}(\theta) \vec{\psi} }  {\Psi_F\left(\theta\right)} \\ 
\equiv &  \int D\theta \hspace{2pt} p(\theta) \expval{H_{E,ff,\text{fer}}(\theta)}{\Psi_F\left(\theta\right)}  \\ 
    \end{aligned}
\end{equation}
with 
\begin{equation} \label{eq: general form of f_theta}
    \begin{aligned}
        f_{\mathbf{x},i}(\theta)
        &=  \frac{1}{i} \left( \partial_{\theta_{\mathbf{x},i}}  e^{i \xi(\theta)} \right) e^{-i \xi(\theta)}.
    \end{aligned}
\end{equation}

The form of $f_{\mathbf{x},i}(\theta)$ above is for a general gauge-field dependent fermionic Gaussian state characterized by some $\xi(\theta)$. 
To get an expression explicitly diagonal in $\theta$ we insert our ansatz $\xi(\theta)=V(\theta) \tilde{\xi} V^{\dagger}(\theta)$ defined in eq.~\eqref{eq: xi_theta_definition} which is based on the eigendecomposition of the gauge-matter Hamiltonian, $h_{GM}(\theta)=V(\theta) \Lambda (\theta) V^{\dagger}(\theta)$. 
We obtain (see Appendix~\ref{sec:app_el_energy}  for the derivation):
\begin{equation}
    \begin{aligned}
         f_{\mathbf{x},i}(\theta)
        &= \vec{\psi}^{\dagger}  V(\theta) \left( \alpha^{\mathbf{x},i}(\theta) -  e^{i \tilde{\xi} } \alpha^{\mathbf{x},i}(\theta) e^{-i \tilde{\xi}}\right) V^{\dagger}(\theta) \vec{\psi}
    \end{aligned}
\end{equation}
with $\alpha^{\mathbf{x},i}(\theta)= -i V^{\dagger}(\theta) \partial_{\theta_{\mathbf{x},i}} V(\theta)$. 
We can find an explicit expression for $\alpha^{\mathbf{x},i}(\theta)$ which amounts to finding the derivatives of the eigenvectors of $h_{GM}(\theta)$:
\begin{align}
    \alpha^{\mathbf{x},i}_{kl}(\theta)
    &=  \frac{  V^{\dagger}_{k \mathbf{x}}(\theta) e^{i\theta_{\mathbf{x},i}} V(\theta)_{\mathbf{x}+\mathbf{e}_i  l}  - h.c.}{\lambda_l(\theta) - \lambda_k(\theta)} 
\end{align}
where $\lambda_i(\theta)$ are the eigenvalues of $h_{GM}(\theta)$. 
The final expression for $H_{E,ff,\text{fer}} (\theta)$ is thus diagonal in $\theta$ but still a quartic fermionic operator. 
This form of the electric Hamiltonian intuitively illustrates that the gauge field mediates interactions between the fermions.

In the following we want to evaluate these fermionic interactions w.r.t. the fermionic state $\ket{\Psi_F(\theta)}$ as shown in the last row in eq.~\eqref{eq:electric_ham_in_gauge} to compute the local electric energy $H_{E,ff,\text{loc}} (\theta)$ that can then be measured in our Monte Carlo simulation. 

As a side note we want to mention that $\ket{\Psi_F(\theta)}$ in its general form defined in eq.~\eqref{eq: general fermionic ansatz} does not need to be Gaussian as one can also choose a Non-Guassian reference state $\ket{\Psi_0}$. 
This might be useful if one is particularly interested in the strong-coupling regime (from the high-energy physics perspective one is usually interested in the weak-coupling region where the continuum limit is located). 
In the strong-coupling regime the electric field is strongly suppressed and the Hilbert space effectively reduces to a fermionic Fock space. 
Such models can be tackled by other many-body methods (e.g. tensor networks) which are not suitable for lattice gauge theories with infinte-dimensional local Hilbert spaces. 
One could combine our ansatz with such methods by carrying out the unitary transformation given by the fermionic Gaussian operator $U_{GS}(\theta)$ (acting on top of $\ket{\Psi_0}$) so that the remaining expression can be evaluated w.r.t. the reference state $\ket{\Psi_0}$ whose fermionic correlation functions could be computed with another method (in Appendix~\ref{sec:app_heisenberg} we demonstrate this for two fermion flavors at half-filling where the effective model is the Heisenberg model).

If we focus, however, on the case of one fermion flavor and the Gaussian reference state $\ket{D}$ as defined in eq.~\eqref{eq:psi_fermionic}, we need to evaluate a fermionic Gaussian state for every gauge field configuration $\theta$. 
The fermionic expectation values in eq.~\eqref{eq:electric_ham_in_gauge} can then be computed as 
\begin{equation} \label{eq: eval with fermionc Gaussian}
    \begin{aligned}
&\expval{ \vec{\psi}^{\dagger} f_{\mathbf{x},i}(\theta) \vec{\psi} \vec{\psi}^{\dagger} f_{\mathbf{x},i}(\theta) \vec{\psi} }  {\Psi_F\left(\theta\right)}    \\
   =& \Tr \left( \left( \mathds{1} - \Gamma_{\psi \psi^{\dagger}}(\theta)  \right) f_{\mathbf{x},i}(\theta)  \right)^2 \\
   +& \Tr \left( \left( \mathds{1} - \Gamma_{\psi \psi^{\dagger}}(\theta)  \right) f_{\mathbf{x},i}(\theta) \Gamma_{\psi \psi^{\dagger}}(\theta) f_{\mathbf{x},i}(\theta) \right)     
    \end{aligned}
\end{equation} 
where $\Gamma_{\psi \psi^{\dagger}}(\theta)= V(\theta) \tilde{\Gamma} V(\theta)^{\dagger}$ is the covariance matrix of the Gaussian state $\ket{\Psi_F\left(\theta\right)}$ and $\tilde{\Gamma} = e^{i \tilde{\xi}}  V(\theta)^{\dagger} \Gamma_0 V(\theta) e^{-i \tilde{\xi}}$ with $\Gamma_0$ the covariance matrix of the reference state $\ket{D}$ and $\tilde{\xi}$ containing the variational parameters (see eq.~\eqref{eq: xi_theta_definition}). 
Inserting the expectation values above in eq.~\eqref{eq:electric_ham_in_gauge} gives $H_{E,ff,\text{loc}} (\theta)$.

The last remaining part of the electric Hamiltonian, the crossterm $\expval{H_E}_{fg}$, involves a quadratic expression in the fermions coming from $\ket{\Psi_F(\theta)}$ and a derivative in $b(\theta)$ coming from $\Psi_G(\theta)$ and is thus easier to compute than the quartic expressions in the pure fermionic contribution (for the explicit form see Appendix~\ref{sec:app_el_energy}). 

Other parts of the Hamiltonian are easier to evaluate since they are already diagonal in the gauge field basis. 
For the sake of completeness we will provide them here briefly. 
First, the magnetic part which is directly suitable for Monte Carlo sampling:
\begin{equation}
    \begin{aligned}
        \expval{H_B}&=g_{\text{mag}} \int D\theta \sum_{\mathbf{p}} \left(1-\cos(\theta_{\mathbf{p}})\right)  p(\theta) \\
        &\equiv \int D\theta H_{B,\text{loc}} (\theta) p(\theta) \\
    \end{aligned}
\end{equation}
The gauge-matter interactions are already diagonal and only quadratic in the fermions:
\begin{equation}
    \begin{aligned}
        \expval{H_{GM}}&=-t \int D\theta \vec{\psi}^{\dagger}  h_{GM}(\theta) \vec{\psi} p(\theta) \\
        &=-t \int D\theta \Tr \left( \left( \mathds{1} - \Gamma_{\psi \psi^{\dagger}}(\theta)  \right) h_{GM}(\theta) \right) p(\theta) \\
        &\equiv \int D\theta H_{GM,\text{loc}} (\theta) p(\theta) \\
    \end{aligned}
\end{equation}
where the quadratic expressions in the fermions are evaluated in analogy to the electric part of the Hamiltonian. 
In the same fashion are other purely fermionic parts evaluated such as the  mass term $H_M$.

In terms of computational cost the local electric energy $H_{E,\text{loc}}(\theta)$ is the most difficult part to evaluate. 
Naively, one expects the required number of operations for evaluating it to be $\mathcal{O}(N^4)$ ($N$ the number of lattice sites) but with the chosen parametrization of $\tilde{\xi}$ it can be shown to be $\mathcal{O}(N^3)$ (see Appendix~\ref{sec:tilde_xi}).

\subsubsection{Monte Carlo algorithm} \label{sec: update}
In the following we show how to efficiently evaluate an observable $O$ with our ansatz $\ket{\Psi}$ in a Monte Carlo simulation given an expression for $O_{\text{loc}}(\theta)$. 
The expectation value of $O$ is computed as an average over $N$ samples $\theta_i$ drawn from the probability distribution $p(\theta)$:  
\begin{equation} \label{eq: monte carlo average}
\frac{\expval{O}{\Psi}}{\braket{\Psi}}= \int D \theta \hspace{2pt} O_{\text{loc}}(\theta)   p(\theta) \approx \frac{1}{N}\sum_{i=1}^N O_{\text{loc}}(\theta_i)  
\end{equation}
The samples $\theta_i$ are generated by a Markov chain $\theta_1 \to \hdots \to \theta_i \to \hdots \to \theta_N$ using Metropolis algorithm~\cite{metropolis_equation_1953}. 

One iteration in this procedure, i.e. $\theta_i \to \theta_{i+1}$, is described as follows: starting from $\theta_i$ a new configuration $\theta'$ is proposed according to some update scheme. 
In our case this involves sweeping through every link of the lattice and performing local updates on the gauge variables $\theta_{\mathbf{x},i}$. 
At the same time, we also perform global updates to switch between different monopole-like configurations which is hard to achieve with local updates (for details on the update scheme see Appendix~\ref{sec:app_montecarlo}).
Recalling from eq.~\eqref{eq: prob_distribution} the form $p(\theta) \sim e^{-S(\theta)}$ of our probability distribution, we compute the transition probability $p(\theta \rightarrow \theta')=e^{-S(\theta')}/e^{- S(\theta_i)}=e^{-\Delta S}$. 
In the acceptance step, a random number $u$ between zero and one is generated and the new configuration is accepted if $e^{-\Delta S} \geq u$, i.e. $\theta_{i+1}=\theta'$. Otherwise, the configuration $\theta'$ is rejected and $\theta_{i+1}=\theta_i$.
In the first phase of the Monte Carlo simulation (the warm-up phase) these iterations are performed to equilibrate the system (i.e. reach configurations with sufficiently low weight $S(\theta)$) and only after that the configurations $\theta_i$ are used to compute the expectation value in eq.~\eqref{eq: monte carlo average}.

The numerical cost of performing Metropolis algorithm depends on computing the transition probability between the old configuration $\theta$ and the proposed configuration $\theta'$. 
For local updates they differ only in a single link variable $\theta_{\mathbf{x},i}$, respectively two plaquette variables $\theta_{\mathbf{p}}$. 
The vector $b(\theta)$, constructed out of $\sin(\theta_{\mathbf{p}})$ and $\cos(\theta_{\mathbf{p}})$, is thus changed in four places. 
Since $S(\theta)$ is bilinear in $b(\theta)$, the cost of computing $\Delta S$ is of order $\mathcal{O}(N)$ where $N$ is the number of lattice sites. 
Sweeping through the lattice with this procedure is thus of order $\mathcal{O}(N^2)$. For the global updates the transition probability requires $\mathcal{O}(N^2)$ operations but is only performed $\mathcal{O}(1)$ times so that the cost of a full update is $\mathcal{O}(N^2)$.

Having such a low cost for updates has several advantages: we can perform multiple local and global updates to further decorrelate expensive measurements. 
The acceptance probability in our simulations stays on a high level throughout the whole coupling region (see Appendix~\ref{sec:app_montecarlo}).
Moreover, if we parallelize the Monte Carlo simulation with multiple runners there is practically no overhead due to the warm-up phase. 

\subsection{Adaption of variational parameters}
In the last section we described the evaluation with our Monte Carlo scheme for a fixed set of variational parameters. 
To study ground states and dynamical phenomena we need to adjust the variational parameters accordingly. 
Here, we focus on the study of ground state properties but the discussion can be readily extended to time-evolution phenomena as we use an imaginary time-evoultion procedure (called stochastic reconfiguration in the variational Monte Carlo language~\cite{sorella_wave_2005}) to find the optimal set of parameters. 
We project the equations of motion onto the tangent plane of our variational manifold. 
For every variational parameter $\gamma_i$, either fermionic (in  $\widetilde{\xi}$) or pure gauge (in $\alpha$ and $\beta$) we define a corresponding tangent vector $\ket{\Psi_{i}} \equiv \mathbb{P}_{\Psi} \left( \partial_{\gamma_i} \ket{\Psi} \right)$ where $\mathbb{P}_{\Psi}$ ensures orthogonality to $\ket{\Psi}$: 
\begin{align} \label{projectionvarmanifold}
    \mathbb{P}_{\Psi}(\ket{\psi}) \equiv \ket{\psi}-\braket{\Psi}{\psi}\ket{\Psi}
\end{align}
All tangent vectors in our ansatz are linearly independent which allows to invert the Gram matrix $G_{ij} \equiv \braket{\Psi_i}{\Psi_j}$. 
This can be intuitively explained by considering the different types of tangent vectors: the ones corresponding to the fermionic parameters are related to the single-particle eigenstates of the
gauge-matter Hamiltonian and are therefore orthogonal. 
The tangent vectors corresponding to the pure gauge part are quadratic (for $\alpha$) or linear (for $\beta$) in the entries of the vector $b(\theta)$ which are related to the different plaquette variables $\theta_{\mathbf{p}}$, thus leading to linearly independent tangent vectors.
The imaginary time evolution of the variational parameters can then be expressed in the following way:
\begin{align}
- \dot{\gamma}_{i}&= \frac{1}{2} \sum_{j}  (G^{-1})_{ij}  \frac{\partial E}{\partial \gamma_j}  
\end{align}
with $E \equiv \frac{ \expval{H}{\Psi}}{\braket{\Psi}}$ the variational energy (whose evaluation was described in the previous section) and $\dot{\gamma}\equiv\pdv{\gamma}{\tau}$. 

The gradient of the variational energy and the Gram matrix need to be measured in a Monte Carlo simulation. 
The cost of both can be shown to scale in the same way as the cost of computing the variational energy (see Appendix~\ref{sec:gram}). 
Summarizing, the computational complexity of our variational Monte Carlo algorithm is $\mathcal{O}(N^2)$ for the update procedure and $\mathcal{O}(N^3)$ for the measurement procedure, thus allowing for an efficient implementation.

\section{Benchmarking of the variational method\label{sec:validity}} 
Now, we have all the ingredients to apply our variational method:
We constructed a gauge-invariant state and showed how it can be efficiently evaluated for a fixed set of parameters using Monte Carlo sampling.
Additionally, we have a scheme to adapt the parameters using stochastic reconfiguration.

In the following section, we investigate the validity of the variational method. 
It will be threefold: 
first, to confirm the analytical arguments about gauge invariance of the ansatz given in the previous section we will show numerically that our state is gauge-invariant up to machine precision. 
Secondly, we investigate different limiting cases of $\text{cQED}_3$ where the ground states are known.
In the last part, we benchmark our results for the $N_{f}=2$ case at half-filling (in the sector of exactly one fermion per lattice site) with a recent Monte Carlo simulation~\cite{xu_monte_2019}.

\subsection{Gauge invariance} \label{sec:gaugeinvarince}
\begin{figure}[t]
    \centering
    \includegraphics[width=\columnwidth]{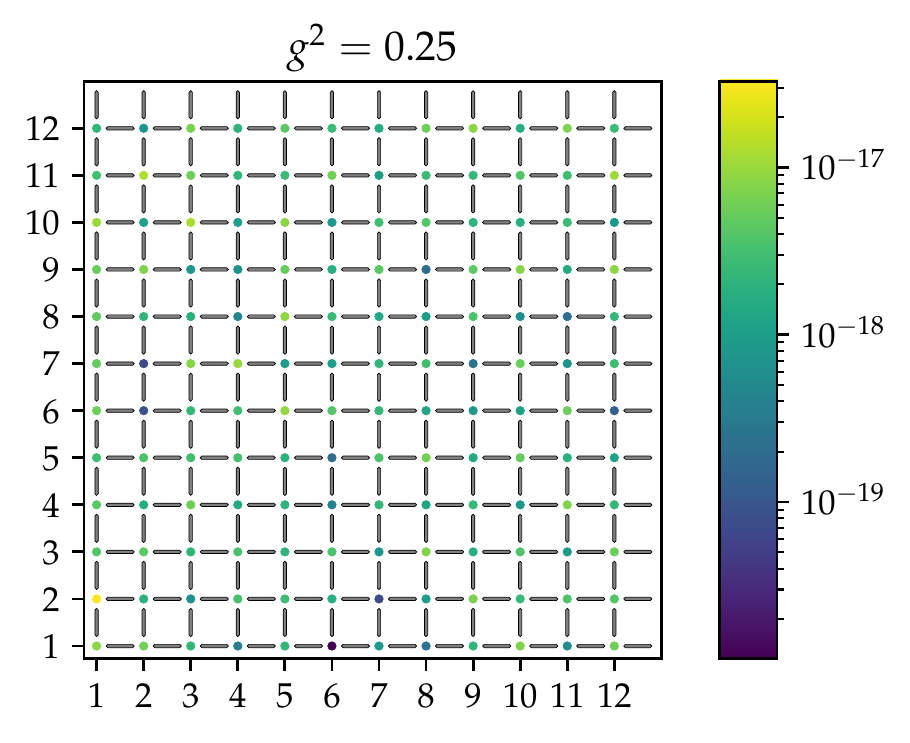}
    \caption{Gauss law violation $\expval{G_{\mathbf{x}}} - q_{\mathbf{x}}$ for a $12 \times 12$ lattice at $g^2=0.25$, $t=1$ and $g_{\text{mag}}=-1$ with a random choice of variational parameters and a sampling size of $N=10$. The Gauss law violation is of the order of machine precision even for a small sampling size, demonstrating that the ansatz is inherently gauge-invariant.}
    \label{fig:gauge_invariance}
\end{figure}

To also show numerically that gauge invariance is manifest in our ansatz we compute the expectation value of the Gauss law operator $\expval{G_{\mathbf{x}}}$ (as defined in eq.~\eqref{eq:gauss_operator}) for every site $\mathbf{x}$ and plot $\expval{G_{\mathbf{x}}} - q_{\mathbf{x}}$ for the whole lattice since this quantity needs to be zero for a physical, gauge-invariant state, see eq.~\eqref{eq:gauss_phys}. 
We choose different variational parameters, different lattice sizes and different sampling sizes but the violation of the Gauss law is always found to be of the order of machine precision, i.e. $ \expval{G_{\mathbf{x}}} - q_{\mathbf{x}} \lesssim 10^{-16}$. 
One such configuration for a very small sampling size of $N=10$ and a system size $L=12$ is illustrated in Fig.~\ref{fig:gauge_invariance}. 

\subsection{Limiting cases\label{sec:limiting}} 
It is useful to consider the limiting cases of compact QED with fermionic matter and convince ourselves that the ground state properties can be captured accurately by our method. In the following, we consider massless fermions without chemical potentials.

\begin{figure}[t]
    \centering
    \includegraphics[width=\columnwidth]{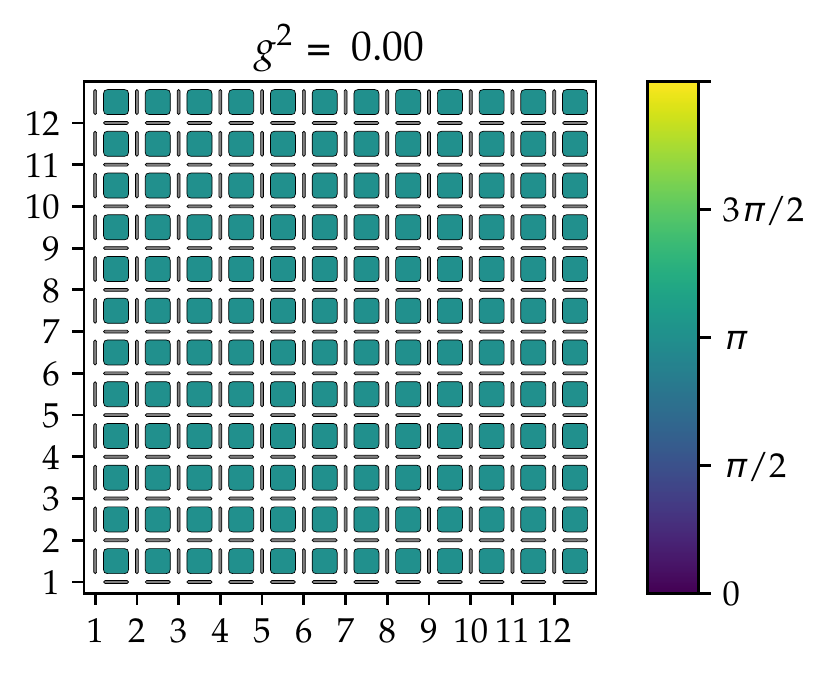}
    \caption{The flux $\theta_{\mathbf{p}}$ per plaquette is shown for the variational ground state at $g^2=0.0$, $t=1$ and $g_{\text{mag}}=-1$ for a $12 \times 12$ lattice. The average deviation of $\theta_{\mathbf{p}}$ from $\pi$ is on the order of $10^{-4}$. The global loops $\theta_1$ and $\theta_2$ (winding around the axes of the lattice) also acquire a $\pi$-flux with a similar deviation as the plaquette fluxes.}
    \label{fig:pi-flux}
\end{figure}

We first study the limit $g^2 \to 0$ while keeping $t$ and $g_{\text{mag}}$ fixed: it is well known that in this limit the gauge field forms a $\pi$-flux pattern and the fermions fill up the lower band at half-filling~\cite{PhysRevLett.73.2158}. 
A typical problem in mean-field theory is to investigate the stability of the $\pi$-flux pattern against gauge field fluctuations. 
This can be studied naturally in our ansatz by watching the parameter flow upon increasing the electric coupling constant $g^2$. 
The $\pi$-flux state itself is naturally incorporated in our ansatz since we can fix the gauge field to a certain configuration by tuning the $\beta$-parameters to a very high value such that the constraint $\cos{\theta_{\mathbf{p}}}=-1$ is enforced for all plaquettes. 
In addition, since we have periodic boundary conditions, we also need to choose the optimal flux configuration for the global non-contractible loops which depends on the size of the lattice. 
To accomplish that it is important to have a global update in our update scheme since these global changes in the configuration can not be captured by only updating plaquettes locally. Finding the $\pi$-flux state is in general a useful test for our update scheme since the probability distribution needs to approximate a delta distribution for which a good update scheme is required.
The fermionic part is obtained by tuning the variational parameters of the fermions in such a way that for all flux configurations the lower half of the band is occupied (which corresponds to choosing all fermionic parameters $\xi_i=1$ as described in Appendix~\ref{sec:tilde_xi}). 
The result of our variational optimization is an accurate representation of the $\pi$-flux state with an average deviation on the order of $10^{-8}$ from $\cos{\theta_{\mathbf{p}}}=-1$, respectively an average deviation on the order of $10^{-4}$ from $\theta_{\mathbf{p}}=\pi$ (depicted in Fig.~\ref{fig:pi-flux}).

Next we consider the opposite limit to the $\pi$-flux state, the strong-coupling limit with large $g^2$. 
In this limit, the electric energy dominates and some fluctuations are introduced in second-order perturbation theory by the gauge-matter Hamiltonian. 
For one fermion flavor this perturbation does not have a large effect and the ground state is described by a Gaussian state. For two fermion flavors, however, one can have correlated hopping processes which at half-filling give rise to the Heisenberg Hamiltonian. 
Both cases can be captured by construction in our ansatz since we design the fermionic part of the ansatz in such a way that our gauge-field dependent Gaussian operator acts on a strong-coupling reference state $\ket{\Psi_0}$ (see eq.~\eqref{eq: general fermionic ansatz}) and we can choose that reference state according to our needs. We can either choose a Gaussian state for $\ket{\Psi_0}$ or include more advanced methods, e.g. to approximate the Heisenberg ground state we can include spin wave theory in $\ket{\Psi_0}$ (see Appendix~\ref{sec:app_heisenberg}). 

\begin{figure}[t]
    \centering
    \includegraphics[width=\columnwidth]{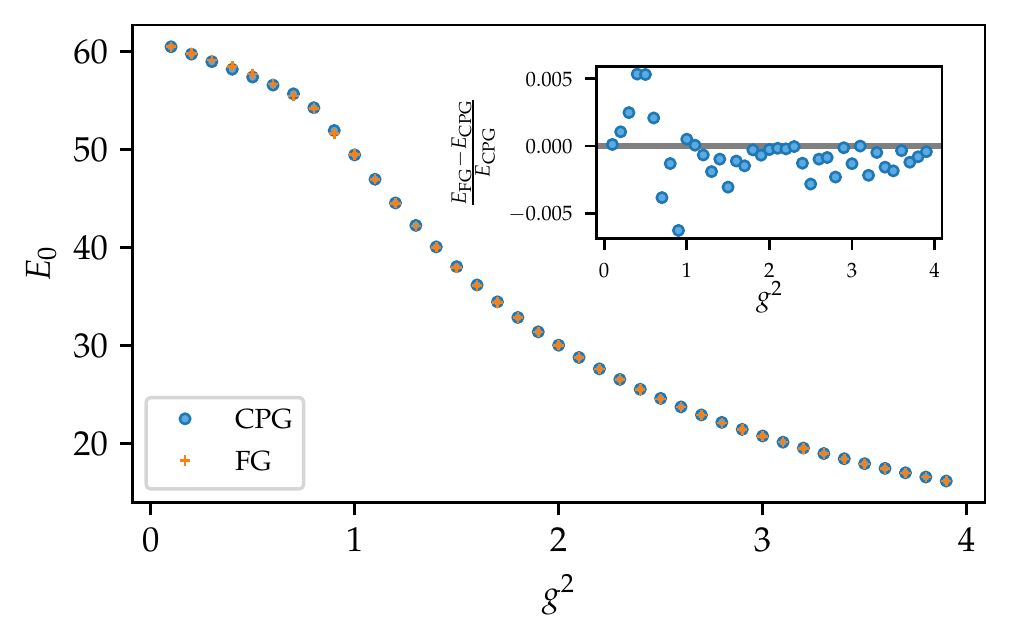}
    \caption{Benchmark for pure gauge compact $\text{QED}_3$ of our ansatz (denoted by FG) against the variational method in ref.~\cite{bender_real-time_2020} based on complex periodic Gaussian states (denoted by CPG): we compare the ground state energy $E_0$ for an $8 \times 8$ lattice over the whole coupling region and compute the relative error (see inset). }
    \label{fig:puregauge_benchmark}
\end{figure}

We also benchmark for the limiting case that the gauge-matter interactions vanish ($t=0$) so that fermions and gauge-field decouple and we obtain the standard pure gauge compact QED described by the Kogut-Susskind Hamiltonian $H_{KS}=H_E+H_B$ with $g_{\text{mag}}=\frac{1}{g^2}$~\cite{kogut_hamiltonian_1975}. 
We therefore only consider the pure gauge part of our ansatz (setting our fermionic variational parameters to zero, $\widetilde{\xi}=0$). 
We benchmark our ansatz against a recent work~\cite{bender_real-time_2020} which has given good ground state and real-time dynamics of compact QED (the results were recently confirmed by another variational study~\cite{luo_gauge_2022}). 
We compare the ground state energy of both methods for an $L=8 \times 8$ for the whole coupling region of $g^2$ (since $g^2$ is the only coupling constant in pure gauge compact QED). 
We find that our results agree very well for the whole coupling region (with a maximal difference of half a percent) while our method performs a tiny bit better at large couplings where the method in ref.~\cite{bender_real-time_2020} gives minimally better results for small $g^2$. 
The benchmark is illustrated in Fig.~\ref{fig:puregauge_benchmark}.

\subsection{Benchmark against Euclidean Monte Carlo \label{sec:benchmark}}
Benchmarking for $\text{cQED}_3$ including dynamical fermions is in general difficult since in most scenarios a sign-problem occurs so that no Euclidean Monte Carlo studies exist. 
However, it  was shown to be absent for an even fermion number at zero chemical potential~\cite{xu_monte_2019}. 
This was exploited in order to perform determinantal Monte Carlo simulations. 
Thus, it is natural to compare our ansatz with the Monte Carlo simulations for the case of $N_{f}=2$ fermionic species at half-filling. 
The analysis in ref.~\cite{xu_monte_2019} revolves around the question of whether a confinement-deconfinement transition takes place and what the nature of this phase transitions is. 

\begin{figure}[t]
    \centering
    \includegraphics[width=\columnwidth]{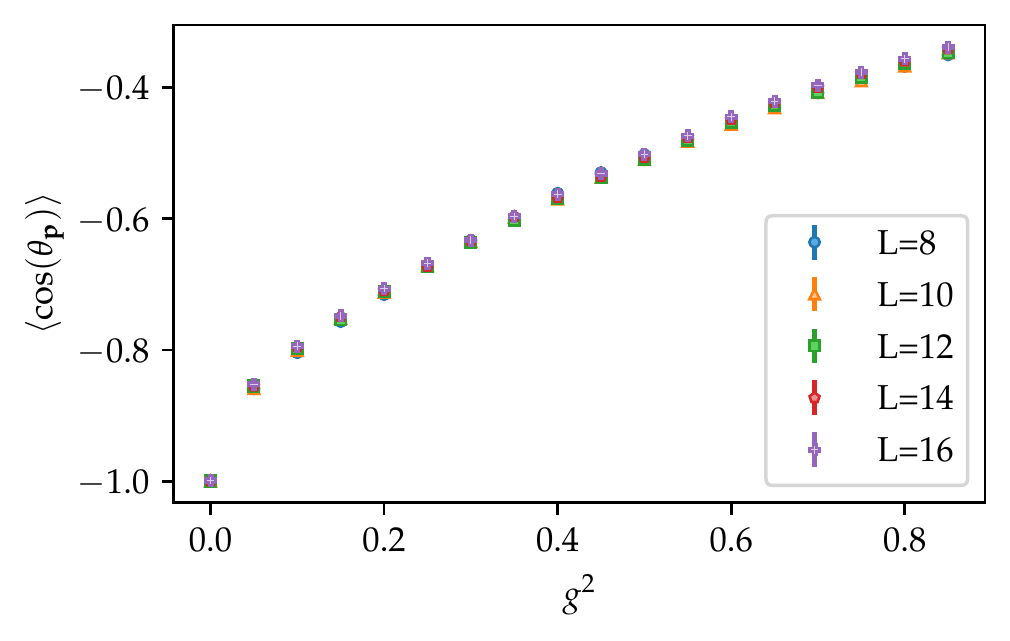}
    \caption{Benchmark for compact $\text{QED}_3$ coupled to $N_f=2$ species of dynamical fermions at half-filling: the shown averaged flux energy per plaquette $\cos(\theta_{\mathbf{p}})$ in the variational ground state is to be compared with results obtained in an Euclidean Monte Carlo study shown in Fig.~13 in ref.~\cite{xu_monte_2019}. The data agrees over the whole coupling region, showing no evidence of a discontinuous phase transition.}
    \label{fig:flux_per_plaquette}
\end{figure}

We fix the magnetic coupling and the gauge-matter coupling to $g_{\text{mag}}=-1$ and $t=1$ and will mostly vary the electric coupling $g^2$. 
The first observable that is compared is the flux energy per plaquette $\cos(\theta_{\mathbf{p}})$ averaged over the whole lattice.
Our results are shown in Fig.~\ref{fig:flux_per_plaquette}. 
We see agreement over the whole coupling region of $g^2$ with Fig.~13 in ref.~\cite{xu_monte_2019}. Note that in ref.~\cite{xu_monte_2019} a different convention for the electric coupling is used differing by a factor 4. 
Thus, the upper end of the coupling ranges is the same while our lower end goes further down to $g^2=0$. 
Since we also do not observe finite-size effects it supports the claim in ref.~\cite{xu_monte_2019} that there is no discontinuous phase transition taking place. 

In the second part we study fermionic observables, related to the fermionic correlations of the ground state. 
These are used in ref.~\cite{xu_monte_2019} to probe a phase transition between a deconfined U(1) spin-liquid and a confined phase exhibiting antiferromagnetic order (AFM). 
The observable that is computed is the spin structure factor $\chi_S(\mathbf{k})$:
\begin{equation} \label{eq: spin correlations}
    \chi_S(\mathbf{k})= \frac{1}{L^4} \sum_{\mathbf{x},\mathbf{y}} \sum_{\alpha,\beta=1,2} \expval{ S^{\alpha}_{\beta} (\mathbf{x}) S^{\beta}_{\alpha}(\mathbf{y})} e^{i \mathbf{k} \left(\mathbf{x} - \mathbf{y} \right)}
\end{equation}
with $S^{\alpha}_{\beta} (\mathbf{x})= \psi_{\mathbf{x},\alpha}^{\dagger} \psi_{\mathbf{x},\beta} - 1/2 \delta_{\alpha \beta} \sum_{\gamma} \psi_{\mathbf{x},\gamma}^{\dagger} \psi_{\mathbf{x},\gamma} $. From the spin structure factor one can compute the AFM correlation ratio defined as 
\begin{equation}
    r_{\text{AFM}}= 1 - \frac{\chi_S((\mathbf{\pi},\mathbf{\pi}) + \delta \mathbf{k})}{\chi_S((\mathbf{\pi},\mathbf{\pi}))}
\end{equation}
which quantifies the strength of AFM order ($\delta \mathbf{k}=(2\pi/L,0)$ denotes the smallest momentum vector). 
The question addressed in ref.~\cite{xu_monte_2019} is whether in the thermodynamic limit AFM order persists down to $g^2=0$, in other words whether the $\pi$-flux state is stable against gauge-field fluctuations. 
\begin{figure*}[t]
    \centering
    \includegraphics{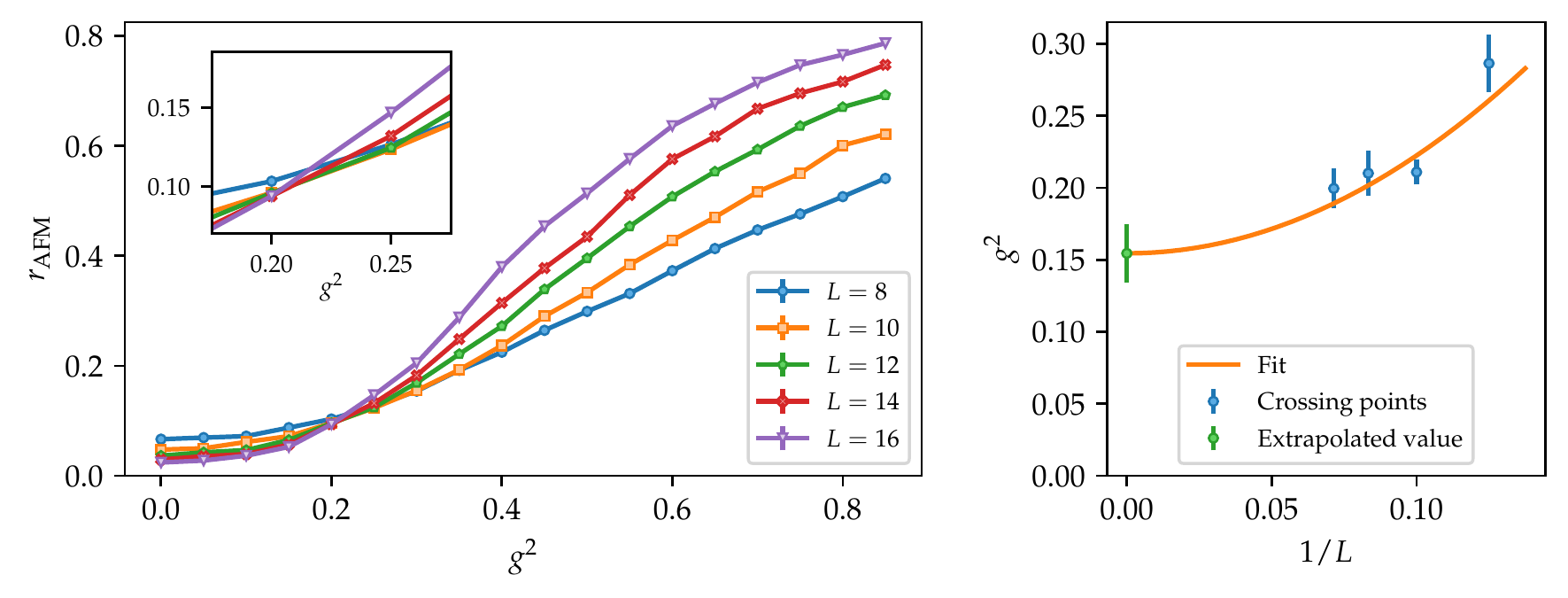}
    \caption{Benchmark for compact $\text{QED}_3$ coupled to $N_f=2$ species of dynamical fermions at half-filling: we compute the AFM correlation ratio $r_{\text{AFM}}$ in the variational ground state for lattice size up to $16 \times 16$ (left). The AFM correlation ratio is computed from the spin correlations and quantifies the strength of antiferromagnetic order. The crossing points are extracted and extrapolated to the thermodynamic limit, resulting in $g^2_{c,\infty}=0.15(2)$ (right). This is to be compared with the Euclidean Monte Carlo study in ref.~\cite{xu_monte_2019} where also a non-zero coupling was extrapolated but at a higher value of $g^2_{c,\infty,\text{EMC}}=0.40(5)$.} 
    \label{fig:afm-ratio}
\end{figure*}
The AFM correlation ratio is computed up to lattice sizes of $16 \times 16$ and the crossing points between neighboring lattice sizes are extracted. 
The crossing points are extrapolated to the thermodynamic limit, resulting in $g^2_{c,\infty}=0.15(2)$. 
The procedure is shown in Fig.~\ref{fig:afm-ratio} which is to be compared with the Euclidean Monte Carlo study in ref.~\cite{xu_monte_2019} where the extrapoled value is $g^2_{c,\infty,\text{EMC}}=0.40(5)$. 
We thus obtain qualitatively similar results in the sense that both extrapolated values are significantly larger than zero and indicate a possible phase transition but the value in our method is lower compared to ref.~\cite{xu_monte_2019}.

\begin{figure}[t]
    \centering
    \includegraphics[width=\columnwidth]{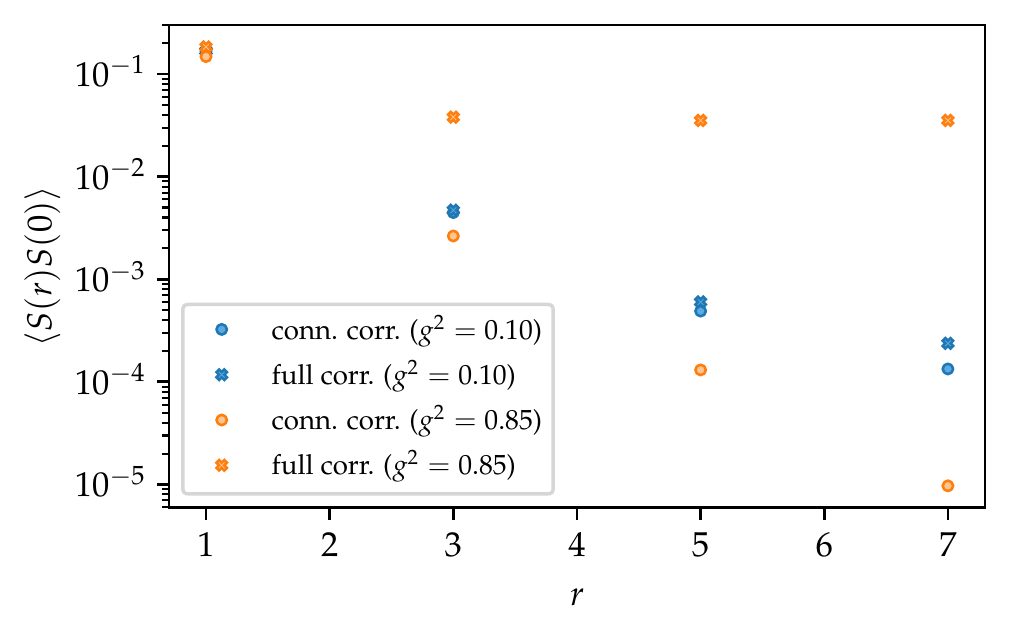}
    \caption{Benchmark for compact $\text{QED}_3$ coupled to $N_f=2$ species of dynamical fermions at half-filling: we compute the decay of spin correlations (both for the full correlation function and the connected correlation function) in the variatonal ground state for a $16\times 16$ lattice at weak coupling ($g^2=0.1$) and at stronger coupling ($g^2=0.85$). Note that we only use odd distances in $r$ to avoid oscillations. At strong coupling (where one expects behaviour similar to the Heisenberg model) the full correlations decay to a constant while the connected part decays exponentially. At weak coupling the decay is rather algebraically, similar to the decay shown in the Euclidean Monte Carlo study in Fig.~4 in ref.~\cite{xu_monte_2019}.} 
    \label{fig:spincorrelations}
\end{figure}

Another interesting quantity are the spin-spin correlations as defined in eq.~\eqref{eq: spin correlations}. We compute the decay of spin correlations on a $16 \times 16$ lattice both in the weak-coupling region ($g^2=0.1$) and in a more strongly-coupled region ($g^2=0.85)$. The result for both the full correlation function and only the connected part is shown in Fig.~\ref{fig:spincorrelations}.
At stronger coupling the correlation function decays to a constant value which is lower than predicted by the Heisenberg model (as to be expected since $g^2=0.85$ is still too small for a Heisenberg description). The connected correlation function decays exponentially as expected.
At weak coupling the connected correlation function rather decays algebraically, as expected for a gapless spin liquid. The form of the decay is very similar to one in the Euclidean Monte Carlo study (see Fig.~4 in ref.~\cite{xu_monte_2019}).
 
We can thus, at least qualitatively, support the claim in ref.~\cite{xu_monte_2019} that there is indeed a deconfined phase which, however, only persists up to a smaller coupling of $g^2_{c,\infty}=0.15(2)$ in our case. 
One should note though that for the extrapolation of the AFM correlation ratio and also the computation of the spin structure factor is very sensitive to errors (as also mentioned in ref.~\cite{xu_monte_2019}) so that a quantitative difference can be expected.

\section{Sign-problem affected regimes\label{sec:signproblem}}  
\begin{figure*}[t]
   \centering
    \includegraphics[width=\textwidth]{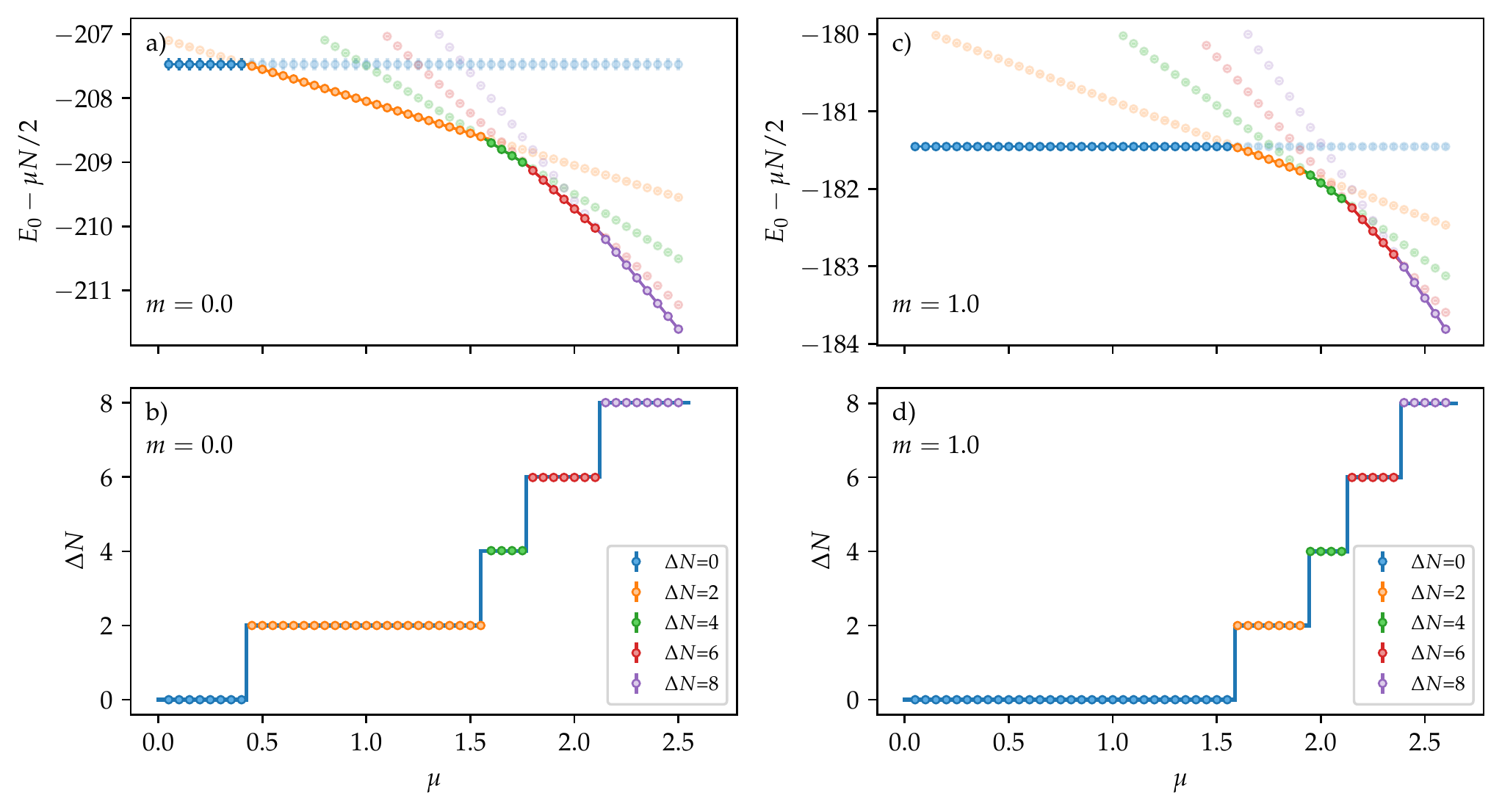}
    \caption{Finite chemical potential study for compact $\text{QED}_3$ coupled to $N_f=2$ species of dynamical fermions: we compute the variational ground state energies (corrected by an overall constant $\mu N/2$) on an $8 \times 8$ lattice for different isospin numbers $\Delta N$ depending on the chemical potential difference $\mu$ between the two species. We study both the case of massless and massive fermions (see top row, (a,c)). By computing the crossing points between the ground state energies we can extract the phase transitions between neighboring $\Delta N$ phases (see bottom row, (b,d)).}
    \label{fig:signproblem}
\end{figure*}
In this section, we access regimes where the sign-problem is present in order to demonstrate that our method does not suffer from the sign-poblem. 
Having benchmarked our ansatz for the scenario of two flavors of fermions at half-filling, i.e. zero chemical potential, it is natural to study this configuration at finite chemical potential. 

We specifically want to look at a scenario that has been used in one dimension with tensor networks~\cite{banuls_density_2017} to demonstrate overcoming the sign-problem and extend it to two dimensions. 
In the referenced work the authors study density-induced phase transitions due to varying flavor-dependent chemical potentials. Analogously to ref.~\cite{banuls_density_2017}, we look at the case of massless and massive fermions. 

We fix the parameters in the Hamiltonian given in eq.~\eqref{eq:full_hamiltonian} to the values $t=1$, $g_{\text{mag}}=-1$ and $g^2=0.2$, similar to the benchmarked case in the previous section. Only the staggered mass $m$ and the chemical potentials $\mu_1$ and $\mu_2$ will be changed.
To make this explicit we rewrite the Hamiltonian as
\begin{equation}
\begin{aligned}
    H=& H_E + H_B + H_{GM} + H_M(m) \\
    =&  H_0(m)  + \mu_{1} N_1 + \mu_{2} N_2 \\
    =&  H_0(m)  + \frac{\mu_{+}}{2} N - \frac{\mu_{-}}{2} \Delta N
\end{aligned} 
\end{equation} 
with the conserved quantities $N_1=\sum_{\mathbf{x}} \psi_{\mathbf{x},1}^{\dagger} \psi_{\mathbf{x},1}$ and $N_2=\sum_{\mathbf{x}} \psi_{\mathbf{x},2}^{\dagger} \psi_{\mathbf{x},2}$. Alternatively, one can also use the total number of fermions $N=N_1+N_2$ and their imbalance (sometimes called isopsin number) $\Delta N=N_1-N_2$ as conserved quantities. Respectively, one defines the chemical potentials $\mu_+=(\mu_1+\mu_2)$ and $\mu_-=(\mu_1-\mu_2)$. The rest of the Hamiltonian is contained in $H_0(m)$ which only depends on $m$.

The Hamiltonian is block-diagonal and different sectors are labelled with $N$ and $\Delta N$. 
In analogy to ref.~\cite{banuls_density_2017}, we fix the total number of fermions $N$ to the number of lattice sites and study the nature of the ground state (characterized by $\Delta N$) depending on $\mu_-$, the isospin chemical potential. Since the energy (up to a constant) only depends on $\mu_-$, we set $\mu_2=0$ so that $\mu_+=\mu_-=\mu_1\equiv \mu$.
The ground state energy for each sector can then be written as 
\begin{equation} \label{eq:groundstate_energy_chempot}
\begin{aligned}
    &E_{\Delta N, N}(\mu, m) = N \frac{\mu}{2} - \Delta N  \frac{\mu}{2} + E_{0,\Delta N, N} (m)
\end{aligned} 
\end{equation}
where $E_{0,\Delta N, N} (m)$ is the ground state w.r.t. $H_{0}(m)$ for fixed $N$ and $\Delta N$.
 
We ran our simulations on an $8 \times 8$ lattice where we saw that finite-size effects were negligible for our purposes. 
To detect the phase transitions between different $\Delta N$ phases we compute the variational ground state energy for a given $\Delta N$ to determine $E_{0,\Delta N, N} (m)$. We plot the ground state energy of every $\Delta N$-sector subtracted by the constant given by the total number of fermions, i.e. $E_{\Delta N, N} (\mu, m) - N \frac{\mu}{2}$.
The crossing points between different $\Delta N$ energies give us the location of the phase transitions. 

The result of that procedure for the massless case is shown in Fig.~\ref{fig:signproblem}(a) which then allows us to plot the $\Delta N$ phase transitions, illustrated in Fig.~\ref{fig:signproblem}(b).
For the massive case we choose a staggered mass of $m=1.0$ and repeat the same procedure as in the massless case. 
The result for the ground state energy is shown in Fig.~\ref{fig:signproblem}(c) whereas the phase transitions are shown in Fig.~\ref{fig:signproblem}(d). 

When comparing the massless and massive case it becomes clear that the phase transitions are all shifted to higher values of $\mu$. However, the extent of this shift depends strongly on the isospin number. 
While phase transitions between small $\Delta N$ are severely affected by the staggered mass term (in particular the transition between $\Delta N=0$ and $\Delta N=2$ which shifted from $\mu=0.5$ to $\mu=1.5$), phase transitions for larger $\Delta N$ are relatively unaffected (e.g. the phase transition between $\Delta N=6$ and $\Delta N=8$ shifts only slightly from $\mu=2.2$ to $\mu=2.3$). 
This is in agreement with the results of the tensor-network study in one dimension~\cite{banuls_density_2017}.  

The reason for this behaviour lies in the different changes in ground state energy $E_{0,\Delta N, N} (m)$ for different $\Delta N$ if we go from $H_0(m=0)$ to $H_0(m=1)$. 
Qualitatively, this can be explained with the fact that for larger isospin numbers $\Delta N$ the imbalance in occupation between even and odd sites (the lattice analogue of the chiral condensate) becomes smaller and thus gets more penalized by a staggered mass term. 
Hence, the phase transitions shift to higher values in chemical potential. 
Since this effect is stronger for smaller isospin number, it mostly affects transitions between such phases.

\section{Conclusion\label{sec:conclusion}} 
In summary, we have presented a variational, sign-problem-free Monte Carlo method to study higher-dimensional lattice gauge theories with dynamical fermions without truncating the gauge field Hilbert space and applied it to (2+1)-dimensional compact QED with dynamical fermions.

We benchmarked the ansatz against limiting cases of the model, against other variational methods~\cite{bender_real-time_2020} and against a Euclidean Monte Carlo study~\cite{xu_monte_2019}. 
To access sign-problem affected regimes we study the model at finite chemical potential, namely (in analogy to a tensor-network study in one dimension~\cite{banuls_density_2017}) we detect density-induced phase transition for both the case of massless and massive staggered fermions.

The variational ansatz is formulated in the gauge field basis of $U(1)$, consisting of two parts:
first, a Jastrow-type ansatz state is constructed out of the gauge field plaquette variables, thus readily gauge-invariant. It can describe the ground state of pure gauge compact QED. 
Secondly, a gauge-fermion part is introduced that is an infinite superposition of gauge-field dependent fermionic Gaussian states. For every gauge field configuration a fermionic Gaussian state is defined in such a way that the integral over all gauge field configurations is gauge-invariant and still efficiently tractable, i.e. the number of variational parameters scales polynomially in system size and not exponentially. 
Such a construction can be achieved by making the variational parameters of a fermionic Gaussian state gauge-field dependent and use a parametrization based on the eigendecomposition of the gauge-matter Hamiltonian.
This requires exact diagonalization at every measurement step in the sampling algorithm but since its scaling is  $\mathcal{O}(N^3)$ in system size, the method is efficient and we can reach large lattice sizes. 

In the future it would be interesting to also study other higher-dimensional lattice gauge theories such as three-dimensional or non-Abelian lattice gauge theories. 
The former would not require any change in the ansatz while the non-Abelian nature of the gauge group only requires changes in the pure gauge part of the ansatz. 
The fermionic part could stay the same since the eigendecomposition of the gauge-matter Hamiltonian can still be carried out efficiently. 
The pure gauge part would need to be changed since the plaquette operators on which the Jastrow wavefunction is based would not be gauge-invariant anymore but this could be remedied by using traces of plaquette operators and potentially other closed loops.

Also, in light of more and more powerful quantum devices which require for the simulation of lattice gauge theories some kind of truncation in the gauge field Hilbert space, the presented method could help in studying and thus controlling the errors caused by such truncations.

\acknowledgements

We thank Mari Carmen Ba\~nuls, Snir Gazit and Fakher Assaad for fruitful discussions.
This work was partly supported by the German Federal Ministry of Education and Research (BMBF) through the funded project EQUAHUMO
(Grant No. 13N16066) and THEQUCO within the funding program quantum
technologies—from basic research to market, in association
to the Munich Quantum Valley.

\bibliography{bibliography.bib}

\begin{thebibliography}{60}%
\makeatletter
\providecommand \@ifxundefined [1]{%
 \@ifx{#1\undefined}
}%
\providecommand \@ifnum [1]{%
 \ifnum #1\expandafter \@firstoftwo
 \else \expandafter \@secondoftwo
 \fi
}%
\providecommand \@ifx [1]{%
 \ifx #1\expandafter \@firstoftwo
 \else \expandafter \@secondoftwo
 \fi
}%
\providecommand \natexlab [1]{#1}%
\providecommand \enquote  [1]{``#1''}%
\providecommand \bibnamefont  [1]{#1}%
\providecommand \bibfnamefont [1]{#1}%
\providecommand \citenamefont [1]{#1}%
\providecommand \href@noop [0]{\@secondoftwo}%
\providecommand \href [0]{\begingroup \@sanitize@url \@href}%
\providecommand \@href[1]{\@@startlink{#1}\@@href}%
\providecommand \@@href[1]{\endgroup#1\@@endlink}%
\providecommand \@sanitize@url [0]{\catcode `\\12\catcode `\$12\catcode
  `\&12\catcode `\#12\catcode `\^12\catcode `\_12\catcode `\%12\relax}%
\providecommand \@@startlink[1]{}%
\providecommand \@@endlink[0]{}%
\providecommand \url  [0]{\begingroup\@sanitize@url \@url }%
\providecommand \@url [1]{\endgroup\@href {#1}{\urlprefix }}%
\providecommand \urlprefix  [0]{URL }%
\providecommand \Eprint [0]{\href }%
\providecommand \doibase [0]{http://dx.doi.org/}%
\providecommand \selectlanguage [0]{\@gobble}%
\providecommand \bibinfo  [0]{\@secondoftwo}%
\providecommand \bibfield  [0]{\@secondoftwo}%
\providecommand \translation [1]{[#1]}%
\providecommand \BibitemOpen [0]{}%
\providecommand \bibitemStop [0]{}%
\providecommand \bibitemNoStop [0]{.\EOS\space}%
\providecommand \EOS [0]{\spacefactor3000\relax}%
\providecommand \BibitemShut  [1]{\csname bibitem#1\endcsname}%
\let\auto@bib@innerbib\@empty
\bibitem [{\citenamefont {Gross}\ and\ \citenamefont
  {Wilczek}(1973)}]{gross_asymptotically_1973}%
  \BibitemOpen
  \bibfield  {author} {\bibinfo {author} {\bibfnamefont {D.~J.}\ \bibnamefont
  {Gross}}\ and\ \bibinfo {author} {\bibfnamefont {F.}~\bibnamefont
  {Wilczek}},\ }\href {\doibase 10.1103/PhysRevD.8.3633} {\bibfield  {journal}
  {\bibinfo  {journal} {Physical Review D}\ }\textbf {\bibinfo {volume} {8}},\
  \bibinfo {pages} {3633} (\bibinfo {year} {1973})}\BibitemShut {NoStop}%
\bibitem [{\citenamefont {Peskin}\ and\ \citenamefont
  {Schroeder}(1995)}]{peskin_introduction_1995}%
  \BibitemOpen
  \bibfield  {author} {\bibinfo {author} {\bibfnamefont {M.~E.}\ \bibnamefont
  {Peskin}}\ and\ \bibinfo {author} {\bibfnamefont {D.~V.}\ \bibnamefont
  {Schroeder}},\ }\href@noop {} {\emph {\bibinfo {title} {An introduction to
  quantum field theory}}}\ (\bibinfo  {publisher} {Addison-Wesley Pub. Co},\
  \bibinfo {address} {Reading, Mass},\ \bibinfo {year} {1995})\BibitemShut
  {NoStop}%
\bibitem [{\citenamefont {Wilson}(1974)}]{wilson_confinement_1974}%
  \BibitemOpen
  \bibfield  {author} {\bibinfo {author} {\bibfnamefont {K.~G.}\ \bibnamefont
  {Wilson}},\ }\href {\doibase 10.1103/PhysRevD.10.2445} {\bibfield  {journal}
  {\bibinfo  {journal} {Physical Review D}\ }\textbf {\bibinfo {volume} {10}},\
  \bibinfo {pages} {2445} (\bibinfo {year} {1974})}\BibitemShut {NoStop}%
\bibitem [{\citenamefont {Kogut}\ and\ \citenamefont
  {Susskind}(1975{\natexlab{a}})}]{kogut_hamiltonian_1975}%
  \BibitemOpen
  \bibfield  {author} {\bibinfo {author} {\bibfnamefont {J.}~\bibnamefont
  {Kogut}}\ and\ \bibinfo {author} {\bibfnamefont {L.}~\bibnamefont
  {Susskind}},\ }\href {\doibase 10.1103/PhysRevD.11.395} {\bibfield  {journal}
  {\bibinfo  {journal} {Physical Review D}\ }\textbf {\bibinfo {volume} {11}},\
  \bibinfo {pages} {395} (\bibinfo {year} {1975}{\natexlab{a}})}\BibitemShut
  {NoStop}%
\bibitem [{\citenamefont {Affleck}\ and\ \citenamefont
  {Marston}(1988)}]{affleck_large-_1988}%
  \BibitemOpen
  \bibfield  {author} {\bibinfo {author} {\bibfnamefont {I.}~\bibnamefont
  {Affleck}}\ and\ \bibinfo {author} {\bibfnamefont {J.~B.}\ \bibnamefont
  {Marston}},\ }\href {\doibase 10.1103/PhysRevB.37.3774} {\bibfield  {journal}
  {\bibinfo  {journal} {Physical Review B}\ }\textbf {\bibinfo {volume} {37}},\
  \bibinfo {pages} {3774} (\bibinfo {year} {1988})}\BibitemShut {NoStop}%
\bibitem [{\citenamefont {Rantner}\ and\ \citenamefont
  {Wen}(2001)}]{rantner_electron_2001}%
  \BibitemOpen
  \bibfield  {author} {\bibinfo {author} {\bibfnamefont {W.}~\bibnamefont
  {Rantner}}\ and\ \bibinfo {author} {\bibfnamefont {X.-G.}\ \bibnamefont
  {Wen}},\ }\href {\doibase 10.1103/PhysRevLett.86.3871} {\bibfield  {journal}
  {\bibinfo  {journal} {Physical Review Letters}\ }\textbf {\bibinfo {volume}
  {86}},\ \bibinfo {pages} {3871} (\bibinfo {year} {2001})}\BibitemShut
  {NoStop}%
\bibitem [{\citenamefont {{FLAG Working Group}}\ \emph
  {et~al.}(2014)\citenamefont {{FLAG Working Group}}, \citenamefont {Aoki},
  \citenamefont {Aoki}, \citenamefont {Bernard}, \citenamefont {Blum},
  \citenamefont {Colangelo}, \citenamefont {Della~Morte}, \citenamefont
  {Dürr}, \citenamefont {El-Khadra}, \citenamefont {Fukaya}, \citenamefont
  {Horsley}, \citenamefont {Jüttner}, \citenamefont {Kaneko}, \citenamefont
  {Laiho}, \citenamefont {Lellouch}, \citenamefont {Leutwyler}, \citenamefont
  {Lubicz}, \citenamefont {Lunghi}, \citenamefont {Necco}, \citenamefont
  {Onogi}, \citenamefont {Pena}, \citenamefont {Sachrajda}, \citenamefont
  {Sharpe}, \citenamefont {Simula}, \citenamefont {Sommer}, \citenamefont
  {Van~de Water}, \citenamefont {Vladikas}, \citenamefont {Wenger},\ and\
  \citenamefont {Wittig}}]{flag_working_group_review_2014}%
  \BibitemOpen
  \bibfield  {author} {\bibinfo {author} {\bibnamefont {{FLAG Working Group}}},
  \bibinfo {author} {\bibfnamefont {S.}~\bibnamefont {Aoki}}, \bibinfo {author}
  {\bibfnamefont {Y.}~\bibnamefont {Aoki}}, \bibinfo {author} {\bibfnamefont
  {C.}~\bibnamefont {Bernard}}, \bibinfo {author} {\bibfnamefont
  {T.}~\bibnamefont {Blum}}, \bibinfo {author} {\bibfnamefont {G.}~\bibnamefont
  {Colangelo}}, \bibinfo {author} {\bibfnamefont {M.}~\bibnamefont
  {Della~Morte}}, \bibinfo {author} {\bibfnamefont {S.}~\bibnamefont {Dürr}},
  \bibinfo {author} {\bibfnamefont {A.~X.}\ \bibnamefont {El-Khadra}}, \bibinfo
  {author} {\bibfnamefont {H.}~\bibnamefont {Fukaya}}, \bibinfo {author}
  {\bibfnamefont {R.}~\bibnamefont {Horsley}}, \bibinfo {author} {\bibfnamefont
  {A.}~\bibnamefont {Jüttner}}, \bibinfo {author} {\bibfnamefont
  {T.}~\bibnamefont {Kaneko}}, \bibinfo {author} {\bibfnamefont
  {J.}~\bibnamefont {Laiho}}, \bibinfo {author} {\bibfnamefont
  {L.}~\bibnamefont {Lellouch}}, \bibinfo {author} {\bibfnamefont
  {H.}~\bibnamefont {Leutwyler}}, \bibinfo {author} {\bibfnamefont
  {V.}~\bibnamefont {Lubicz}}, \bibinfo {author} {\bibfnamefont
  {E.}~\bibnamefont {Lunghi}}, \bibinfo {author} {\bibfnamefont
  {S.}~\bibnamefont {Necco}}, \bibinfo {author} {\bibfnamefont
  {T.}~\bibnamefont {Onogi}}, \bibinfo {author} {\bibfnamefont
  {C.}~\bibnamefont {Pena}}, \bibinfo {author} {\bibfnamefont {C.~T.}\
  \bibnamefont {Sachrajda}}, \bibinfo {author} {\bibfnamefont {S.~R.}\
  \bibnamefont {Sharpe}}, \bibinfo {author} {\bibfnamefont {S.}~\bibnamefont
  {Simula}}, \bibinfo {author} {\bibfnamefont {R.}~\bibnamefont {Sommer}},
  \bibinfo {author} {\bibfnamefont {R.~S.}\ \bibnamefont {Van~de Water}},
  \bibinfo {author} {\bibfnamefont {A.}~\bibnamefont {Vladikas}}, \bibinfo
  {author} {\bibfnamefont {U.}~\bibnamefont {Wenger}}, \ and\ \bibinfo {author}
  {\bibfnamefont {H.}~\bibnamefont {Wittig}},\ }\href {\doibase
  10.1140/epjc/s10052-014-2890-7} {\bibfield  {journal} {\bibinfo  {journal}
  {The European Physical Journal C}\ }\textbf {\bibinfo {volume} {74}}
  (\bibinfo {year} {2014}),\ 10.1140/epjc/s10052-014-2890-7}\BibitemShut
  {NoStop}%
\bibitem [{\citenamefont {Troyer}\ and\ \citenamefont
  {Wiese}(2005)}]{troyer_computational_2005}%
  \BibitemOpen
  \bibfield  {author} {\bibinfo {author} {\bibfnamefont {M.}~\bibnamefont
  {Troyer}}\ and\ \bibinfo {author} {\bibfnamefont {U.-J.}\ \bibnamefont
  {Wiese}},\ }\href {\doibase 10.1103/PhysRevLett.94.170201} {\bibfield
  {journal} {\bibinfo  {journal} {Physical Review Letters}\ }\textbf {\bibinfo
  {volume} {94}} (\bibinfo {year} {2005}),\
  10.1103/PhysRevLett.94.170201}\BibitemShut {NoStop}%
\bibitem [{\citenamefont {Zohar}\ \emph {et~al.}(2016)\citenamefont {Zohar},
  \citenamefont {Cirac},\ and\ \citenamefont {Reznik}}]{zohar_quantum_2016}%
  \BibitemOpen
  \bibfield  {author} {\bibinfo {author} {\bibfnamefont {E.}~\bibnamefont
  {Zohar}}, \bibinfo {author} {\bibfnamefont {J.~I.}\ \bibnamefont {Cirac}}, \
  and\ \bibinfo {author} {\bibfnamefont {B.}~\bibnamefont {Reznik}},\ }\href
  {\doibase 10.1088/0034-4885/79/1/014401} {\bibfield  {journal} {\bibinfo
  {journal} {Reports on Progress in Physics}\ }\textbf {\bibinfo {volume}
  {79}},\ \bibinfo {pages} {014401} (\bibinfo {year} {2016})}\BibitemShut
  {NoStop}%
\bibitem [{\citenamefont {Bañuls}\ \emph {et~al.}(2020)\citenamefont
  {Bañuls}, \citenamefont {Blatt}, \citenamefont {Catani}, \citenamefont
  {Celi}, \citenamefont {Cirac}, \citenamefont {Dalmonte}, \citenamefont
  {Fallani}, \citenamefont {Jansen}, \citenamefont {Lewenstein}, \citenamefont
  {Montangero}, \citenamefont {Muschik}, \citenamefont {Reznik}, \citenamefont
  {Rico}, \citenamefont {Tagliacozzo}, \citenamefont {Van~Acoleyen},
  \citenamefont {Verstraete}, \citenamefont {Wiese}, \citenamefont {Wingate},
  \citenamefont {Zakrzewski},\ and\ \citenamefont
  {Zoller}}]{banuls_simulating_2020}%
  \BibitemOpen
  \bibfield  {author} {\bibinfo {author} {\bibfnamefont {M.~C.}\ \bibnamefont
  {Bañuls}}, \bibinfo {author} {\bibfnamefont {R.}~\bibnamefont {Blatt}},
  \bibinfo {author} {\bibfnamefont {J.}~\bibnamefont {Catani}}, \bibinfo
  {author} {\bibfnamefont {A.}~\bibnamefont {Celi}}, \bibinfo {author}
  {\bibfnamefont {J.~I.}\ \bibnamefont {Cirac}}, \bibinfo {author}
  {\bibfnamefont {M.}~\bibnamefont {Dalmonte}}, \bibinfo {author}
  {\bibfnamefont {L.}~\bibnamefont {Fallani}}, \bibinfo {author} {\bibfnamefont
  {K.}~\bibnamefont {Jansen}}, \bibinfo {author} {\bibfnamefont
  {M.}~\bibnamefont {Lewenstein}}, \bibinfo {author} {\bibfnamefont
  {S.}~\bibnamefont {Montangero}}, \bibinfo {author} {\bibfnamefont {C.~A.}\
  \bibnamefont {Muschik}}, \bibinfo {author} {\bibfnamefont {B.}~\bibnamefont
  {Reznik}}, \bibinfo {author} {\bibfnamefont {E.}~\bibnamefont {Rico}},
  \bibinfo {author} {\bibfnamefont {L.}~\bibnamefont {Tagliacozzo}}, \bibinfo
  {author} {\bibfnamefont {K.}~\bibnamefont {Van~Acoleyen}}, \bibinfo {author}
  {\bibfnamefont {F.}~\bibnamefont {Verstraete}}, \bibinfo {author}
  {\bibfnamefont {U.-J.}\ \bibnamefont {Wiese}}, \bibinfo {author}
  {\bibfnamefont {M.}~\bibnamefont {Wingate}}, \bibinfo {author} {\bibfnamefont
  {J.}~\bibnamefont {Zakrzewski}}, \ and\ \bibinfo {author} {\bibfnamefont
  {P.}~\bibnamefont {Zoller}},\ }\href {\doibase 10.1140/epjd/e2020-100571-8}
  {\bibfield  {journal} {\bibinfo  {journal} {The European Physical Journal D}\
  }\textbf {\bibinfo {volume} {74}},\ \bibinfo {pages} {165} (\bibinfo {year}
  {2020})}\BibitemShut {NoStop}%
\bibitem [{\citenamefont {Martinez}\ \emph {et~al.}(2016)\citenamefont
  {Martinez}, \citenamefont {Muschik}, \citenamefont {Schindler}, \citenamefont
  {Nigg}, \citenamefont {Erhard}, \citenamefont {Heyl}, \citenamefont {Hauke},
  \citenamefont {Dalmonte}, \citenamefont {Monz}, \citenamefont {Zoller},\ and\
  \citenamefont {Blatt}}]{martinez_real-time_2016}%
  \BibitemOpen
  \bibfield  {author} {\bibinfo {author} {\bibfnamefont {E.~A.}\ \bibnamefont
  {Martinez}}, \bibinfo {author} {\bibfnamefont {C.~A.}\ \bibnamefont
  {Muschik}}, \bibinfo {author} {\bibfnamefont {P.}~\bibnamefont {Schindler}},
  \bibinfo {author} {\bibfnamefont {D.}~\bibnamefont {Nigg}}, \bibinfo {author}
  {\bibfnamefont {A.}~\bibnamefont {Erhard}}, \bibinfo {author} {\bibfnamefont
  {M.}~\bibnamefont {Heyl}}, \bibinfo {author} {\bibfnamefont {P.}~\bibnamefont
  {Hauke}}, \bibinfo {author} {\bibfnamefont {M.}~\bibnamefont {Dalmonte}},
  \bibinfo {author} {\bibfnamefont {T.}~\bibnamefont {Monz}}, \bibinfo {author}
  {\bibfnamefont {P.}~\bibnamefont {Zoller}}, \ and\ \bibinfo {author}
  {\bibfnamefont {R.}~\bibnamefont {Blatt}},\ }\href {\doibase
  10.1038/nature18318} {\bibfield  {journal} {\bibinfo  {journal} {Nature}\
  }\textbf {\bibinfo {volume} {534}},\ \bibinfo {pages} {516} (\bibinfo {year}
  {2016})}\BibitemShut {NoStop}%
\bibitem [{\citenamefont {Görg}\ \emph {et~al.}(2019)\citenamefont {Görg},
  \citenamefont {Sandholzer}, \citenamefont {Minguzzi}, \citenamefont
  {Desbuquois}, \citenamefont {Messer},\ and\ \citenamefont
  {Esslinger}}]{gorg_realization_2019}%
  \BibitemOpen
  \bibfield  {author} {\bibinfo {author} {\bibfnamefont {F.}~\bibnamefont
  {Görg}}, \bibinfo {author} {\bibfnamefont {K.}~\bibnamefont {Sandholzer}},
  \bibinfo {author} {\bibfnamefont {J.}~\bibnamefont {Minguzzi}}, \bibinfo
  {author} {\bibfnamefont {R.}~\bibnamefont {Desbuquois}}, \bibinfo {author}
  {\bibfnamefont {M.}~\bibnamefont {Messer}}, \ and\ \bibinfo {author}
  {\bibfnamefont {T.}~\bibnamefont {Esslinger}},\ }\href {\doibase
  10.1038/s41567-019-0615-4} {\bibfield  {journal} {\bibinfo  {journal} {Nature
  Physics}\ }\textbf {\bibinfo {volume} {15}},\ \bibinfo {pages} {1161}
  (\bibinfo {year} {2019})}\BibitemShut {NoStop}%
\bibitem [{\citenamefont {Schweizer}\ \emph {et~al.}(2019)\citenamefont
  {Schweizer}, \citenamefont {Grusdt}, \citenamefont {Berngruber},
  \citenamefont {Barbiero}, \citenamefont {Demler}, \citenamefont {Goldman},
  \citenamefont {Bloch},\ and\ \citenamefont
  {Aidelsburger}}]{schweizer_floquet_2019}%
  \BibitemOpen
  \bibfield  {author} {\bibinfo {author} {\bibfnamefont {C.}~\bibnamefont
  {Schweizer}}, \bibinfo {author} {\bibfnamefont {F.}~\bibnamefont {Grusdt}},
  \bibinfo {author} {\bibfnamefont {M.}~\bibnamefont {Berngruber}}, \bibinfo
  {author} {\bibfnamefont {L.}~\bibnamefont {Barbiero}}, \bibinfo {author}
  {\bibfnamefont {E.}~\bibnamefont {Demler}}, \bibinfo {author} {\bibfnamefont
  {N.}~\bibnamefont {Goldman}}, \bibinfo {author} {\bibfnamefont
  {I.}~\bibnamefont {Bloch}}, \ and\ \bibinfo {author} {\bibfnamefont
  {M.}~\bibnamefont {Aidelsburger}},\ }\href {\doibase
  10.1038/s41567-019-0649-7} {\bibfield  {journal} {\bibinfo  {journal} {Nature
  Physics}\ }\textbf {\bibinfo {volume} {15}},\ \bibinfo {pages} {1168}
  (\bibinfo {year} {2019})}\BibitemShut {NoStop}%
\bibitem [{\citenamefont {Yang}\ \emph {et~al.}(2020)\citenamefont {Yang},
  \citenamefont {Sun}, \citenamefont {Ott}, \citenamefont {Wang}, \citenamefont
  {Zache}, \citenamefont {Halimeh}, \citenamefont {Yuan}, \citenamefont
  {Hauke},\ and\ \citenamefont {Pan}}]{yang_observation_2020}%
  \BibitemOpen
  \bibfield  {author} {\bibinfo {author} {\bibfnamefont {B.}~\bibnamefont
  {Yang}}, \bibinfo {author} {\bibfnamefont {H.}~\bibnamefont {Sun}}, \bibinfo
  {author} {\bibfnamefont {R.}~\bibnamefont {Ott}}, \bibinfo {author}
  {\bibfnamefont {H.-Y.}\ \bibnamefont {Wang}}, \bibinfo {author}
  {\bibfnamefont {T.~V.}\ \bibnamefont {Zache}}, \bibinfo {author}
  {\bibfnamefont {J.~C.}\ \bibnamefont {Halimeh}}, \bibinfo {author}
  {\bibfnamefont {Z.-S.}\ \bibnamefont {Yuan}}, \bibinfo {author}
  {\bibfnamefont {P.}~\bibnamefont {Hauke}}, \ and\ \bibinfo {author}
  {\bibfnamefont {J.-W.}\ \bibnamefont {Pan}},\ }\href {\doibase
  10.1038/s41586-020-2910-8} {\bibfield  {journal} {\bibinfo  {journal}
  {Nature}\ }\textbf {\bibinfo {volume} {587}},\ \bibinfo {pages} {392}
  (\bibinfo {year} {2020})}\BibitemShut {NoStop}%
\bibitem [{\citenamefont {Mil}\ \emph {et~al.}(2020)\citenamefont {Mil},
  \citenamefont {Zache}, \citenamefont {Hegde}, \citenamefont {Xia},
  \citenamefont {Bhatt}, \citenamefont {Oberthaler}, \citenamefont {Hauke},
  \citenamefont {Berges},\ and\ \citenamefont
  {Jendrzejewski}}]{mil_scalable_2020}%
  \BibitemOpen
  \bibfield  {author} {\bibinfo {author} {\bibfnamefont {A.}~\bibnamefont
  {Mil}}, \bibinfo {author} {\bibfnamefont {T.~V.}\ \bibnamefont {Zache}},
  \bibinfo {author} {\bibfnamefont {A.}~\bibnamefont {Hegde}}, \bibinfo
  {author} {\bibfnamefont {A.}~\bibnamefont {Xia}}, \bibinfo {author}
  {\bibfnamefont {R.~P.}\ \bibnamefont {Bhatt}}, \bibinfo {author}
  {\bibfnamefont {M.~K.}\ \bibnamefont {Oberthaler}}, \bibinfo {author}
  {\bibfnamefont {P.}~\bibnamefont {Hauke}}, \bibinfo {author} {\bibfnamefont
  {J.}~\bibnamefont {Berges}}, \ and\ \bibinfo {author} {\bibfnamefont
  {F.}~\bibnamefont {Jendrzejewski}},\ }\href {\doibase
  10.1126/science.aaz5312} {\bibfield  {journal} {\bibinfo  {journal}
  {Science}\ }\textbf {\bibinfo {volume} {367}},\ \bibinfo {pages} {1128}
  (\bibinfo {year} {2020})}\BibitemShut {NoStop}%
\bibitem [{\citenamefont {Tagliacozzo}\ \emph {et~al.}(2013)\citenamefont
  {Tagliacozzo}, \citenamefont {Celi}, \citenamefont {Orland}, \citenamefont
  {Mitchell},\ and\ \citenamefont {Lewenstein}}]{tagliacozzo_simulation_2013}%
  \BibitemOpen
  \bibfield  {author} {\bibinfo {author} {\bibfnamefont {L.}~\bibnamefont
  {Tagliacozzo}}, \bibinfo {author} {\bibfnamefont {A.}~\bibnamefont {Celi}},
  \bibinfo {author} {\bibfnamefont {P.}~\bibnamefont {Orland}}, \bibinfo
  {author} {\bibfnamefont {M.~W.}\ \bibnamefont {Mitchell}}, \ and\ \bibinfo
  {author} {\bibfnamefont {M.}~\bibnamefont {Lewenstein}},\ }\href {\doibase
  10.1038/ncomms3615} {\bibfield  {journal} {\bibinfo  {journal} {Nature
  Communications}\ }\textbf {\bibinfo {volume} {4}},\ \bibinfo {pages} {2615}
  (\bibinfo {year} {2013})}\BibitemShut {NoStop}%
\bibitem [{\citenamefont {Zohar}\ \emph {et~al.}(2017)\citenamefont {Zohar},
  \citenamefont {Farace}, \citenamefont {Reznik},\ and\ \citenamefont
  {Cirac}}]{zohar_digital_2017}%
  \BibitemOpen
  \bibfield  {author} {\bibinfo {author} {\bibfnamefont {E.}~\bibnamefont
  {Zohar}}, \bibinfo {author} {\bibfnamefont {A.}~\bibnamefont {Farace}},
  \bibinfo {author} {\bibfnamefont {B.}~\bibnamefont {Reznik}}, \ and\ \bibinfo
  {author} {\bibfnamefont {J.~I.}\ \bibnamefont {Cirac}},\ }\href {\doibase
  10.1103/PhysRevA.95.023604} {\bibfield  {journal} {\bibinfo  {journal}
  {Physical Review A}\ }\textbf {\bibinfo {volume} {95}} (\bibinfo {year}
  {2017}),\ 10.1103/PhysRevA.95.023604}\BibitemShut {NoStop}%
\bibitem [{\citenamefont {Bender}\ \emph {et~al.}(2018)\citenamefont {Bender},
  \citenamefont {Zohar}, \citenamefont {Farace},\ and\ \citenamefont
  {Cirac}}]{bender_digital_2018}%
  \BibitemOpen
  \bibfield  {author} {\bibinfo {author} {\bibfnamefont {J.}~\bibnamefont
  {Bender}}, \bibinfo {author} {\bibfnamefont {E.}~\bibnamefont {Zohar}},
  \bibinfo {author} {\bibfnamefont {A.}~\bibnamefont {Farace}}, \ and\ \bibinfo
  {author} {\bibfnamefont {J.~I.}\ \bibnamefont {Cirac}},\ }\href {\doibase
  10.1088/1367-2630/aadb71} {\bibfield  {journal} {\bibinfo  {journal} {New
  Journal of Physics}\ }\textbf {\bibinfo {volume} {20}},\ \bibinfo {pages}
  {093001} (\bibinfo {year} {2018})}\BibitemShut {NoStop}%
\bibitem [{\citenamefont {Gustafson}(2021)}]{gustafson_prospects_2021}%
  \BibitemOpen
  \bibfield  {author} {\bibinfo {author} {\bibfnamefont {E.~J.}\ \bibnamefont
  {Gustafson}},\ }\href {\doibase 10.1103/PhysRevD.103.114505} {\bibfield
  {journal} {\bibinfo  {journal} {Physical Review D}\ }\textbf {\bibinfo
  {volume} {103}},\ \bibinfo {pages} {114505} (\bibinfo {year}
  {2021})}\BibitemShut {NoStop}%
\bibitem [{\citenamefont {Bhattacharya}\ \emph {et~al.}(2021)\citenamefont
  {Bhattacharya}, \citenamefont {Buser}, \citenamefont {Chandrasekharan},
  \citenamefont {Gupta},\ and\ \citenamefont
  {Singh}}]{bhattacharya_qubit_2021}%
  \BibitemOpen
  \bibfield  {author} {\bibinfo {author} {\bibfnamefont {T.}~\bibnamefont
  {Bhattacharya}}, \bibinfo {author} {\bibfnamefont {A.~J.}\ \bibnamefont
  {Buser}}, \bibinfo {author} {\bibfnamefont {S.}~\bibnamefont
  {Chandrasekharan}}, \bibinfo {author} {\bibfnamefont {R.}~\bibnamefont
  {Gupta}}, \ and\ \bibinfo {author} {\bibfnamefont {H.}~\bibnamefont
  {Singh}},\ }\href {\doibase 10.1103/PhysRevLett.126.172001} {\bibfield
  {journal} {\bibinfo  {journal} {Physical Review Letters}\ }\textbf {\bibinfo
  {volume} {126}},\ \bibinfo {pages} {172001} (\bibinfo {year}
  {2021})}\BibitemShut {NoStop}%
\bibitem [{\citenamefont {Ciavarella}\ \emph {et~al.}(2021)\citenamefont
  {Ciavarella}, \citenamefont {Klco},\ and\ \citenamefont
  {Savage}}]{ciavarella_trailhead_2021}%
  \BibitemOpen
  \bibfield  {author} {\bibinfo {author} {\bibfnamefont {A.}~\bibnamefont
  {Ciavarella}}, \bibinfo {author} {\bibfnamefont {N.}~\bibnamefont {Klco}}, \
  and\ \bibinfo {author} {\bibfnamefont {M.~J.}\ \bibnamefont {Savage}},\
  }\href {\doibase 10.1103/PhysRevD.103.094501} {\bibfield  {journal} {\bibinfo
   {journal} {Physical Review D}\ }\textbf {\bibinfo {volume} {103}},\ \bibinfo
  {pages} {094501} (\bibinfo {year} {2021})}\BibitemShut {NoStop}%
\bibitem [{\citenamefont {Zohar}\ \emph {et~al.}(2013)\citenamefont {Zohar},
  \citenamefont {Cirac},\ and\ \citenamefont {Reznik}}]{zohar_simulating_2013}%
  \BibitemOpen
  \bibfield  {author} {\bibinfo {author} {\bibfnamefont {E.}~\bibnamefont
  {Zohar}}, \bibinfo {author} {\bibfnamefont {J.~I.}\ \bibnamefont {Cirac}}, \
  and\ \bibinfo {author} {\bibfnamefont {B.}~\bibnamefont {Reznik}},\ }\href
  {\doibase 10.1103/PhysRevLett.110.055302} {\bibfield  {journal} {\bibinfo
  {journal} {Physical Review Letters}\ }\textbf {\bibinfo {volume} {110}}
  (\bibinfo {year} {2013}),\ 10.1103/PhysRevLett.110.055302}\BibitemShut
  {NoStop}%
\bibitem [{\citenamefont {Ott}\ \emph {et~al.}(2021)\citenamefont {Ott},
  \citenamefont {Zache}, \citenamefont {Jendrzejewski},\ and\ \citenamefont
  {Berges}}]{ott_scalable_2021}%
  \BibitemOpen
  \bibfield  {author} {\bibinfo {author} {\bibfnamefont {R.}~\bibnamefont
  {Ott}}, \bibinfo {author} {\bibfnamefont {T.~V.}\ \bibnamefont {Zache}},
  \bibinfo {author} {\bibfnamefont {F.}~\bibnamefont {Jendrzejewski}}, \ and\
  \bibinfo {author} {\bibfnamefont {J.}~\bibnamefont {Berges}},\ }\href
  {\doibase 10.1103/PhysRevLett.127.130504} {\bibfield  {journal} {\bibinfo
  {journal} {Physical Review Letters}\ }\textbf {\bibinfo {volume} {127}},\
  \bibinfo {pages} {130504} (\bibinfo {year} {2021})}\BibitemShut {NoStop}%
\bibitem [{\citenamefont {Raychowdhury}\ and\ \citenamefont
  {Stryker}(2020)}]{raychowdhury_loop_2020}%
  \BibitemOpen
  \bibfield  {author} {\bibinfo {author} {\bibfnamefont {I.}~\bibnamefont
  {Raychowdhury}}\ and\ \bibinfo {author} {\bibfnamefont {J.~R.}\ \bibnamefont
  {Stryker}},\ }\href {\doibase 10.1103/PhysRevD.101.114502} {\bibfield
  {journal} {\bibinfo  {journal} {Physical Review D}\ }\textbf {\bibinfo
  {volume} {101}},\ \bibinfo {pages} {114502} (\bibinfo {year}
  {2020})}\BibitemShut {NoStop}%
\bibitem [{\citenamefont {Bender}\ and\ \citenamefont
  {Zohar}(2020)}]{bender_gauge_2020}%
  \BibitemOpen
  \bibfield  {author} {\bibinfo {author} {\bibfnamefont {J.}~\bibnamefont
  {Bender}}\ and\ \bibinfo {author} {\bibfnamefont {E.}~\bibnamefont {Zohar}},\
  }\href {\doibase 10.1103/PhysRevD.102.114517} {\bibfield  {journal} {\bibinfo
   {journal} {Physical Review D}\ }\textbf {\bibinfo {volume} {102}},\ \bibinfo
  {pages} {114517} (\bibinfo {year} {2020})}\BibitemShut {NoStop}%
\bibitem [{\citenamefont {Haase}\ \emph {et~al.}(2021)\citenamefont {Haase},
  \citenamefont {Dellantonio}, \citenamefont {Celi}, \citenamefont {Paulson},
  \citenamefont {Kan}, \citenamefont {Jansen},\ and\ \citenamefont
  {Muschik}}]{haase_resource_2021}%
  \BibitemOpen
  \bibfield  {author} {\bibinfo {author} {\bibfnamefont {J.~F.}\ \bibnamefont
  {Haase}}, \bibinfo {author} {\bibfnamefont {L.}~\bibnamefont {Dellantonio}},
  \bibinfo {author} {\bibfnamefont {A.}~\bibnamefont {Celi}}, \bibinfo {author}
  {\bibfnamefont {D.}~\bibnamefont {Paulson}}, \bibinfo {author} {\bibfnamefont
  {A.}~\bibnamefont {Kan}}, \bibinfo {author} {\bibfnamefont {K.}~\bibnamefont
  {Jansen}}, \ and\ \bibinfo {author} {\bibfnamefont {C.~A.}\ \bibnamefont
  {Muschik}},\ }\href {\doibase 10.22331/q-2021-02-04-393} {\bibfield
  {journal} {\bibinfo  {journal} {{Quantum}}\ }\textbf {\bibinfo {volume}
  {5}},\ \bibinfo {pages} {393} (\bibinfo {year} {2021})}\BibitemShut {NoStop}%
\bibitem [{\citenamefont {Bañuls}\ \emph {et~al.}(2013)\citenamefont
  {Bañuls}, \citenamefont {Cichy}, \citenamefont {Cirac},\ and\ \citenamefont
  {Jansen}}]{banuls_mass_2013}%
  \BibitemOpen
  \bibfield  {author} {\bibinfo {author} {\bibfnamefont {M.}~\bibnamefont
  {Bañuls}}, \bibinfo {author} {\bibfnamefont {K.}~\bibnamefont {Cichy}},
  \bibinfo {author} {\bibfnamefont {J.}~\bibnamefont {Cirac}}, \ and\ \bibinfo
  {author} {\bibfnamefont {K.}~\bibnamefont {Jansen}},\ }\href {\doibase
  10.1007/JHEP11(2013)158} {\bibfield  {journal} {\bibinfo  {journal} {Journal
  of High Energy Physics}\ }\textbf {\bibinfo {volume} {2013}} (\bibinfo {year}
  {2013}),\ 10.1007/JHEP11(2013)158}\BibitemShut {NoStop}%
\bibitem [{\citenamefont {Buyens}\ \emph {et~al.}(2014)\citenamefont {Buyens},
  \citenamefont {Haegeman}, \citenamefont {Van~Acoleyen}, \citenamefont
  {Verschelde},\ and\ \citenamefont {Verstraete}}]{buyens_matrix_2014}%
  \BibitemOpen
  \bibfield  {author} {\bibinfo {author} {\bibfnamefont {B.}~\bibnamefont
  {Buyens}}, \bibinfo {author} {\bibfnamefont {J.}~\bibnamefont {Haegeman}},
  \bibinfo {author} {\bibfnamefont {K.}~\bibnamefont {Van~Acoleyen}}, \bibinfo
  {author} {\bibfnamefont {H.}~\bibnamefont {Verschelde}}, \ and\ \bibinfo
  {author} {\bibfnamefont {F.}~\bibnamefont {Verstraete}},\ }\href {\doibase
  10.1103/PhysRevLett.113.091601} {\bibfield  {journal} {\bibinfo  {journal}
  {Physical Review Letters}\ }\textbf {\bibinfo {volume} {113}} (\bibinfo
  {year} {2014}),\ 10.1103/PhysRevLett.113.091601}\BibitemShut {NoStop}%
\bibitem [{\citenamefont {Rico}\ \emph {et~al.}(2014)\citenamefont {Rico},
  \citenamefont {Pichler}, \citenamefont {Dalmonte}, \citenamefont {Zoller},\
  and\ \citenamefont {Montangero}}]{rico_tensor_2014}%
  \BibitemOpen
  \bibfield  {author} {\bibinfo {author} {\bibfnamefont {E.}~\bibnamefont
  {Rico}}, \bibinfo {author} {\bibfnamefont {T.}~\bibnamefont {Pichler}},
  \bibinfo {author} {\bibfnamefont {M.}~\bibnamefont {Dalmonte}}, \bibinfo
  {author} {\bibfnamefont {P.}~\bibnamefont {Zoller}}, \ and\ \bibinfo {author}
  {\bibfnamefont {S.}~\bibnamefont {Montangero}},\ }\href {\doibase
  10.1103/PhysRevLett.112.201601} {\bibfield  {journal} {\bibinfo  {journal}
  {Physical Review Letters}\ }\textbf {\bibinfo {volume} {112}} (\bibinfo
  {year} {2014}),\ 10.1103/PhysRevLett.112.201601}\BibitemShut {NoStop}%
\bibitem [{\citenamefont {Kühn}\ \emph {et~al.}(2015)\citenamefont {Kühn},
  \citenamefont {Zohar}, \citenamefont {Cirac},\ and\ \citenamefont
  {Bañuls}}]{kuhn_non-abelian_2015}%
  \BibitemOpen
  \bibfield  {author} {\bibinfo {author} {\bibfnamefont {S.}~\bibnamefont
  {Kühn}}, \bibinfo {author} {\bibfnamefont {E.}~\bibnamefont {Zohar}},
  \bibinfo {author} {\bibfnamefont {J.~I.}\ \bibnamefont {Cirac}}, \ and\
  \bibinfo {author} {\bibfnamefont {M.~C.}\ \bibnamefont {Bañuls}},\ }\href
  {\doibase 10.1007/JHEP07(2015)130} {\bibfield  {journal} {\bibinfo  {journal}
  {Journal of High Energy Physics}\ }\textbf {\bibinfo {volume} {2015}},\
  \bibinfo {pages} {1} (\bibinfo {year} {2015})}\BibitemShut {NoStop}%
\bibitem [{\citenamefont {Pichler}\ \emph {et~al.}(2016)\citenamefont
  {Pichler}, \citenamefont {Dalmonte}, \citenamefont {Rico}, \citenamefont
  {Zoller},\ and\ \citenamefont {Montangero}}]{pichler_real-time_2016}%
  \BibitemOpen
  \bibfield  {author} {\bibinfo {author} {\bibfnamefont {T.}~\bibnamefont
  {Pichler}}, \bibinfo {author} {\bibfnamefont {M.}~\bibnamefont {Dalmonte}},
  \bibinfo {author} {\bibfnamefont {E.}~\bibnamefont {Rico}}, \bibinfo {author}
  {\bibfnamefont {P.}~\bibnamefont {Zoller}}, \ and\ \bibinfo {author}
  {\bibfnamefont {S.}~\bibnamefont {Montangero}},\ }\href {\doibase
  10.1103/PhysRevX.6.011023} {\bibfield  {journal} {\bibinfo  {journal} {Phys.
  Rev. X}\ }\textbf {\bibinfo {volume} {6}},\ \bibinfo {pages} {011023}
  (\bibinfo {year} {2016})}\BibitemShut {NoStop}%
\bibitem [{\citenamefont {Bañuls}\ \emph
  {et~al.}(2017{\natexlab{a}})\citenamefont {Bañuls}, \citenamefont {Cichy},
  \citenamefont {Cirac}, \citenamefont {Jansen},\ and\ \citenamefont
  {Kühn}}]{banuls_efficient_2017}%
  \BibitemOpen
  \bibfield  {author} {\bibinfo {author} {\bibfnamefont {M.~C.}\ \bibnamefont
  {Bañuls}}, \bibinfo {author} {\bibfnamefont {K.}~\bibnamefont {Cichy}},
  \bibinfo {author} {\bibfnamefont {J.~I.}\ \bibnamefont {Cirac}}, \bibinfo
  {author} {\bibfnamefont {K.}~\bibnamefont {Jansen}}, \ and\ \bibinfo {author}
  {\bibfnamefont {S.}~\bibnamefont {Kühn}},\ }\href {\doibase
  10.1103/PhysRevX.7.041046} {\bibfield  {journal} {\bibinfo  {journal}
  {Physical Review X}\ }\textbf {\bibinfo {volume} {7}} (\bibinfo {year}
  {2017}{\natexlab{a}}),\ 10.1103/PhysRevX.7.041046}\BibitemShut {NoStop}%
\bibitem [{\citenamefont {Bañuls}\ \emph
  {et~al.}(2017{\natexlab{b}})\citenamefont {Bañuls}, \citenamefont {Cichy},
  \citenamefont {Cirac}, \citenamefont {Jansen},\ and\ \citenamefont
  {Kühn}}]{banuls_density_2017}%
  \BibitemOpen
  \bibfield  {author} {\bibinfo {author} {\bibfnamefont {M.~C.}\ \bibnamefont
  {Bañuls}}, \bibinfo {author} {\bibfnamefont {K.}~\bibnamefont {Cichy}},
  \bibinfo {author} {\bibfnamefont {J.~I.}\ \bibnamefont {Cirac}}, \bibinfo
  {author} {\bibfnamefont {K.}~\bibnamefont {Jansen}}, \ and\ \bibinfo {author}
  {\bibfnamefont {S.}~\bibnamefont {Kühn}},\ }\href {\doibase
  10.1103/PhysRevLett.118.071601} {\bibfield  {journal} {\bibinfo  {journal}
  {Phys. Rev. Lett.}\ }\textbf {\bibinfo {volume} {118}},\ \bibinfo {pages}
  {071601} (\bibinfo {year} {2017}{\natexlab{b}})}\BibitemShut {NoStop}%
\bibitem [{\citenamefont {Silvi}\ \emph {et~al.}(2017)\citenamefont {Silvi},
  \citenamefont {Rico}, \citenamefont {Dalmonte}, \citenamefont {Tschirsich},\
  and\ \citenamefont {Montangero}}]{silvi_finite-density_2017}%
  \BibitemOpen
  \bibfield  {author} {\bibinfo {author} {\bibfnamefont {P.}~\bibnamefont
  {Silvi}}, \bibinfo {author} {\bibfnamefont {E.}~\bibnamefont {Rico}},
  \bibinfo {author} {\bibfnamefont {M.}~\bibnamefont {Dalmonte}}, \bibinfo
  {author} {\bibfnamefont {F.}~\bibnamefont {Tschirsich}}, \ and\ \bibinfo
  {author} {\bibfnamefont {S.}~\bibnamefont {Montangero}},\ }\href {\doibase
  10.22331/q-2017-04-25-9} {\bibfield  {journal} {\bibinfo  {journal}
  {Quantum}\ }\textbf {\bibinfo {volume} {1}},\ \bibinfo {pages} {9} (\bibinfo
  {year} {2017})}\BibitemShut {NoStop}%
\bibitem [{\citenamefont {Buyens}\ \emph {et~al.}(2017)\citenamefont {Buyens},
  \citenamefont {Haegeman}, \citenamefont {Hebenstreit}, \citenamefont
  {Verstraete},\ and\ \citenamefont {Van~Acoleyen}}]{buyens_real-time_2017}%
  \BibitemOpen
  \bibfield  {author} {\bibinfo {author} {\bibfnamefont {B.}~\bibnamefont
  {Buyens}}, \bibinfo {author} {\bibfnamefont {J.}~\bibnamefont {Haegeman}},
  \bibinfo {author} {\bibfnamefont {F.}~\bibnamefont {Hebenstreit}}, \bibinfo
  {author} {\bibfnamefont {F.}~\bibnamefont {Verstraete}}, \ and\ \bibinfo
  {author} {\bibfnamefont {K.}~\bibnamefont {Van~Acoleyen}},\ }\href {\doibase
  10.1103/PhysRevD.96.114501} {\bibfield  {journal} {\bibinfo  {journal}
  {Physical Review D}\ }\textbf {\bibinfo {volume} {96}} (\bibinfo {year}
  {2017}),\ 10.1103/PhysRevD.96.114501}\BibitemShut {NoStop}%
\bibitem [{\citenamefont {Silvi}\ \emph {et~al.}(2019)\citenamefont {Silvi},
  \citenamefont {Sauer}, \citenamefont {Tschirsich},\ and\ \citenamefont
  {Montangero}}]{silvi_tensor_2019}%
  \BibitemOpen
  \bibfield  {author} {\bibinfo {author} {\bibfnamefont {P.}~\bibnamefont
  {Silvi}}, \bibinfo {author} {\bibfnamefont {Y.}~\bibnamefont {Sauer}},
  \bibinfo {author} {\bibfnamefont {F.}~\bibnamefont {Tschirsich}}, \ and\
  \bibinfo {author} {\bibfnamefont {S.}~\bibnamefont {Montangero}},\ }\href
  {\doibase 10.1103/PhysRevD.100.074512} {\bibfield  {journal} {\bibinfo
  {journal} {Physical Review D}\ }\textbf {\bibinfo {volume} {100}},\ \bibinfo
  {pages} {074512} (\bibinfo {year} {2019})}\BibitemShut {NoStop}%
\bibitem [{\citenamefont {Bruckmann}\ \emph {et~al.}(2019)\citenamefont
  {Bruckmann}, \citenamefont {Jansen},\ and\ \citenamefont
  {Kühn}}]{bruckmann_o3_2019}%
  \BibitemOpen
  \bibfield  {author} {\bibinfo {author} {\bibfnamefont {F.}~\bibnamefont
  {Bruckmann}}, \bibinfo {author} {\bibfnamefont {K.}~\bibnamefont {Jansen}}, \
  and\ \bibinfo {author} {\bibfnamefont {S.}~\bibnamefont {Kühn}},\ }\href
  {\doibase 10.1103/PhysRevD.99.074501} {\bibfield  {journal} {\bibinfo
  {journal} {Physical Review D}\ }\textbf {\bibinfo {volume} {99}},\ \bibinfo
  {pages} {074501} (\bibinfo {year} {2019})}\BibitemShut {NoStop}%
\bibitem [{\citenamefont {Chandrasekharan}\ and\ \citenamefont
  {Wiese}(1997)}]{chandrasekharan_quantum_1997}%
  \BibitemOpen
  \bibfield  {author} {\bibinfo {author} {\bibfnamefont {S.}~\bibnamefont
  {Chandrasekharan}}\ and\ \bibinfo {author} {\bibfnamefont {U.-J.}\
  \bibnamefont {Wiese}},\ }\href {\doibase 10.1016/S0550-3213(97)80041-7}
  {\bibfield  {journal} {\bibinfo  {journal} {Nuclear Physics B}\ }\textbf
  {\bibinfo {volume} {492}},\ \bibinfo {pages} {455} (\bibinfo {year}
  {1997})}\BibitemShut {NoStop}%
\bibitem [{\citenamefont {Tschirsich}\ \emph {et~al.}(2019)\citenamefont
  {Tschirsich}, \citenamefont {Montangero},\ and\ \citenamefont
  {Dalmonte}}]{tschirsich_phase_2019}%
  \BibitemOpen
  \bibfield  {author} {\bibinfo {author} {\bibfnamefont {F.}~\bibnamefont
  {Tschirsich}}, \bibinfo {author} {\bibfnamefont {S.}~\bibnamefont
  {Montangero}}, \ and\ \bibinfo {author} {\bibfnamefont {M.}~\bibnamefont
  {Dalmonte}},\ }\href {\doibase 10.21468/SciPostPhys.6.3.028} {\bibfield
  {journal} {\bibinfo  {journal} {SciPost Physics}\ }\textbf {\bibinfo {volume}
  {6}} (\bibinfo {year} {2019}),\ 10.21468/SciPostPhys.6.3.028}\BibitemShut
  {NoStop}%
\bibitem [{\citenamefont {Magnifico}\ \emph {et~al.}(2021)\citenamefont
  {Magnifico}, \citenamefont {Felser}, \citenamefont {Silvi},\ and\
  \citenamefont {Montangero}}]{magnifico_lattice_2021}%
  \BibitemOpen
  \bibfield  {author} {\bibinfo {author} {\bibfnamefont {G.}~\bibnamefont
  {Magnifico}}, \bibinfo {author} {\bibfnamefont {T.}~\bibnamefont {Felser}},
  \bibinfo {author} {\bibfnamefont {P.}~\bibnamefont {Silvi}}, \ and\ \bibinfo
  {author} {\bibfnamefont {S.}~\bibnamefont {Montangero}},\ }\href {\doibase
  10.1038/s41467-021-23646-3} {\bibfield  {journal} {\bibinfo  {journal}
  {Nature Communications}\ }\textbf {\bibinfo {volume} {12}},\ \bibinfo {pages}
  {3600} (\bibinfo {year} {2021})}\BibitemShut {NoStop}%
\bibitem [{\citenamefont {Montangero}\ \emph {et~al.}(2022)\citenamefont
  {Montangero}, \citenamefont {Rico},\ and\ \citenamefont
  {Silvi}}]{montangero2022loop}%
  \BibitemOpen
  \bibfield  {author} {\bibinfo {author} {\bibfnamefont {S.}~\bibnamefont
  {Montangero}}, \bibinfo {author} {\bibfnamefont {E.}~\bibnamefont {Rico}}, \
  and\ \bibinfo {author} {\bibfnamefont {P.}~\bibnamefont {Silvi}},\
  }\href@noop {} {\bibfield  {journal} {\bibinfo  {journal} {Philosophical
  Transactions of the Royal Society A}\ }\textbf {\bibinfo {volume} {380}},\
  \bibinfo {pages} {20210065} (\bibinfo {year} {2022})}\BibitemShut {NoStop}%
\bibitem [{\citenamefont {Tagliacozzo}\ \emph {et~al.}(2014)\citenamefont
  {Tagliacozzo}, \citenamefont {Celi},\ and\ \citenamefont
  {Lewenstein}}]{tagliacozzo_tensor_2014}%
  \BibitemOpen
  \bibfield  {author} {\bibinfo {author} {\bibfnamefont {L.}~\bibnamefont
  {Tagliacozzo}}, \bibinfo {author} {\bibfnamefont {A.}~\bibnamefont {Celi}}, \
  and\ \bibinfo {author} {\bibfnamefont {M.}~\bibnamefont {Lewenstein}},\
  }\href {\doibase 10.1103/PhysRevX.4.041024} {\bibfield  {journal} {\bibinfo
  {journal} {Phys. Rev. X}\ }\textbf {\bibinfo {volume} {4}},\ \bibinfo {pages}
  {041024} (\bibinfo {year} {2014})}\BibitemShut {NoStop}%
\bibitem [{\citenamefont {Emonts}\ \emph {et~al.}(2020)\citenamefont {Emonts},
  \citenamefont {Bañuls}, \citenamefont {Cirac},\ and\ \citenamefont
  {Zohar}}]{emonts_variational_2020}%
  \BibitemOpen
  \bibfield  {author} {\bibinfo {author} {\bibfnamefont {P.}~\bibnamefont
  {Emonts}}, \bibinfo {author} {\bibfnamefont {M.~C.}\ \bibnamefont {Bañuls}},
  \bibinfo {author} {\bibfnamefont {I.}~\bibnamefont {Cirac}}, \ and\ \bibinfo
  {author} {\bibfnamefont {E.}~\bibnamefont {Zohar}},\ }\href {\doibase
  10.1103/PhysRevD.102.074501} {\bibfield  {journal} {\bibinfo  {journal}
  {Physical Review D}\ }\textbf {\bibinfo {volume} {102}},\ \bibinfo {pages}
  {074501} (\bibinfo {year} {2020})}\BibitemShut {NoStop}%
\bibitem [{\citenamefont {Robaina}\ \emph {et~al.}(2021)\citenamefont
  {Robaina}, \citenamefont {Bañuls},\ and\ \citenamefont
  {Cirac}}]{robaina_simulating_2021}%
  \BibitemOpen
  \bibfield  {author} {\bibinfo {author} {\bibfnamefont {D.}~\bibnamefont
  {Robaina}}, \bibinfo {author} {\bibfnamefont {M.~C.}\ \bibnamefont
  {Bañuls}}, \ and\ \bibinfo {author} {\bibfnamefont {J.~I.}\ \bibnamefont
  {Cirac}},\ }\href {\doibase 10.1103/PhysRevLett.126.050401} {\bibfield
  {journal} {\bibinfo  {journal} {Physical Review Letters}\ }\textbf {\bibinfo
  {volume} {126}},\ \bibinfo {pages} {050401} (\bibinfo {year}
  {2021})}\BibitemShut {NoStop}%
\bibitem [{\citenamefont {Bender}\ \emph {et~al.}(2020)\citenamefont {Bender},
  \citenamefont {Emonts}, \citenamefont {Zohar},\ and\ \citenamefont
  {Cirac}}]{bender_real-time_2020}%
  \BibitemOpen
  \bibfield  {author} {\bibinfo {author} {\bibfnamefont {J.}~\bibnamefont
  {Bender}}, \bibinfo {author} {\bibfnamefont {P.}~\bibnamefont {Emonts}},
  \bibinfo {author} {\bibfnamefont {E.}~\bibnamefont {Zohar}}, \ and\ \bibinfo
  {author} {\bibfnamefont {J.~I.}\ \bibnamefont {Cirac}},\ }\href {\doibase
  10.1103/PhysRevResearch.2.043145} {\bibfield  {journal} {\bibinfo  {journal}
  {Physical Review Research}\ }\textbf {\bibinfo {volume} {2}},\ \bibinfo
  {pages} {043145} (\bibinfo {year} {2020})}\BibitemShut {NoStop}%
\bibitem [{\citenamefont {Luo}\ \emph {et~al.}(2022)\citenamefont {Luo},
  \citenamefont {Yuan}, \citenamefont {Stokes},\ and\ \citenamefont
  {Clark}}]{luo_gauge_2022}%
  \BibitemOpen
  \bibfield  {author} {\bibinfo {author} {\bibfnamefont {D.}~\bibnamefont
  {Luo}}, \bibinfo {author} {\bibfnamefont {S.}~\bibnamefont {Yuan}}, \bibinfo
  {author} {\bibfnamefont {J.}~\bibnamefont {Stokes}}, \ and\ \bibinfo {author}
  {\bibfnamefont {B.~K.}\ \bibnamefont {Clark}},\ }\href {\doibase
  10.48550/arXiv.2211.03198} {\  (\bibinfo {year} {2022}),\
  10.48550/arXiv.2211.03198}\BibitemShut {NoStop}%
\bibitem [{\citenamefont {Chen}\ \emph {et~al.}(2022)\citenamefont {Chen},
  \citenamefont {Luo}, \citenamefont {Hu},\ and\ \citenamefont
  {Clark}}]{chen_simulating_2022}%
  \BibitemOpen
  \bibfield  {author} {\bibinfo {author} {\bibfnamefont {Z.}~\bibnamefont
  {Chen}}, \bibinfo {author} {\bibfnamefont {D.}~\bibnamefont {Luo}}, \bibinfo
  {author} {\bibfnamefont {K.}~\bibnamefont {Hu}}, \ and\ \bibinfo {author}
  {\bibfnamefont {B.~K.}\ \bibnamefont {Clark}},\ }\href {\doibase
  10.48550/arXiv.2212.06835} {\  (\bibinfo {year} {2022}),\
  10.48550/arXiv.2212.06835}\BibitemShut {NoStop}%
\bibitem [{\citenamefont {Polyakov}(1977)}]{polyakov_quark_1977}%
  \BibitemOpen
  \bibfield  {author} {\bibinfo {author} {\bibfnamefont {A.~M.}\ \bibnamefont
  {Polyakov}},\ }\href {\doibase DOI: 10.1016/0550-3213(77)90086-4} {\bibfield
  {journal} {\bibinfo  {journal} {Nucl. Phys. B}\ }\textbf {\bibinfo {volume}
  {120}},\ \bibinfo {pages} {429 } (\bibinfo {year} {1977})}\BibitemShut
  {NoStop}%
\bibitem [{\citenamefont {Gazit}\ \emph {et~al.}(2017)\citenamefont {Gazit},
  \citenamefont {Randeria},\ and\ \citenamefont
  {Vishwanath}}]{gazit_emergent_2017}%
  \BibitemOpen
  \bibfield  {author} {\bibinfo {author} {\bibfnamefont {S.}~\bibnamefont
  {Gazit}}, \bibinfo {author} {\bibfnamefont {M.}~\bibnamefont {Randeria}}, \
  and\ \bibinfo {author} {\bibfnamefont {A.}~\bibnamefont {Vishwanath}},\
  }\href {\doibase 10.1038/nphys4028} {\bibfield  {journal} {\bibinfo
  {journal} {Nature Physics}\ }\textbf {\bibinfo {volume} {13}},\ \bibinfo
  {pages} {484} (\bibinfo {year} {2017})}\BibitemShut {NoStop}%
\bibitem [{\citenamefont {Xu}\ \emph {et~al.}(2019)\citenamefont {Xu},
  \citenamefont {Qi}, \citenamefont {Zhang}, \citenamefont {Assaad},
  \citenamefont {Xu},\ and\ \citenamefont {Meng}}]{xu_monte_2019}%
  \BibitemOpen
  \bibfield  {author} {\bibinfo {author} {\bibfnamefont {X.~Y.}\ \bibnamefont
  {Xu}}, \bibinfo {author} {\bibfnamefont {Y.}~\bibnamefont {Qi}}, \bibinfo
  {author} {\bibfnamefont {L.}~\bibnamefont {Zhang}}, \bibinfo {author}
  {\bibfnamefont {F.~F.}\ \bibnamefont {Assaad}}, \bibinfo {author}
  {\bibfnamefont {C.}~\bibnamefont {Xu}}, \ and\ \bibinfo {author}
  {\bibfnamefont {Z.~Y.}\ \bibnamefont {Meng}},\ }\href {\doibase
  10.1103/PhysRevX.9.021022} {\bibfield  {journal} {\bibinfo  {journal}
  {Physical Review X}\ }\textbf {\bibinfo {volume} {9}},\ \bibinfo {pages}
  {021022} (\bibinfo {year} {2019})}\BibitemShut {NoStop}%
\bibitem [{\citenamefont {Glasser}\ \emph {et~al.}(2018)\citenamefont
  {Glasser}, \citenamefont {Pancotti}, \citenamefont {August}, \citenamefont
  {Rodriguez},\ and\ \citenamefont {Cirac}}]{glasser_neural-network_2018}%
  \BibitemOpen
  \bibfield  {author} {\bibinfo {author} {\bibfnamefont {I.}~\bibnamefont
  {Glasser}}, \bibinfo {author} {\bibfnamefont {N.}~\bibnamefont {Pancotti}},
  \bibinfo {author} {\bibfnamefont {M.}~\bibnamefont {August}}, \bibinfo
  {author} {\bibfnamefont {I.~D.}\ \bibnamefont {Rodriguez}}, \ and\ \bibinfo
  {author} {\bibfnamefont {J.~I.}\ \bibnamefont {Cirac}},\ }\href {\doibase
  10.1103/PhysRevX.8.011006} {\bibfield  {journal} {\bibinfo  {journal}
  {Physical Review X}\ }\textbf {\bibinfo {volume} {8}} (\bibinfo {year}
  {2018}),\ 10.1103/PhysRevX.8.011006}\BibitemShut {NoStop}%
\bibitem [{\citenamefont {Sorella}(2005)}]{sorella_wave_2005}%
  \BibitemOpen
  \bibfield  {author} {\bibinfo {author} {\bibfnamefont {S.}~\bibnamefont
  {Sorella}},\ }\href {\doibase 10.1103/PhysRevB.71.241103} {\bibfield
  {journal} {\bibinfo  {journal} {Physical Review B}\ }\textbf {\bibinfo
  {volume} {71}} (\bibinfo {year} {2005}),\
  10.1103/PhysRevB.71.241103}\BibitemShut {NoStop}%
\bibitem [{\citenamefont {Lieb}(1994)}]{PhysRevLett.73.2158}%
  \BibitemOpen
  \bibfield  {author} {\bibinfo {author} {\bibfnamefont {E.~H.}\ \bibnamefont
  {Lieb}},\ }\href {\doibase 10.1103/PhysRevLett.73.2158} {\bibfield  {journal}
  {\bibinfo  {journal} {Phys. Rev. Lett.}\ }\textbf {\bibinfo {volume} {73}},\
  \bibinfo {pages} {2158} (\bibinfo {year} {1994})}\BibitemShut {NoStop}%
\bibitem [{\citenamefont {Kogut}\ and\ \citenamefont
  {Susskind}(1975{\natexlab{b}})}]{PhysRevD.11.395}%
  \BibitemOpen
  \bibfield  {author} {\bibinfo {author} {\bibfnamefont {J.}~\bibnamefont
  {Kogut}}\ and\ \bibinfo {author} {\bibfnamefont {L.}~\bibnamefont
  {Susskind}},\ }\href {\doibase 10.1103/PhysRevD.11.395} {\bibfield  {journal}
  {\bibinfo  {journal} {Phys. Rev. D}\ }\textbf {\bibinfo {volume} {11}},\
  \bibinfo {pages} {395} (\bibinfo {year} {1975}{\natexlab{b}})}\BibitemShut
  {NoStop}%
\bibitem [{\citenamefont {Villain}(1975)}]{villain_theory_1975}%
  \BibitemOpen
  \bibfield  {author} {\bibinfo {author} {\bibfnamefont {J.}~\bibnamefont
  {Villain}},\ }\href {\doibase 10.1051/jphys:01975003606058100} {\bibfield
  {journal} {\bibinfo  {journal} {Journal de Physique}\ }\textbf {\bibinfo
  {volume} {36}},\ \bibinfo {pages} {581} (\bibinfo {year} {1975})}\BibitemShut
  {NoStop}%
\bibitem [{\citenamefont {Shi}\ \emph {et~al.}(2018)\citenamefont {Shi},
  \citenamefont {Demler},\ and\ \citenamefont
  {Ignacio~Cirac}}]{shi_variational_2018}%
  \BibitemOpen
  \bibfield  {author} {\bibinfo {author} {\bibfnamefont {T.}~\bibnamefont
  {Shi}}, \bibinfo {author} {\bibfnamefont {E.}~\bibnamefont {Demler}}, \ and\
  \bibinfo {author} {\bibfnamefont {J.}~\bibnamefont {Ignacio~Cirac}},\ }\href
  {\doibase 10.1016/j.aop.2017.11.014} {\bibfield  {journal} {\bibinfo
  {journal} {Annals of Physics}\ }\textbf {\bibinfo {volume} {390}},\ \bibinfo
  {pages} {245} (\bibinfo {year} {2018})}\BibitemShut {NoStop}%
\bibitem [{\citenamefont {Sorella}(2001)}]{sorella_generalized_2001}%
  \BibitemOpen
  \bibfield  {author} {\bibinfo {author} {\bibfnamefont {S.}~\bibnamefont
  {Sorella}},\ }\href {\doibase 10.1103/PhysRevB.64.024512} {\bibfield
  {journal} {\bibinfo  {journal} {Physical Review B}\ }\textbf {\bibinfo
  {volume} {64}} (\bibinfo {year} {2001}),\
  10.1103/PhysRevB.64.024512}\BibitemShut {NoStop}%
\bibitem [{\citenamefont {Metropolis}\ \emph {et~al.}(1953)\citenamefont
  {Metropolis}, \citenamefont {Rosenbluth}, \citenamefont {Rosenbluth},
  \citenamefont {Teller},\ and\ \citenamefont
  {Teller}}]{metropolis_equation_1953}%
  \BibitemOpen
  \bibfield  {author} {\bibinfo {author} {\bibfnamefont {N.}~\bibnamefont
  {Metropolis}}, \bibinfo {author} {\bibfnamefont {A.~W.}\ \bibnamefont
  {Rosenbluth}}, \bibinfo {author} {\bibfnamefont {M.~N.}\ \bibnamefont
  {Rosenbluth}}, \bibinfo {author} {\bibfnamefont {A.~H.}\ \bibnamefont
  {Teller}}, \ and\ \bibinfo {author} {\bibfnamefont {E.}~\bibnamefont
  {Teller}},\ }\href {\doibase 10.1063/1.1699114} {\bibfield  {journal}
  {\bibinfo  {journal} {The Journal of Chemical Physics}\ }\textbf {\bibinfo
  {volume} {21}},\ \bibinfo {pages} {1087} (\bibinfo {year}
  {1953})}\BibitemShut {NoStop}%
\bibitem [{\citenamefont {Heisenberg}(1928)}]{heisenberg_zur_1928}%
  \BibitemOpen
  \bibfield  {author} {\bibinfo {author} {\bibfnamefont {W.}~\bibnamefont
  {Heisenberg}},\ }\href {\doibase 10.1007/BF01328601} {\bibfield  {journal}
  {\bibinfo  {journal} {Zeitschrift fur Physik}\ }\textbf {\bibinfo {volume}
  {49}},\ \bibinfo {pages} {619} (\bibinfo {year} {1928})}\BibitemShut
  {NoStop}%
\bibitem [{\citenamefont {Holstein}\ and\ \citenamefont
  {Primakoff}(1940)}]{holstein_field_1940}%
  \BibitemOpen
  \bibfield  {author} {\bibinfo {author} {\bibfnamefont {T.}~\bibnamefont
  {Holstein}}\ and\ \bibinfo {author} {\bibfnamefont {H.}~\bibnamefont
  {Primakoff}},\ }\href {\doibase 10.1103/PhysRev.58.1098} {\bibfield
  {journal} {\bibinfo  {journal} {Physical Review}\ }\textbf {\bibinfo {volume}
  {58}},\ \bibinfo {pages} {1098} (\bibinfo {year} {1940})}\BibitemShut
  {NoStop}%
\end{thebibliography}%

\clearpage
\appendix
\onecolumngrid

\section{Details on the computation of the local electric energy \texorpdfstring{$H_{E,\text{loc}}(\theta)$}{H E loc(theta)}\label{sec:app_el_energy}}
The most difficult observables to compute in our variational Monte Carlo scheme are electric quantites, with the electric energy being its most prominent representative. In section~\ref{sec: evaluating expectation values}, we discussed the computation of $\expval{H_E}$ which according to eq.~\eqref{eq: electric_energy} consists of three parts: $\expval{H_E}=\expval{H_E}_{ff}+\expval{H_E}_{gg}+\expval{H_E}_{fg}$. 
To evaluate them one needs the local quantity $H_{E,\text{loc}}(\theta)$ that can sampled in a Monte Carlo simulation (see eq.~\eqref{eq: local_quantity O_loc}).
In the body of the manuscript we derived $H_{E,gg,\text{loc}}(\theta)$ and gave the final result for $H_{E,ff,\text{loc}}(\theta)$. In the following we present some details on the derivation of $H_{E,ff,\text{loc}}(\theta)$ and give the final result for the crossterm $H_{E,fg,\text{loc}}(\theta)$. 

The computation of  $H_{E,ff,\text{loc}}(\theta)$ involves deriving the fermionic Gaussian operator $U_{GS}(\theta)$ defined in eq.~\eqref{eq:psi_fermionic} w.r.t. $\theta_{\mathbf{x},i}$ which results in the form of $f_{\mathbf{x},i}(\theta)$ given in eq.~\eqref{eq: general form of f_theta}:  
\begin{equation}
\begin{aligned}
&\frac{1}{i} \left(\partial_{\theta_{\mathbf{x},i}} U_{GS}(\theta) \right) U_{GS}^{\dagger}(\theta)\\
&= \int_0^1 dt \exp \left( it \sum_{\mathbf{x},\mathbf{y}} \psi^{\dagger}_{\mathbf{x}} \xi(\theta)_{\mathbf{x}\mathbf{y}} \psi_{\mathbf{y}}   \right) \sum_{\mathbf{x},\mathbf{y}} \psi^{\dagger}_{\mathbf{x}} \frac{\partial \xi(\theta)_{\mathbf{x}\mathbf{y}}}{\partial \theta_{\mathbf{x},i }}  \psi_{\mathbf{y}} \exp \left( - it \sum_{\mathbf{x},\mathbf{y}} \psi^{\dagger}_{\mathbf{x}} \xi(\theta)_{\mathbf{x}\mathbf{y}} \psi_{\mathbf{y}}   \right) \\
&= \sum_{\mathbf{x},\mathbf{y},\mathbf{x'},\mathbf{y'}} \psi^{\dagger}_{\mathbf{x}} \int_0^1 dt \left[\exp \left( it  \xi(\theta)    \right) \right]_{\mathbf{x}\mathbf{x'}} \left[\frac{\partial \xi(\theta)}{\partial \theta_{\mathbf{x},i }} \right]_{\mathbf{x'}\mathbf{y'}} \left[\exp \left( - it  \xi(\theta)   \right) \right]_{\mathbf{y'}\mathbf{y}} \psi_{\mathbf{y}} \\
&=\vec{\psi}^{\dagger} \int_0^1 dt \exp \left( it  \xi(\theta)    \right) \frac{\partial \xi(\theta)}{\partial \theta_{\mathbf{x},i }} \exp \left( - it  \xi(\theta)   \right) \vec{\psi} \\
&=  \vec{\psi}^{\dagger} \frac{1}{i} \left( \partial_{\theta_{\mathbf{x},i}}  e^{i \xi(\theta)}  \right) e^{-i \xi(\theta)} \vec{\psi} \equiv \vec{\psi}^{\dagger} f_{\mathbf{x},i}(\theta) \vec{\psi}
\end{aligned}
\end{equation}
where $\vec{\psi}$ is a vector of the fermionic annihilation operators $\psi_{\mathbf{x}}$ and we used the identity: 
\begin{equation}
\partial_\theta e^{M(\theta)}  = \int_0^1 dt e^{t M(\theta)} \left(\partial_{\theta} M(\theta) \right) e^{-t M(\theta)} e^{M(\theta)} .
\end{equation}
If we carry out the second derivative $\partial_{\theta_{\mathbf{x},i}}$ we obtain an additional term corresponding to the derivative of $f_{\mathbf{x},i}(\theta)$ which vanishes. Thus, we remain with another contribution $\vec{\psi}^{\dagger} f_{\mathbf{x},i}(\theta) \vec{\psi}$ due to the derivative of $U_{GS}(\theta)$, resulting in the fermionic operator $H_{E,ff,\text{fer}}(\theta)$ in eq.~\eqref{eq:electric_ham_in_gauge}. 
 Inserting $\xi(\theta)=V(\theta) \tilde{\xi} V^{\dagger}(\theta)$ defined in eq.~\eqref{eq: xi_theta_definition} we derive an explicit expression for $f_{\mathbf{x},i}(\theta)$:
\begin{equation}
\begin{aligned}
 f_{\mathbf{x},i}(\theta)=&-i      \partial_{\theta_{\mathbf{x},i}} \left( e^{iV(\theta) \tilde{\xi} V^{\dagger}(\theta)} \right) e^{-iV(\theta) \tilde{\xi} V^{\dagger}(\theta)} \\
=& V(\theta)   \left( \frac{1}{i}  V^{\dagger}(\theta) \partial_{\theta_{\mathbf{x},i}} V(\theta)  +   e^{i \tilde{\xi} } \frac{1}{i} \underbrace{\partial_{\theta_{\mathbf{x},i}} V^{\dagger}(\theta) V(\theta)}_{=-V^{\dagger}(\theta) \partial_{\theta_{\mathbf{x},i}} V(\theta)} e^{-i \tilde{\xi}} \right)   V^{\dagger}(\theta)  \\
=& V(\theta) \left( \alpha^{\mathbf{x},i}(\theta) - e^{i \tilde{\xi} } \alpha^{\mathbf{x},i}(\theta) e^{-i \tilde{\xi}} \right) V^{\dagger}(\theta)  
\end{aligned}
\end{equation}
with $V(\theta)$ defined by the eigendecomposition of $h_{GM}(\theta)=V(\theta) \Lambda (\theta) V^{\dagger}(\theta)$ and $\alpha^{\mathbf{x},i}(\theta)= -i V^{\dagger}(\theta) \partial_{\theta_{\mathbf{x},i}} V(\theta)$ containing the derivatives of the eigenvectors of $h_{GM}(\theta)$. An explicit expression for $\alpha^{\mathbf{x},i}(\theta)$ can be derived by using a connection with the derivative of $h_{GM}(\theta)$:
\begin{equation}
\begin{aligned}
\alpha^{\mathbf{x},i}_{kl}(\theta)&=-i \frac{V^{\dagger}_{k\mathbf{x'}}(\theta) \frac{\partial  h_{GM}(\theta)_{\mathbf{x'} \mathbf{x''}}}  {\partial \theta_{\mathbf{x},i}} V(\theta)_{\mathbf{x''}l} }{\lambda_l(\theta) - \lambda_k(\theta)} \\
&=  \frac{ V^{\dagger}_{k\mathbf{x}}(\theta) e^{i\theta_{\mathbf{x},i}} V(\theta)_{\mathbf{x}+\mathbf{e}_i l}  - h.c.}{\lambda_l(\theta) - \lambda_k(\theta)}.
\end{aligned}
\end{equation}
Since the derivative of $h_{GM}(\theta)$ is non-zero only for sites adjacent to the link $\mathbf{x},i$ the expression for $\alpha^{\mathbf{x},i}(\theta)$ simplifies significantly. 
With the diagonal expressions for $ f_{\mathbf{x},i}(\theta)$ as given above, $H_{E,ff,\text{loc}}(\theta)$ is straightforwardly computed as explained in the body of the manuscript.

The derivation of $H_{E,fg,\text{loc}}(\theta)$ involves a derivative of $\Psi_G(\theta)$ which is expressed via the vector $b_{\mathbf{x},i}(\theta)$ defined in eq.~\eqref{eq: definition b_x,i} and a derivative of $\ket{\Psi_F(\theta)}$ resulting in a quadratic fermionic operator as discussed above:
 
\begin{equation}
\begin{aligned}
\expval{H_E}_{fg}=& \frac{g^2}{2} \sum_{\mathbf{x},i} \int D\theta \hspace{2pt} p(\theta)  2 \expval{  \vec{\psi}^{\dagger}  f_{\mathbf{x},i}(\theta) \vec{\psi} }  {\Psi_F(\theta)}  \left(i b^T (\theta) \alpha b_{\mathbf{x},i}(\theta) + i\beta^T b_{\mathbf{x},i}(\theta) \right) \\
=& \frac{g^2}{2} \sum_{\mathbf{x},i} \int D\theta \hspace{2pt} p(\theta) 2 \Tr \left( \left( \mathds{1} - \Gamma_{\psi \psi^{\dagger}}(\theta)  \right) f_{\mathbf{x},i}(\theta)   \right)  \left( i  b^T (\theta) \alpha b_{\mathbf{x},i}(\theta) + i  \beta^T b_{\mathbf{x},i}(\theta) \right) \\
\equiv& \int D\theta \hspace{2pt} p(\theta) H_{E,fg,\text{loc}}(\theta)  \\
\end{aligned}
\end{equation}
where we used the covariance matrix of the fermionic Gaussian state $\Gamma_{\psi \psi^{\dagger}}(\theta)$ as defined in eq.~\eqref{eq: eval with fermionc Gaussian}. 
For ground state studies as considered in this manuscript the variational parameters $\alpha$ and $\beta$ are chosen real such that the electric energy of the cross term vanishes.

\section{Details on the structure of \texorpdfstring{$\tilde{\xi}$}{xi} \label{sec:tilde_xi}}
In the following we provide details on the parametrization of the matrix $\tilde{\xi}_{ij}$ containing the fermionic variational parameters. 
Recall that $\tilde{\xi}$ is formulated in the eigenbasis of the gauge-matter Hamiltonian, $h_{GM}(\theta)_{\mathbf{x}\mathbf{y}}=V(\theta)_{\mathbf{x}i} \Lambda (\theta)_i V(\theta)^{\dagger}_{i \mathbf{y}}$. 
It allows to control the fermionic state in terms of eigenstates of the gauge-matter Hamiltonian.

In principle one can keep all parameters variational, however, one can simplify the structure of $\tilde{\xi}$ by considering the structure of the eigenstates as already discussed in section~\ref{sec:state_construction}. 
This can be emphasized by considering the two limits of the fermionic Hamiltonian $H_{\text{fer}}=H_E+H_{GM}$, i.e. the strong-coupling limit ($g^2 \to \infty$, $H_E$ dominates) and the weak-coupling limit ($g^2 \to 0$, $H_{GM}$ dominates) and how this fermionic state looks in terms of the eigenbasis of the gauge-matter Hamiltonian.  

In the strong-coupling limit the ground state is in a positional eigenstate, e.g. for one flavor all odd sites are occupied, $\ket{D}=\prod_{\mathbf{x} \in \mathcal{O}} \psi_{\mathbf{x}} \ket{0}$. 
This is already incorporated in the ansatz by setting the whole matrix $\tilde{\xi}$ to zero so that only the strong coupling reference state $\ket{D}$ remains but it is instructive to think of this state in terms of eigenstates of the gauge-matter Hamiltonian. 
Following the discussion in section~\ref{sec:state_construction} we can rewrite the state as 
\begin{equation}
\begin{aligned}
 \ket{D}& = \int D\theta \prod_i \frac{1}{\sqrt{2}} \left( \psi_{i+}^{\dagger}(\theta) - \psi_{i-}^{\dagger}(\theta) \right)  \ket{0}  
\end{aligned} 
\end{equation}
where we used the labeling of the eigenstates as in section~\ref{sec:state_construction} where $\psi_{i+}^{\dagger}(\theta) \ket{0}$ denotes the single-particle eigenstate of $H_{GM}$ with eigenvalue $\lambda_i(\theta)$ and, respectively,  $-\lambda_i(\theta)$ for $\psi_{i-}^{\dagger}(\theta) \ket{0}$.
On the other hand, in the weak-coupling limit the ground state is described by the occupation of all eigenstates with negative eigenvalue, i.e. the lower band,
\begin{equation}
\begin{aligned}
 \ket{\Psi_{0,GM}} = \int D\theta \prod_i   \psi_{i-}^{\dagger}(\theta) \ket{0} \\
\end{aligned} 
\end{equation}
Thus, one can smoothly transform from the strong coupling ground state to the weak coupling ground state by performing for every pair $i$ of single-particle eigenstates ($\psi_{i+}^{\dagger}(\theta)\ket{0}$ and $\psi_{i-}^{\dagger}(\theta)\ket{0}$) the transformation
 \begin{equation}
  \frac{1}{\sqrt{2}} \left( \psi_{i+}^{\dagger}(\theta) - \psi_{i-}^{\dagger}(\theta) \right)  \ket{0}    \to  \psi_{i-}^{\dagger}(\theta) \ket{0} .
 \end{equation}
Viewed in terms of the covariance matrix for the single-particle eigenstates $\psi_{i+}^{\dagger}(\theta)\ket{0}$ and $\psi_{i-}^{\dagger}(\theta)\ket{0}$, this amounts to $ \frac{1}{2} \left(\mathds{1} -  \sigma_x \right) \to \frac{1}{2} \left(\mathds{1} +  \sigma_z \right) $. 
This can be incorporated into $\tilde{\xi}$ by choosing the submatrix of $\tilde{\xi}$ related to the $\psi_{i+}^{\dagger}(\theta)\ket{0}$ and $\psi_{i-}^{\dagger}(\theta)\ket{0}$ eigenstates (a $2 \times 2$-matrix) as $\tilde{\xi}|_i = -\frac{\pi}{4} \sigma_y  \xi_i$ (where $\xi_i$ is a variational parameter) so that changing $\xi_i$ from zero to one smoothly transforms from the weak-coupling to the strong-coupling ground state. 
We thus end up with $N/2$ fermionic variational parameters ($N=L^2$ the number of lattice sites) and a block-diagonal form of $\tilde{\xi}$: 
\begin{equation}
\tilde{\xi} = 
\begin{pmatrix}
\tilde{\xi}|_1 & 0 & \hdots & 0 \\
0 & \tilde{\xi}|_2 & \ddots & \vdots \\
\vdots & \ddots & \ddots & 0 \\
 0 & \hdots & 0 & \tilde{\xi}|_{N/2} \\
\end{pmatrix}
\end{equation}

For the case of $N_f=2$ fermion flavors (and potentially even more flavors) one can choose a similar block-diagonal structure where one now blocks the single-particle eigenstates of both flavors together, i.e. $\psi_{1,i+}^{\dagger}(\theta)\ket{0},\psi_{1,i-}^{\dagger}(\theta)\ket{0},\psi_{2,i+}^{\dagger}(\theta)\ket{0},\psi_{2,i-}^{\dagger}(\theta)\ket{0}.$ The individual blocks $\tilde{\xi}|_i$ are then $4 \times 4$-matrices (or more generally $2N_f \times 2N_f$-matrices). 
The variational parametrization of these blocks is kept general as this allows to control certain properties between the two species, e.g. the imbalance $\Delta N$ between the two species as was used for the study of sign-problem affected regimes in section~\ref{sec:signproblem}. 
The block-diagonal structure allows even for multiple fermion flavors to compute the local energy $H_{\text{loc}}(\theta)$, in particular $H_{E,\text{loc}}(\theta)$, (see section~\ref{sec: measurement_procedure}) with a computational cost of only $\mathcal{O}(N^3)$.

\section{Choosing a non-Gaussian reference state for the strong-coupling limit\label{sec:app_heisenberg}}
Our fermionic state construction is based on a gauge-field dependent fermionic Gaussian operator $U_{GS}(\theta)$ acting on a strong-coupling reference state $\ket{\Psi_0}$ (see eq.~\eqref{eq: general fermionic ansatz}). 
In this section we show that this reference state can also be chosen non-Gaussian, using the example of $N_f=2$ fermionic species at half-filling as discussed in the body of the manuscript.
In the strong-coupling limit ($g^2 >> 1$) the lattice gauge theory reduces to an effective fermionic theory where the electric field vanishes to zeroth order. 
In second-order perturbation theory one can have virtual hopping processes between the two fermion species in opposite directions. 
This gives effectively rise to fermionic interactions as can be seen in our ansatz by the appearance of quartic expressions in the fermions (see eq.~\eqref{eq:electric_ham_in_gauge}). 
In the considered sector of one fermion per site this allows a mapping to a spin Hamiltonian, the Heisenberg model~\cite{heisenberg_zur_1928}. 
This can be incorporated in the ansatz by using a good approximation of the ground state of the effective fermionic theory (the Heisenberg model in our case) as reference state $\ket{\Psi_0}$. 
This ground state approximation can be obtained using any method of choice, e.g. tensor networks or spin wave theory (as was chosen in our case). 

The most difficult terms to evaluate in this scenario involve again quartic fermionic operators of the form 
\begin{equation} \label{eq: heisenberg_expval_psi0}
\begin{aligned}
\expval{   \vec{\psi}^{\dagger} f_{\mathbf{x},i,N_f=2}(\theta) \vec{\psi} \vec{\psi}^{\dagger} f_{\mathbf{x},i,N_f=2}(\theta) \vec{\psi} } {\Psi_0} \\
\end{aligned}
\end{equation}
with $\vec{\psi}=(\vec{\psi}_1,\vec{\psi}_2)^T$ now containing annihilation operators of both fermionic species. 
In the expression above we already performed the unitary transformation defined by $U_{GS}(\theta)$ resulting in a slightly different form of $f_{\mathbf{x},i}(\theta)$ compared to eq.~\eqref{eq: general form of f_theta}:
\begin{equation}
\begin{aligned}
f_{\mathbf{x},i,N_f=2}(\theta)=& \left({\begin{array}{cc}
	V(\theta) & 0 \\
	0 & V(\theta)\\
	\end{array} }\right)   \left[ e^{-i \tilde{\xi}}\left({\begin{array}{cc}
	\alpha^{\mathbf{x},i}(\theta) & 0 \\
	0 & \alpha^{\mathbf{x},i}(\theta)\\
	\end{array} }\right) e^{i \tilde{\xi} }- \left({\begin{array}{cc}
	\alpha^{\mathbf{x},i}(\theta) & 0 \\
	0 & \alpha^{\mathbf{x},i}(\theta)\\
	\end{array} }\right) \right]  \left({\begin{array}{cc}
	V^{\dagger}(\theta) & 0 \\
	0 & V(\theta)^{\dagger}\\
	\end{array} } \right) \\
 \equiv &  \left({\begin{array}{cc}
	f(\theta)_{11} & f(\theta)_{12} \\
	f(\theta)^{\dagger}_{12} & f(\theta)_{22}\\
	\end{array} }\right) 
 \end{aligned}
 \end{equation}
For ease of notation we dropped the subscripts for the submatrices of $f(\theta)$ in the last row. 
The fermionic operator $\vec{\psi}^{\dagger} f_{\mathbf{x},i,N_f=2}(\theta) \vec{\psi}$ can be written in the individual components of the fermionic species as
\begin{equation}
\vec{\psi}^{\dagger} f_{\mathbf{x},i,N_f=2}(\theta) \vec{\psi}= \vec{\psi}_1^{\dagger}f(\theta)_{11}\vec{\psi}_1+\vec{\psi}_2^{\dagger}f(\theta)_{22}\vec{\psi}_2+\vec{\psi}_1^{\dagger}f(\theta)_{12}\vec{\psi}_2+\vec{\psi}_2^{\dagger}f(\theta)_{12}^{\dagger}\vec{\psi}_1 .
\end{equation}
The fermionic expressions appearing in eq.~\eqref{eq: heisenberg_expval_psi0} then take the general form (explicitly writing out the site dependence): $\vec{\psi}_{\alpha \mathbf{x}}^{\dagger} f(\theta)_{\alpha \alpha ', \mathbf{x} \mathbf{x}'} \vec{\psi}_{\alpha ' \mathbf{x}'} \vec{\psi}_{\beta \mathbf{y}}^{\dagger} f(\theta)_{\beta  \beta ' , \mathbf{y} \mathbf{y}'} \vec{\psi}_{ \beta ' \mathbf{y}'} $.
Since the expression in eq.~\eqref{eq: heisenberg_expval_psi0} is evaluated w.r.t. the strong coupling vacuum ${\ket{\Psi_0}}$ we can project this expression onto the spin subspace with $\sum_{\alpha=1,2} \psi_{\alpha \mathbf{x}}^{\dagger} \psi_{\alpha \mathbf{x}}=1$ for all $\mathbf{x}$. 
This simplifies the expression since only combinations of fermionic operators remain that respect the single occupancy constraint, i.e. either $\mathbf{x}=\mathbf{x}'$ and $\mathbf{y}=\mathbf{y}'$ or $\mathbf{x}=\mathbf{y}'$ and $\mathbf{x}'=\mathbf{y}$. 
If we further keep only contributions that are known to be non-zero for the Heisenberg ground state, we can express the expectation value in eq.~\eqref{eq: heisenberg_expval_psi0} in terms of the spin correlations:
\begin{equation}
\begin{aligned}
&\expval{ \vec{\psi}^{\dagger} f_{\mathbf{x},i,N_f=2}(\theta) \vec{\psi} \vec{\psi}^{\dagger}f_{\mathbf{x},i,N_f=2}(\theta) \vec{\psi}} {\Psi_0} \\
=& \sum_{\mathbf{x},\mathbf{y}} \frac{1}{4} \left( 
|f(\theta)_{11,\mathbf{x}\mathbf{y}}|^2+|f(\theta)_{22,\mathbf{x}\mathbf{y}}|^2+ 2 |f(\theta)_{12,\mathbf{x}\mathbf{y}}|^2 + f(\theta)_{11,\mathbf{x}\mathbf{x}}f(\theta)_{11,\mathbf{y}\mathbf{y}} +f(\theta)_{22,\mathbf{x}\mathbf{x}}f(\theta)_{22,\mathbf{y}\mathbf{y}}+2f(\theta)_{11,\mathbf{x}\mathbf{x}}f(\theta)_{22,\mathbf{y}\mathbf{y}}
\right)\\
+&\sum_{\mathbf{x},\mathbf{y}} \expval{S^z_{\mathbf{x}}} \left( 
 f(\theta)_{11,\mathbf{x}\mathbf{x}} f(\theta)_{11,\mathbf{y}\mathbf{y}}-  f(\theta)_{22,\mathbf{x}\mathbf{x}} f(\theta)_{22,\mathbf{y}\mathbf{y}}+  f(\theta)_{11,\mathbf{x}\mathbf{x}} f(\theta)_{22,\mathbf{y}\mathbf{y}}- f(\theta)_{22,\mathbf{x}\mathbf{x}} f(\theta)_{11,\mathbf{y}\mathbf{y}}\right)\\
+&\sum_{\mathbf{x},\mathbf{y}} \expval{S^z_{\mathbf{x}} S^z_\mathbf{y}} \left( 
2 |f(\theta)_{12,\mathbf{x}\mathbf{y}}|^2-|f(\theta)_{11,\mathbf{x}\mathbf{y}}|^2-|f(\theta)_{22,\mathbf{x}\mathbf{y}}|^2 +f(\theta)_{11,\mathbf{x}\mathbf{x}}f(\theta)_{11,\mathbf{y}\mathbf{y}}+f(\theta)_{22,\mathbf{x}\mathbf{x}}f(\theta)_{22,\mathbf{y}\mathbf{y}}-2f(\theta)_{11,\mathbf{x}\mathbf{x}}f(\theta)_{22,\mathbf{y}\mathbf{y}}  
\right)\\
+&\sum_{\mathbf{x},\mathbf{y}} 2  \expval{S^{+}_{\mathbf{x}} S^{-}_\mathbf{y}} \left( 
f(\theta)_{12,\mathbf{x}\mathbf{x}}\overline{f(\theta)}_{12,\mathbf{y}\mathbf{y}} - f(\theta)_{11,\mathbf{x}\mathbf{y}}f(\theta)_{22,\mathbf{y}\mathbf{x}} 
\right)\\
\end{aligned}
\end{equation}
where all spin correlations are evaluated w.r.t. $\ket{\Psi_0}$. 
We chose spin wave theory to approximate the ground state of the Heisenberg model~\cite{holstein_field_1940}. 
One can even make the parameters of the spin waves variational and thus interpolate between a Gaussian state (the Neel state) and spin wave theory. 

\section{Update scheme of Monte Carlo algorithm \label{sec:app_montecarlo}}
In this section we provide some more details on the update scheme in our Monte Carlo algorithm. 
As the cost of updates is quite low, we are free to perform various types of updates. 

For local updates we perform the update of a link $\theta_{\mathbf{x},i}$ which changes the two plaquette variables that contain the link, for one of them the value of $\theta_{\mathbf{p}}$ is raised, for the other, respectively, lowered. 
One can extend this update scheme if one changes a second link variable in on one of the two plaquettes in such a way that it compensates the change due to $\theta_{\mathbf{x},i}$ and $\theta_{\mathbf{p}}$ is unchanged. 
Thus, only one of the two plaquettes containing the link variable $\theta_{\mathbf{x},i}$ will be updated and another plaquette that is next-nearest neighbor to it. 
Performing this procedure for all possible pairs, we update six pairs of next-nearest neighbor plaquettes. 

The global updates are related to changes in the gauge field configuration that are hard to obtain by iteratively applying local updates. 
It turned out that one such configuration is a change in all plaquettes $\theta_{\mathbf{p}}$ by $2 \pi/N$ and a change in a specific plaquette $\theta_{\mathbf{p}'}$ by $-2 \pi(1-1/N)$ with $N$ the number of plaquettes. 
The corresponding link configuration $\theta_{\mathbf{x},i}$ to create such a change in $\theta_{\mathbf{p}}$ can be computed via the lattice Green's function: 
\begin{equation}
\begin{aligned}
\phi^{\text{glob}}_{\mathbf{p}}&=\frac{1}{\sqrt{N}} \sum_{\mathbf{k}} e^{i \mathbf{k} \mathbf{p}} \frac{\tilde{Q}(\mathbf{k})}{4- 2 \cos \left( k_x \right)- 2 \cos \left( k_x \right)}   \\
\theta_{\mathbf{x},i}^{\text{glob}}&= \epsilon_{ij } \Delta^{(-)}_j \phi^{\text{glob}}_{\mathbf{p}} = \epsilon_{ij } \left( \phi^{\text{glob}}_{\mathbf{p}} - \phi^{\text{glob}}_{\mathbf{p}-\mathbf{e}_j} \right)
\end{aligned}
\end{equation}
where $\tilde{Q}(\mathbf{k})$ is the Fourier transform of $Q(\mathbf{p})$ which contains the desired changes in plaquette variables $\theta_{\mathbf{p}}$ that we want to create. 
In a first step a scalar field $\phi^{\text{glob}}_{\mathbf{p}}$ on the plaquettes is generated from $Q(\mathbf{p})$, from which one can derive the link variables $\theta_{\mathbf{x},i}^{\text{glob}}$ by applying the lattice curl to $\phi^{\text{glob}}$ which involves the plaquettes $\mathbf{p}$ and $\mathbf{p}-\mathbf{e}_j$ that contain the link $\mathbf{x},i$.
With the procedure above various kinds of global updates can be performed by choosing $Q(\mathbf{p})$ appropriately.

\begin{figure}
    \centering
    \includegraphics[width=0.5\columnwidth]{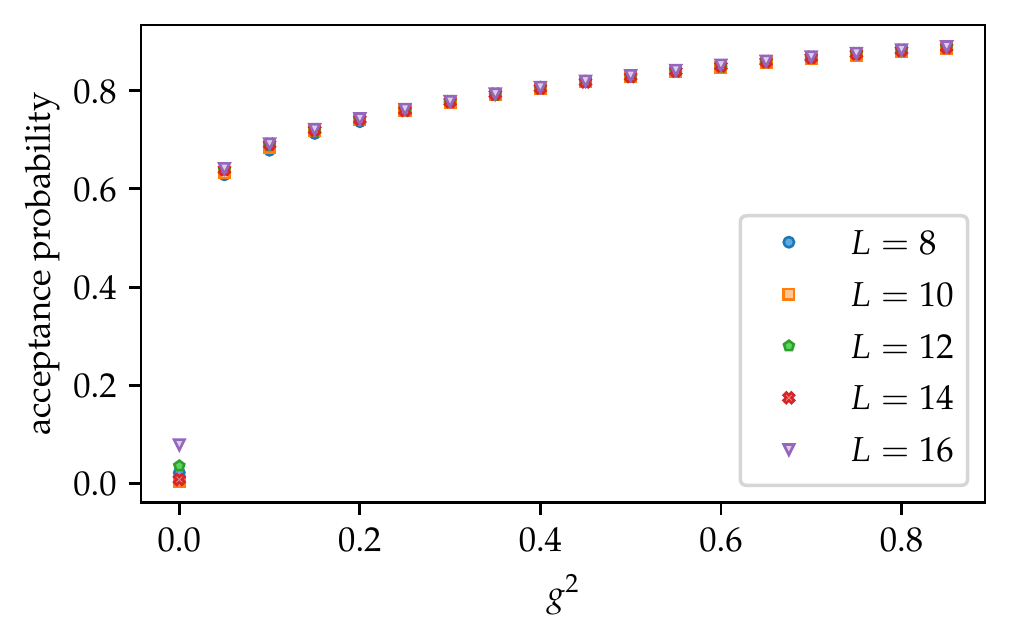}
    \caption{The acceptance probability in the variational ground states of the $N_f=2$ model. The acceptance probability is on a high level down to the lowest non-zero coupling and only for $g^2=0.0$ it sharply drops. This is expected since the ground state there is diagonal in $\theta$ with all plaquettes $\theta_{\mathbf{p}}$ having $\pi$ flux so that our variational probability, while approaching the delta distribution, gets lower and lower in acceptance probability.}
    \label{fig:acceptprob}
\end{figure}

As supporting evidence that our update scheme gives reasonable results we provide in Fig.~\ref{fig:acceptprob} the acceptance probability in the variational ground states of the $N_f=2$ model where we compared our results with Euclidean Monte Carlo results (see \ref{sec:benchmark}). 
Throughout the whole coupling region the acceptance probability is on a high level, except for $g^2=0.0$ where it is expected since the ground state is the $\pi$-flux state and our probability distribution approximates a delta distribution. 

\section{Details on gradient and Gram matrix\label{sec:gram}}
In this section we sketch the computation of the gradient of the variational energy and the Gram matrix using Monte Carlo simulation as required for the adaptation of the parameters with stochastic reconfiguration. 
Starting with the gradient, we first recall the form of the variational energy in terms of the local energy $H_{\text{loc}}(\theta)$: 
\begin{equation}
\begin{aligned}
    \frac{\expval{H}{\Psi}}{\braket{\Psi}}
    &=\frac{\int D\theta \hspace{2pt} H_{\text{loc}}(\theta)  |\Psi_G(\theta)|^2 } {\int D\theta \hspace{2pt} |\Psi_G(\theta)|^2    } = \frac{ \int D\theta \hspace{2pt} H_{\text{loc}}(\theta)   e^{- b^T(\theta) \alpha b(\theta) - 2\beta^T b(\theta)}}{\int D\theta \hspace{2pt} e^{- b^T(\theta) \alpha b - 2\beta^T b(\theta)}} = \int D\theta \hspace{2pt} H_{\text{loc}}(\theta) p(\theta)\\   
\end{aligned}
\end{equation}
The gradient will involve derivatives of $H_{\text{loc}}(\theta)$ w.r.t. all variational parameters and derivatives of $p(\theta)$ but these only w.r.t. the pure gauge parameters $\alpha$ and $\beta$. 
We provide the latter (denoted by the subscript $p(\theta)$) exemplary for the matrix element $\alpha_{\mathbf{p} \mathbf{p}'}$ (the expressions for $\beta$ are analogous but easier since the corresponding term in the exponential is only linear in $b(\theta))$: 
\begin{equation} \label{eq: gradient norm_energy}
 \frac{\partial \expval{E}_{p(\theta)}}{\partial \alpha_{\mathbf{p} \mathbf{p}'}}=\frac{ \int D\theta \hspace{2pt} H_{\text{loc}}(\theta)  \left(- b(\theta)_{\mathbf{p}} b(\theta)_{\mathbf{p}'} \right) e^{- b^T(\theta) \alpha b(\theta) - 2\beta^T b(\theta)}}{\int D\theta \hspace{2pt} e^{- b^T(\theta) \alpha b - 2\beta^T b(\theta)}} -   \frac{\expval{H}{\Psi}}{\braket{\Psi}} \frac{ \int D\theta \hspace{2pt}   \left(- b(\theta)_{\mathbf{p}} b(\theta)_{\mathbf{p}'} \right) e^{- b^T(\theta) \alpha b(\theta) - 2\beta^T b(\theta)}}{\int D\theta \hspace{2pt} e^{- b^T(\theta) \alpha b - 2\beta^T b(\theta)}}\\   
\end{equation}
where the first term is from the derivative of $|\Psi_G(\theta)|^2$ in the numerator and the second term from the denominatior. 
Both can be efficiently evaluated. 
Regarding derivatives of $H_{\text{loc}}(\theta)$, only $H_{E,\text{loc}}(\theta)$ depends on the variational parameters $\alpha$ and $\beta$ through the expression $b^T (\theta) \alpha b_{\mathbf{x},i}(\theta) + \beta^T b_{\mathbf{x},i}(\theta)$. 
Therefore derivatives are easily calculated, e.g. $b (\theta)_{\mathbf{p}} b_{\mathbf{x},i}(\theta)_{\mathbf{p}'}$ for the matrix element $\alpha_{\mathbf{p} \mathbf{p}'}$. 
The derivatives of $H_{\text{loc}}(\theta)$ w.r.t. the fermionic parameters $\xi_i$ are non-zero for $H_E$, $H_{GM}$ and $H_M$ and can be shown to take the schematic form
\begin{equation}
\frac{\partial}{\partial \xi_i} \Tr \left( \tilde{\Gamma} \left ( \tilde{\xi} \right)  \sum_{\mathbf{x},i} A_{\mathbf{x},i}(\theta)  \right) = i \sum_{kl} \left( \frac{\tilde{\xi}}{\partial \xi_i} \right)_{kl} \left( \tilde{\Gamma} \left ( \tilde{\xi} \right)   \sum_{\mathbf{x},i} A_{\mathbf{x},i}(\theta) - \sum_{\mathbf{x},i} A_{\mathbf{x},i}(\theta) \tilde{\Gamma} \left ( \tilde{\xi} \right)   \right )^T_{kl} 
\end{equation}
where $\tilde{\Gamma} \left ( \tilde{\xi} \right)$ is defined in eq.~\eqref{eq: eval with fermionc Gaussian} and $A_{\mathbf{x},i}(\theta)$ is some gauge-field dependent matrix containing a link dependence. 
Since only the right term (that is transposed) needs to be computed in a Monte Carlo simulation and the whole derivative can be post-processed, the computational cost of the derivatives scales the same as the computation of the variational energy.

For the computation of the Gram matrix $G_{ij} \equiv \braket{\Psi_i}{\Psi_j}$ it is useful to look at the tangent vectors first. 
The tangent vectors corresponding to the fermionic parameters $\xi_i$ are related to the single-particle eigenstates of the gauge-matter Hamiltonian and can therefore be shown to be orthogonal. 
The tangent vectors corresponding to $\alpha$  and $\beta$ are quadratic, respectively linear, in the vector $b(\theta)$ (defined for eq.~\eqref{eq: gauge_ansatz}):
\begin{equation}
\begin{aligned}
\ket{\Psi_{\alpha_{\mathbf{p},\mathbf{p'}}}}&=  \int D\theta \hspace{2pt} b(\theta)_{\mathbf{p}}  b(\theta)_{\mathbf{p}'} e^{- \frac{1}{2} b(\theta)^T \alpha b(\theta) - \beta^T b(\theta)} \ket{\Psi_F(\theta)} \ket{\theta} \\
\ket{\Psi_{\beta_{\mathbf{p}}}}&=  \int D\theta \hspace{2pt} b(\theta)_{\mathbf{p}}  e^{- \frac{1}{2} b(\theta)^T \alpha b(\theta) - \beta^T b(\theta)} \ket{\Psi_F(\theta)} \ket{\theta} 
\end{aligned}    
\end{equation}
The local quantity $O_{\text{loc}}(\theta)$ that needs to be sampled in a Monte Carlo simulation thus takes a simple form that is very similar to the gradient of the norm $\braket{\Psi}$ that needs to be computed for the gradient of the variational energy (see eq.~\eqref{eq: gradient norm_energy}). 
Since we use translational invariance to parametrize $\alpha$, the tangent vectors will be related to the Fourier components $\tilde{b}(\theta)_{\mathbf{k}}$ and not $b(\theta)_{\mathbf{p}}  b(\theta)_{\mathbf{p}'}$. 
The part of Gram matrix related to the overlaps between $\alpha$-tangent vectors will thus involve sampling $\tilde{b}(\theta)_{\mathbf{k}} \tilde{b}(\theta)_{\mathbf{k}'}$, thus being of size $\mathcal{O}(N^2)$ and efficiently tractable.
\end{document}